\documentclass[british]{article}
\usepackage[T1]{fontenc}
\usepackage[latin9]{inputenc}
\usepackage{geometry}
\usepackage{color}
\usepackage{babel}
\usepackage{mathrsfs}
\usepackage{bm}
\usepackage{amsmath}
\usepackage{amssymb}
\usepackage{graphicx}
\usepackage[unicode=true]
 {hyperref}

\makeatletter

\newcommand{\lyxdot}{.}

\newcommand{\lyxaddress}[1]{
\par {\raggedright #1
\vspace{1.4em}
\noindent\par}
}

\@ifundefined{date}{}{\date{}}
\usepackage{cite}
\usepackage{upgreek}

\makeatother

\begin{document}

\title{\textbf{Optical Supersymmetry in the Time Domain}}

\author{Carlos Garc\'ia-Meca\textsuperscript{$\ast$,$\dagger$}, Andr\'es
Macho Ortiz\textsuperscript{$\ast$,$\dagger$} and Roberto Llorente
S\'aez}
\maketitle

\lyxaddress{\begin{center}
{\small{}Nanophotonics Technology Centre, Universitat Polit\`ecnica
de Val\`encia, Valencia 46022, Spain}
\par\end{center}}

\noindent {\small{}$\ast$corresponding author: \href{mailto:cargarm2@ntc.upv.es}{cargarm2@ntc.upv.es},
\href{mailto:amachor@ntc.upv.es}{amachor@ntc.upv.es}}{\small \par}

\noindent {\small{}$\dagger$These authors have contributed equally
to this work}{\small \par}

\subsection*{Abstract}

Originally emerged within the context of string and quantum field
theory, and later fruitfully extra-polated to photonics, the algebraic
transformations of quantum-mechanical supersymmetry were\linebreak{}
conceived in the space realm. Here, we introduce a paradigm shift,
demonstrating that Maxwell\textquoteright s equations also possess
an underlying supersymmetry in the time domain. As a result, we obtain
a simple analytic relation between the scattering coefficients of
a large variety of time-varying optical systems and uncover a wide
new class of reflectionless, three-dimensional, all-dielectric, isotropic,
omnidirectional, polarization-independent, non-complex media. Temporal
supersymmetry is also shown to arise in dispersive media supporting
temporal bound states, which allows engineering their momentum spectra
and dispersive properties. These unprecedented features define a promising
design platform for free-space and integrated photonics, enabling
the creation of a number of novel reconfigurable reflectionless devices,
such as frequency-selective, polarization-independent and omnidirectional
invisible\linebreak{}
materials, compact frequency-independent phase shifters, broadband
isolators, and versatile pulse-shape transformers.
\noindent \begin{center}
\newpage{}
\par\end{center}

\subsection*{Introduction}

Supersymmetry (SUSY) emerged within the context of string and quantum
field theory as a mathematical framework for the unification of all
physical interactions of the universe \cite{key-P1,key-P2,key-P3,key-P4,key-P5}.
Attempting to solve fundamental questions about SUSY, scientists subsequently
created the field of supersymmetric quantum mechanics (SUSYQM), a
non-relativistic model for testing the aforementioned theories \cite{key-P6}.
Essentially, the one-dimensional (1D) version of SUSYQM considers
two different quantum-mechanical systems described by the eigenvalue
problems:
\begin{equation}
\widehat{\textrm{H}}_{1,2}\psi^{\left(1,2\right)}\left(x\right)=\Omega^{\left(1,2\right)}\psi^{\left(1,2\right)}\left(x\right),
\end{equation}
with Hamiltonians $\widehat{\textrm{H}}_{1,2}=-\alpha\textrm{d}^{2}/\textrm{d}x^{2}+V_{1,2}\left(x\right)$
and $\alpha\in\mathbb{R}^{+}$. The fundamental idea is to factorize
$\widehat{\textrm{H}}_{1}$ as $\widehat{\textrm{H}}_{1}=\widehat{\textrm{A}}^{+}\widehat{\textrm{A}}^{-}$
{[}$\widehat{\textrm{A}}^{\pm}:=\mp\sqrt{\alpha}\textrm{d}/\textrm{d}x+W\left(x\right)$
are known as the SUSY operators and $W$ as the superpotential{]},
and then construct the second (supersymmetric) Hamiltonian as $\widehat{\textrm{H}}_{2}=\widehat{\textrm{A}}^{-}\widehat{\textrm{A}}^{+}$.
The power of SUSYQM lies in the fact that $\widehat{\textrm{H}}_{2}$
yields the same eigenvalue spectrum and scattering properties as $\widehat{\textrm{H}}_{1}$.
For this reason, while evidence of SUSY in nature remains elusive
\cite{key-P7}, SUSYQM became of great interest in itself, leading
to the discovery of new analytically-solvable potentials, explaining
intriguing aspects of QM (such as the energy spectrum equality shared
by very different systems or the existence of non-trivial reflectionless
potentials), and offering a unique way to generate new families of
isospectral and reflectionless systems \cite{key-P6}. 

Interestingly, under specific circumstances, equation (1) also describes
the dynamics of electromagnetic waves (and of any physical phenomenon
governed by Helmholtz\textquoteright s equation), enabling a straightforward
application of SUSYQM theory in optics \cite{key-P8}. As a result,
notions of this formalism have been very recently utilized to design
ground-breaking photonic devices \cite{key-P9,key-P10,key-P11,key-P12,key-P13}.

Being a 1D theory, we asked ourselves whether a temporal supersymmetry
might exist for time-varying potentials. Nevertheless, to our knowledge,
the SUSYQM formalism has never been applied in the time domain, whether
in QM, optics, or any other field (SUSY quantum field theory is a
multidimensional spacetime theory, but the formalism is considerably
different and more complex than that of SUSYQM). This is probably
due to the fact that the vast majority of 1D SUSY work has been developed
within the realm of QM, and the time derivative in Schr\"{o}dinger\textquoteright s
equation is of first order, preventing a similar decomposition to
that of equation (1) in the time domain (time-dependent potentials
have been considered in SUSYQM, but also using SUSY operators based
on first-order spatial derivatives \cite{key-P14,key-P15}, making
it impossible to exploit the potential of the standard spatial SUSY
(S-SUSY) factorization in the time domain. In fact, none of the results
we will derive here could be obtained with such operators). On the
other hand, only a few works deal with optical SUSY, all focused on
S-SUSY. Remarkably, however, the fact that the temporal derivative
in the electromagnetic wave equation is of second order may enable
a temporal optical version of SUSYQM, which has been overlooked so
far. This paradigm shift would extend the foundations and unique properties
of SUSYQM to the time domain, adding an unprecedented degree of understanding
and control over time-varying photonic systems and opening the door
to a myriad of new applications. Actually, time-varying optical systems
are becoming crucial in a broad range of scenarios, including optical
modulation \cite{key-P16}, isolation and non-reciprocity \cite{key-P17,key-P18},
all-optical signal processing \cite{key-P19,key-P20}, quantum information
\cite{key-P21}, and reconfigurable photonics \cite{key-P22,key-P23}.

Here, we show that Maxwell\textquoteright s equations indeed possess
an underlying time-domain supersymmetry (T-SUSY) for any non-dispersive
optical system characterized by a refractive index of the form:
\begin{equation}
n\left(\mathbf{r},t\right)=n_{\textrm{S}}\left(\mathbf{r}\right)n_{\textrm{T}}\left(t\right)=\sqrt{\varepsilon_{\textrm{S}}\left(\mathbf{r}\right)\mu_{\textrm{S}}\left(\mathbf{r}\right)}\sqrt{\varepsilon_{\textrm{T}}\left(t\right)\mu_{\textrm{T}}\left(t\right)},
\end{equation}
where $\varepsilon_{\textrm{r}}\left(\mathbf{r},t\right)=\varepsilon_{\textrm{S}}\left(\mathbf{r}\right)\varepsilon_{\textrm{T}}\left(t\right)$
is the medium relative permittivity and $\mu_{\textrm{r}}\left(\mathbf{r},t\right)=\mu_{\textrm{S}}\left(\mathbf{r}\right)\mu_{\textrm{T}}\left(t\right)$
its relative permeability (T-SUSY can also be found in dispersive
systems, as discussed below, and in anisotropic and nonlocal media,
as discussed in Supplementary Section 1). In the following, we develop
this idea, illustrating its potential through different applications,
sketched in Fig.\,1.\newpage{}

\subsection*{Results}

Consider a linear, isotropic, heterogeneous, time-varying non-dispersive
medium with $n_{\textrm{T}}^{2}\left(t\right)=\varepsilon_{\textrm{T}}\left(t\right)$.
Applying separation of variables in the electric flux density $D\left(\mathbf{r},t\right)=\phi\left(\mathbf{r}\right)\psi\left(t\right)$
of Maxwell\textquoteright s equations, we find that $\psi\left(t\right)$
exactly obeys the Helmholtz\textquoteright s equation:
\begin{equation}
\left(\frac{\textrm{d}^{2}}{\textrm{d}t^{2}}+\omega_{0}^{2}N^{2}\left(t\right)\right)\psi\left(t\right)=0,
\end{equation}
where $N^{2}\left(t\right):=n_{-}^{2}/n_{\textrm{T}}\left(t\right)$,
$n_{-}:=n_{\textrm{T}}\left(t\rightarrow-\infty\right)$, and $\omega_{0}$
is the angular frequency (central angular frequency) at $t\rightarrow-\infty$
in the monochromatic (non-monochromatic) regime. The same equation
is obtained for $n_{\textrm{T}}^{2}\left(t\right)=\mu_{\textrm{T}}\left(t\right)$
and even for general materials with $n_{\textrm{T}}^{2}\left(t\right)=\varepsilon_{\textrm{T}}\left(t\right)\mu_{\textrm{T}}\left(t\right)$
{[}in which case, $\varepsilon_{\textrm{T}}\left(t\right)$ and/or
$\mu_{\textrm{T}}\left(t\right)$ must vary slowly in time{]}, see
Supplementary Section 1. Equation\,(3) exactly matches equation\,(1)
by taking $\alpha=1$, relabelling $x\rightarrow t$, and identifying
$\Omega-V\left(t\right)\equiv\omega_{0}^{2}N^{2}\left(t\right)$.
Using the eigenvalue $\Omega$ as a degree of freedom, this will allow
us to apply 1D SUSY in the time domain, with two fundamental noteworthy
features: 1) T-SUSY is exact for both all-dielectric and all-magnetic
indices $n_{\textrm{T}}$; 2) T-SUSY is completely uncoupled from
space. Hence, it is valid for all polarizations, all propagation directions
and any 3D spatial medium dependence $n_{\textrm{S}}^{2}\left(\mathbf{r}\right)=\varepsilon_{\textrm{S}}\left(\mathbf{r}\right)\mu_{\textrm{S}}\left(\mathbf{r}\right)$.
This means that we can generate T-SUSY partners of devices such as
waveguides or structures with any desired 3D scattering response while
keeping the spatial properties of interest (e.g., ability of guiding
or reflect/refract the fields in a specific way). Contrarily, 1D SUSYQM
is, by definition, only valid for 1D spatial variations, and only
for a specific polarization in the optical case \cite{key-P9,key-P10,key-P11}.
Moreover, T-SUSY can be used to study temporal scattering in systems
with continuous spectra, as well as time-varying systems supporting
discrete-spectrum bound states. In both cases, its application is
not as straightforward as that of S-SUSY. 

First, unlike in S-SUSY, to develop T-SUSY for wave scattering, the
concept of negative frequencies is essential. This comes from the
differences between spatial and temporal scattering, exemplified in
Fig.\,2 with a simple model having one spatial dimension. As is well
known, when a plane wave traverses a localized spatial variation in
a time-invariant medium, there appear reflected and transmitted waves
of the same frequency (photon energy), with the wave number (photon
momentum) of the incident ($k_{-}$), reflected ($k_{\textrm{R}}$)
and transmitted ($k_{+}$) waves fulfilling the Snell\textquoteright s
relations: $k_{\textrm{R}}=-k_{-}$ and $k_{-}/n_{-}=k_{+}/n_{+}$,
resulting from spatial symmetry breaking, Fig.\,2(a). Less known
is the fact that, when a wave propagates through a homogeneous medium,
reflections also appear under a localized time variation {[}Fig.\,2(c){]}.
In this case, since only time symmetry is broken, momentum is conserved
and photon energy changes, with the frequency of the incident ($\omega_{-}=\omega_{0}$),
reflected ($\omega_{\textrm{R}}$) and transmitted ($\omega_{+}$)
waves obeying the relations $\omega_{R}=-\omega_{+}$ and $n_{+}\omega_{+}=n_{-}\omega_{-}$
\cite{key-P24}. That is, light can exchange energy with the medium.
Notably, the frequencies of the reflected and transmitted waves have
opposite signs. Although the physical meaning of negative-frequency
waves is striking and controversial \cite{key-P25,key-P26}, mathematically,
the Hermiticity of the fields in $k$-$\omega$ space allows reinterpreting
a negative-frequency wave as a counter-propagating positive-frequency
one, leading to the standard use of only-positive frequencies. However,
the introduction of negative frequencies in this work is not a mere
convention. It is a mathematical tool that enables the analysis of
temporal scattering and, more importantly, a necessary ingredient
to relate the reflection and transmission coefficients of T-SUSY index
profiles, which otherwise cannot be decoupled (Supplementary Section
2). Concretely, for a given system with a temporal index $n_{\textrm{T}1}$,
T-SUSY provides a systematic way of generating a superpartner, whose
index is (Supplementary Section 2):

\begin{equation}
n_{\textrm{T}2}\left(t\right)=\frac{n_{2,-}}{\sqrt{\frac{n_{1,-}^{2}}{n_{\textrm{T}1}^{2}\left(t\right)}-\frac{2}{\omega_{0}^{2}}W^{\prime}\left(t\right)}},
\end{equation}
\newpage{}

\noindent where $n_{1,2,\pm}:=n_{\textrm{T}1,2}\left(t\rightarrow\pm\infty\right)$
is assumed to be constant. In this case, equation (3) admits asymptotic
solutions for $n_{\textrm{T}1,2}$ in the form of the following incident,
reflected and transmitted plane waves: $\psi_{\textrm{I}}^{\left(1,2\right)}\left(t\rightarrow-\infty\right)=e^{\textrm{i}\omega_{0}t}$,
$\psi_{\textrm{R}}^{\left(1,2\right)}\left(t\rightarrow\infty\right)=R_{1,2}e^{-\textrm{i}N_{+}\omega_{0}t}$
and $\psi_{\textrm{T}}^{\left(1,2\right)}\left(t\rightarrow\infty\right)=T_{1,2}e^{\textrm{i}N_{+}\omega_{0}t}$,
where $N_{+}:=n_{1,-}/n_{1,+}=n_{2,-}/n_{2,+}$. The combined use
of negative frequencies and T-SUSY then relates the reflection and
transmission coefficients of both media as (Supplementary Section
2):
\begin{equation}
\begin{aligned}R_{1}=\frac{W_{+}+\textrm{i}N_{+}\omega_{0}}{W_{-}-\textrm{i}\omega_{0}}R_{2}, & \ \ \ \ \ \ \ T_{1}=\frac{W_{+}-\textrm{i}N_{+}\omega_{0}}{W_{-}-\textrm{i}\omega_{0}}T_{2},\end{aligned}
\end{equation}
where $W_{\pm}:=W\left(t\rightarrow\pm\infty\right)$, $W$ follows
from the Riccati equation $V_{1,2}\left(t\right)=W^{2}\left(t\right)\mp W^{\prime}\left(t\right)$,
and $V_{1,2}\left(t\right)=\Omega-\omega_{0}^{2}N_{1,2}^{2}\left(t\right)$.
The first important consequence of equation (5) is that it can be
employed to obtain the reflection and transmission coefficients of
a large number of non-trivial time-varying optical media without solving
Maxwell\textquoteright s equations (specifically, of any supersymmetric
partner of a medium with a known response). As a second important
consequence, since $n_{\textrm{T}1}$ and $n_{\textrm{T}2}$ share
the same eigenvalue $\Omega$, $\left|R_{1}\right|=\left|R_{2}\right|$
and $\left|T_{1}\right|=\left|T_{2}\right|$. This allows us to readily
generate families of temporal index profiles exhibiting the same scattering
intensity as another medium (which can have any spatial variation),
synthesizable over the same ($n_{1,-}=n_{2,-}$) or different ($n_{1,-}\neq n_{2,-}$)
background materials, opening up a variety of applications.

As an example, consider the simplest case: a constant refractive index
$n_{1}\left(\mathbf{r},t\right)=n_{1,-}$. Its SUSY partner is $n_{2}\left(\mathbf{r},t\right)=n_{\textrm{T}2}\left(t\right)=n_{2,-}[1+2(\Omega/\omega_{0}^{2}-1)\textrm{sech}^{2}(\sqrt{\Omega-\omega_{0}^{2}}\thinspace t)]^{-1/2}$
{[}see Fig. 2(b){]}. The free parameters $n_{2,-}$ and $\Omega$
allow tailoring the asymptotic value of $n_{\textrm{T}2}$, as well
as its temporal width $\Delta t$ and maximal excursion $\Delta n$
(Supplementary Section 3). Since $n_{\textrm{T}1}$ is constant, $R_{1}=0$.
Therefore, $n_{\textrm{T}2}$ will also be reflectionless as demonstrated
in Fig.\,2(d) for a quasi-monochromatic optical pulse. From our previous
discussion, $n_{\textrm{T}2}$ represents a new class of all-dielectric
(all-magnetic), omnidirectional, isotropic, polarization-independent,
and transparent 3D media with real positive ($>1$) permittivity (permeability).
No known spatially-varying material possesses all these features,
including transformation media \cite{key-P27}, complex-parameter
materials \cite{key-P28}, and S-SUSY media \cite{key-P10}. The only
previously reported time-varying reflectionless media required simultaneously
time-varying permittivity and permeability values \cite{key-P29}.
Our T-SUSY proposal is completely different, since it is valid for
all-magnetic {[}$n_{2}^{2}=\mu_{\textrm{S}}(r)\mu_{\textrm{T}}(t)${]}
and all-dielectric materials {[}$n_{2}^{2}=\varepsilon_{\textrm{S}}(r)\varepsilon_{\textrm{T}}(t)${]},
see Supplementary Section 1. The latter are particularly important,
as implementing temporal permittivity variations is extremely easier
than implementing permeability ones. Another general feature of T-SUSY
is that it is only exact for the design frequency $\omega=\omega_{0}$.
Therefore, $n_{2}$ will be invisible ($R_{2}=0$, $\left|T_{2}\right|=1$)
for all directions and polarizations at $\omega_{0}$, while a reflected
wave will appear for $\omega\neq\omega_{0}$. Moreover, the spectral
span for which $n_{2}$ is almost invisible ($R_{2}\simeq0$ and $\left|T_{2}\right|\simeq1$)
can also be tailored via $\Omega$ (being considerably wide around
$\Omega=\omega_{0}^{2}$), enabling us to generate custom-made transparent
temporal windows within $n_{2}$ only for desired bands {[}Figs.\,1(a)
and 2{]}. Notably, out of the invisible band, all waves are (partially)
retroreflected along the input path, in contrast to spatial retroreflectors,
in which the reflected path is parallel to, but different from, the
input one \cite{key-P30}.

Additionally, $n_{2}$ presents an outstanding unexpected property:
the phases $\Phi_{R_{2}}$ and $\Phi_{T_{2}}$ of the reflected and
transmitted waves ($R_{2}=\left|R_{2}\right|e^{\textrm{i}\Phi_{R_{2}}}$,
$T_{2}=\left|T_{2}\right|e^{\textrm{i}\Phi_{T_{2}}}$) show a frequency-independent
response, also tuneable through $\Omega$ {[}Fig.\,2(e), Fig.\,S3.2{]}.
This makes $n_{2}$ a perfect dynamically-reconfigurable phase shifter:
polarization-independent, flat-frequency, reflectionless, and requiring
a short time variation ($<10\pi/\omega_{0}$) to generate any phase
shift $\in\left[0,\pi\right]$ (allowing a dramatically reduced device
length), paving the way to ideal ultra-compact optical modulators
{[}see Fig.\,1(b) and Supplementary Section 3{]}. Contrariwise, optical-path-based
phase shifters demand slow index variations (and thus long devices)
to be reflectionless, and are inherently frequency-dependent \cite{key-P31,key-P32}.
In contrast, the T-SUSY device can simultaneously introduce the same
phase shift over many spectral channels, which could be useful in,
e.g., wavelength-division multiplexing and frequency combs. Furthermore,
the non-linear behaviour of $\Phi_{T_{2}}$ may be employed to implement
pulse shaping operations (Supplementary Section 3). Additional media
can also be connected to a constant index via T-SUSY (being therefore
reflectionless), such as the hyperbolic Rosen-Morse II (HRMII) potential
{[}Figs.\,3(a), S3.7 and S3.8{]}, which has the advantage of allowing
an independent design control over $\Delta n$ for a fixed $\Delta t\sim40\pi/\omega_{0}$,
enabling a technology-oriented adjustment of the index modulation.

The previous results can be extended via different T-SUSY variants.
Firstly, isospectral T-SUSY deformations provide a root to obtain
$m$-parameter index families $\widetilde{n}_{\textrm{T}}(t;\eta_{1},\ldots,\eta_{m})$
with exactly the same scattering properties in module and phase as
another medium $n_{\textrm{T}}$ (Supplementary Section 2). As an
example, Fig.\,3(a) shows a reflectionless two-parameter family of
the HRMII index. Secondly, for some $n_{\textrm{T}1}$ profiles, the
shape invariance (SI) SUSY strategy allows us to construct T-SUSY
index chains $\{n_{\textrm{T}1}(t;a_{1}),n_{\textrm{T}2}(t;a_{1}),\text{\dots},n_{\textrm{T}m}(t;a_{1})\}$
satisfying the relations $n_{\textrm{T}m}(t;a_{1})\propto n_{\textrm{T}1}(t;a_{m})$,
$R_{m}(a_{1})=R_{1}(a_{m})$, $T_{m}(a_{1})=T_{1}(a_{m})$ and $a_{m}=f(a_{m-1})=(f\circ f)(a_{m-2})=(f\circ f\circ\ldots\circ f)(a_{1})$,
with $f$ a real function. Therefore, we can straightforwardly analyse
or design the temporal scattering properties of a large number of
time varying media. To illustrate the benefits of SI, consider the
following variation of the HRMII index ($\alpha$ is a real parameter):
\begin{equation}
n_{\textrm{T}1}\left(t;a_{1}\right)=\frac{n_{1,-}\left(a_{1}\right)}{\sqrt{1-\frac{2B}{\omega_{0}^{2}}+\frac{a_{1}\left(a_{1}+\alpha\right)}{\omega_{0}^{2}}\textrm{sech}^{2}\left(\alpha t\right)-\frac{2B}{\omega_{0}^{2}}\tanh\left(\alpha t\right)}},
\end{equation}
which is also reflectionless in a wide spectral band {[}Figs.\,3(c)
and S3.10{]}. Since $n_{1,-}\neq n_{1,+}$, the system performs a
frequency down-conversion with $\omega_{+}=(n_{1,-}/n_{1,+})\omega_{-}$,
where $n_{1,-}/n_{1,+}$ can be engineered via the design parameters
$\omega_{0}$ and $B$. A device exhibiting all these properties has
many potential applications. Unfortunately, the exotic shape (reaching
values below $n_{1,-}$) and large maximal excursion of $n_{\textrm{T}1}\left(t;a_{1}\right)$
hampers its experimental implementation. T-SUSY can overcome this
drawback by using SI. Specifically, equation (6) satisfies the SI
condition with $a_{m}=a_{1}-(m-1)\alpha$, allowing us to generate
different index profiles with the same reflectionless band as $n_{\textrm{T}1}\left(t;a_{1}\right)$
{[}Fig.\,3(b,c){]}. Taking $m=6$, we find an index $n_{\textrm{T}6}(t;a_{1})=(n_{6,-}\left(a_{1}\right)/n_{1,-}\left(a_{6}\right))n_{\textrm{T}1}(t;a_{6})$
with a considerably smoother time variation and a significantly lower
$\Delta n$. As illustrated in Fig.\,1(c), $n_{\textrm{T}6}(t;a_{1})$
can be used to build a polarization-independent optical isolator with
an ideally unlimited bandwidth (unlike previous time- and spacetime-modulated
isolators and frequency converters \cite{key-P17,key-P33,key-P34},
which, in addition, usually involve complicated non-omnidirectional
implementations), difficult to achieve by other means (Supplementary
Section 3 includes more details on this device and additional SI examples).
Reflectionless frequency converters can also be designed via transformation
optics, but their implementation requires extremely complex spacetime-varying
bianisotropic materials \cite{key-P35}.

Let us now discuss the discrete-spectrum case. This scenario naturally
arises in optical S-SUSY. Specifically, when applying 1D SUSYQM to
a spatial dimension normal to the propagation direction, the propagation
constant enters the wave equation as an effective energy, which is
quantized by the eigenvalue problem \cite{key-P9,key-P10}. However,
since time is unidimensional, there is no possible quantity playing
the role of an energy in T-SUSY, leading to free-particle systems.
Outstandingly, a discrete-spectrum version of T-SUSY can be developed
for time-varying dispersive media. To this end, we require the concept
of temporal waveguide (TWG): two adjacent temporal index boundaries
(allowed to move at a speed $v_{\textrm{B}}$) defining a (position-dependent)
temporal index window, which can confine and carry optical pulses
by temporal total internal reflection \cite{key-P36,key-P37,key-P38}.
TWGs can be created by inducing a perturbation $\Delta n_{\textrm{eff}}(t-z/v_{\textrm{B}})$
of the effective index $n_{\textrm{eff}}$ of a given mode in a spatial
waveguide. In a co-moving reference frame, the complex envelope of
the electric field associated with a TWG can be written as $A\left(z,\tau\right)=\sum_{n}\psi_{n}\left(\tau\right)e^{\textrm{i}(\Delta\beta_{1}/\beta_{2})\tau}e^{\textrm{i}K_{n}z}$
\cite{key-P37,key-P38}. It is then shown that a TWG supports temporal
bound states $\psi_{n}$ ($n\in\left\{ 0,1,2,\ldots\right\} $) fulfilling
the discrete-spectrum eigenvalue equation:
\begin{equation}
\left(-\frac{\textrm{d}^{2}}{\textrm{d}\tau^{2}}+2\frac{\beta_{\textrm{B}}\left(\tau\right)}{\beta_{2}}\right)\psi_{n}\left(\tau\right)=\left(2\frac{K_{n}}{\beta_{2}}+\frac{\Delta\beta_{1}^{2}}{\beta_{2}^{2}}\right)\psi_{n}\left(\tau\right),
\end{equation}
where $\tau:=t-z/v_{\textrm{B}}$, $\beta_{1}$ and $\beta_{2}$ are
the inverse group velocity and group-velocity dispersion constant
of the perturbed spatial mode, $\Delta\beta_{1}:=\beta_{1}-1/v_{\textrm{B}}$,
$\beta_{\textrm{B}}\left(\tau\right)=k_{0}\Delta n_{\textrm{eff}}\left(\tau\right)$,
$k_{0}=\omega_{0}/c_{0}$, and $\omega_{0}$ is the optical carrier
angular frequency. We can apply T-SUSY to equation (7), as it matches
equation (1) for $\alpha=1$, $x\rightarrow\tau$, $V\left(\tau\right)\equiv2\beta_{\textrm{B}}\left(\tau\right)/\beta_{2}$
and $\Omega_{n}\equiv2K_{n}/\beta_{2}+\Delta\beta_{1}^{2}/\beta_{2}^{2}$,
expanding the TWG landscape and its potential applications.

As an example, consider an analytically-solvable TWG with a step temporal
perturbation $\beta_{\textrm{B}1}$. Its unbroken T-SUSY partner is
$\beta_{\textrm{B}2}\left(\tau\right)=\beta_{\textrm{B}1}\left(\tau\right)-\beta_{2}(\ln\psi_{0}^{\left(1\right)}\left(\tau\right))^{\prime\prime}$,
where $\psi_{0}^{\left(1\right)}$ is the ground state (fundamental
mode) of $\beta_{\textrm{B}1}$ {[}Fig.\,4(a), Supplementary Section
4{]}. From T-SUSY theory, both TWGs have the same energy spectrum.
Moreover, $\widehat{\textrm{A}}^{-}$ ($\widehat{\textrm{A}}^{+}$)
maps each state $\psi_{n}^{\left(1\right)}$ ($\psi_{n}^{\left(2\right)}$)
of $\beta_{\textrm{B}1}$ ($\beta_{\textrm{B}2}$) into a state of
$\beta_{\textrm{B}2}$ ($\beta_{\textrm{B}1}$) having the same eigenvalue
$\Omega_{n}$, with the exception of the ground state $\psi_{0}^{\left(1\right)}$,
which is annihilated by $\widehat{\textrm{A}}^{-}$ ($\widehat{\textrm{A}}^{-}\psi_{0}^{\left(1\right)}=0$)
and thus has no equal-energy counterpart in $\beta_{\textrm{B}2}$.
Particularly, $\psi_{n}^{\left(2\right)}\propto\widehat{\textrm{A}}^{-}\psi_{n+1}^{\left(1\right)}$,
where $\widehat{\textrm{A}}^{-}:=\textrm{d}/\textrm{d}\tau-(\ln\psi_{0}^{\left(1\right)}\left(\tau\right))^{\prime}$.
In T-SUSY, energy is related to phase constant, implying that $\psi_{0}^{\left(1\right)}$
is not phase-matched with any $\psi_{n}^{\left(2\right)}$ and that
$\psi_{n+1}^{\left(1\right)}$ and $\psi_{n}^{\left(2\right)}$ are
perfectly phase-matched. This occurs in an extremely large optical
bandwidth $\Delta\nu\sim0.5$ ($\nu\propto1/\beta_{2}$ is the normalized
frequency), provided that both TWGs are built on a dispersion-flattened
spatial waveguide ($\textrm{d}\beta_{2}/\textrm{d}\omega\simeq0$),
see Fig.\,4(b). Furthermore, Fig.\,4(b) reveals that $\beta_{\textrm{B}2}$
is less dispersive than $\beta_{\textrm{B}1}$, i.e., T SUSY enables
us to engineer the dispersion properties of TWGs.

To further unfold the potential of discrete-spectrum T-SUSY, we propose
the concept of \emph{temporal photonic lantern} (TPL): close-packed
serial T-SUSY TWGs moving at the same speed and supporting linear
combinations of degenerate temporal bound sates (supermodes) \cite{key-P39}.
To verify the TPL concept, we have developed the first version of
coupled-mode theory (CMT) for serial TWGs (Supplementary Section 4).
Fig.\,4(c) shows an example of a TPL supermode. Remarkably, TWGs
can carry soliton-like (shape-invariant) optical pulses in a dispersive
medium, with the advantage of allowing arbitrary pulse amplitude and
duration, as well as a tuneable propagation speed \cite{key-P37,key-P38}.
TPLs extend this ability to a serial combination of modes (each with
arbitrary length, amplitude and node number), yielding solitonic supermodes
with almost any desired shape. Achieving the required perfect phase-matching
between modes of different order in serial TWGs without T-SUSY typically
demands neighbouring TWGs of different width, whilst T-SUSY permits
an independent control over this parameter and generally presents
a much larger normalized phase-matching bandwidth \cite{key-P12},
inherently implying a higher tolerance to fluctuations in $T_{\textrm{B}}$
and $\Delta n_{\textrm{eff}}$. More advanced functionalities emerge
by noting that if only one of the TWGs of the TPL is excited, a periodic
energy transfer between adjacent TWGs occurs {[}Supplementary Section
4, Fig.\,4(d){]}. Using two coupled spatial waveguides (WG1, WG2),
this effect enables the construction of a pulse shape transformer
with unprecedented versatility and reconfigurable capability {[}Fig.\,1(d){]}.
Particularly, the final shape of each pulse propagating along WG1
can be dynamically chosen among a large gamut by launching the appropriate
TPL over WG2. The proposed T-SUSY TPLs could also find application
in optical wavelet transforms, coherent laser control of physicochemical
and QM processes, spectrally-selective nonlinear microscopy, and mathematical
computing \cite{key-P40,key-P41,key-P42,key-P43}. T-SUSY TWG theory
can be extended via temporal analogues of SI, broken SUSY and isospectral
constructions \cite{key-P6}.

\subsection*{Discussion}

Overall, these results generalize the foundations of SUSYQM to the
time domain,\,unveiling\,the\,temporal\linebreak{}
supersymmetric nature of Maxwell\textquoteright s equations (which,
unlike S-SUSY, has no previous direct analogue) and, consequently,
leading to the emergence of an entire field of research within optics,
as well as to a new photonic design toolbox. Compared to S-SUSY, T-SUSY
relaxes the need for controlling the polarization state of light and
the medium spatial index variation, which usually involves complex
fabrication steps \cite{key-P9,key-P11}. A possible T-SUSY technological
difficulty might arise if high temporal index excursions are desired
(e.g. to achieve a large phase variation), which can be circumvented
via strong nonlinear media such as indium tin oxide \cite{key-P44},
or multi-level phase-change materials \cite{key-P45}, such as germanium-antimony-tellurium,
indium antimonide and vanadium dioxide \cite{key-P46,key-P47,key-P48}.
Note that extremely low index perturbations suffice to create T-SUSY
TWGs, technologically feasible in standard optical fibres and waveguides
via traveling-wave electro-optic phase modulators or the cross-phase
modulation effect \cite{key-P26,key-P37,key-P38}. Finally, it is
worth mentioning that the sound pressure satisfies a temporal Helmholtz
equation formally equal to equation (3) (Supplementary\,Section\,6).
Hence, T-SUSY can be directly transferred to acoustics.\newpage{}

\subsection*{Methods}

Numerical simulations of the temporal scattering problem (Figs.\,2
and 3) have been performed by solving Eq.\,(S1.8) of the Supplementary
Section 1 with COMSOL Multiphysics, taking $\omega_{0}=38$ rad/s
and $c_{0}=1$ m/s to guarantee a low computational time. However,
the conclusions derived from Figs.\,2 and 3 are valid for any value
of $\omega_{0}$ and $c_{0}$ (see pages \pageref{sec:3} and \pageref{eq:S3.21}
of the Supplementary Information for more details). On the other hand,
the modal analysis of the TWGs and the TPL (Fig.\,4) has been calculated
with CST Microwave Studio and MATLAB by using the analogy reported
in \cite{key-P37} between a dielectric slab waveguide and a TWG.
Finally, the numerical simulation of Fig.\,4(d), based on the CMT
derived in Supplementary Section 4 for serial TWGs, was performed
in MATLAB. In Supplementary Section 5 we include a detailed discussion
about the numerical methods employed in Fig.\,4.

\subsection*{Acknowledgements}

This work was supported by Spanish National Plan projects TEC2015-73581-JIN
PHUTURE\linebreak{}
(AEI/FEDER, UE) and MINECO/FEDER UE XCORE TEC2015-70858-C2-1-R, and
Generalitat Valenciana Plan project NXTIC AICO/2018/324. A. M.\textquoteright s
work was supported by BES-2013-062952 F.P.I. Grant.

\subsection*{Author contributions}

C.G.-M. conceived the idea of temporal SUSY. C.G.-M. and A.M.O. developed
the theory, performed the numerical simulations and analysed the data.
R.L.S. supervised the work. All authors contributed to write the manuscript.

\newpage{}

\noindent \begin{center}
\newpage{}
\par\end{center}

\noindent \begin{center}
\includegraphics[width=12cm,height=14cm,keepaspectratio]{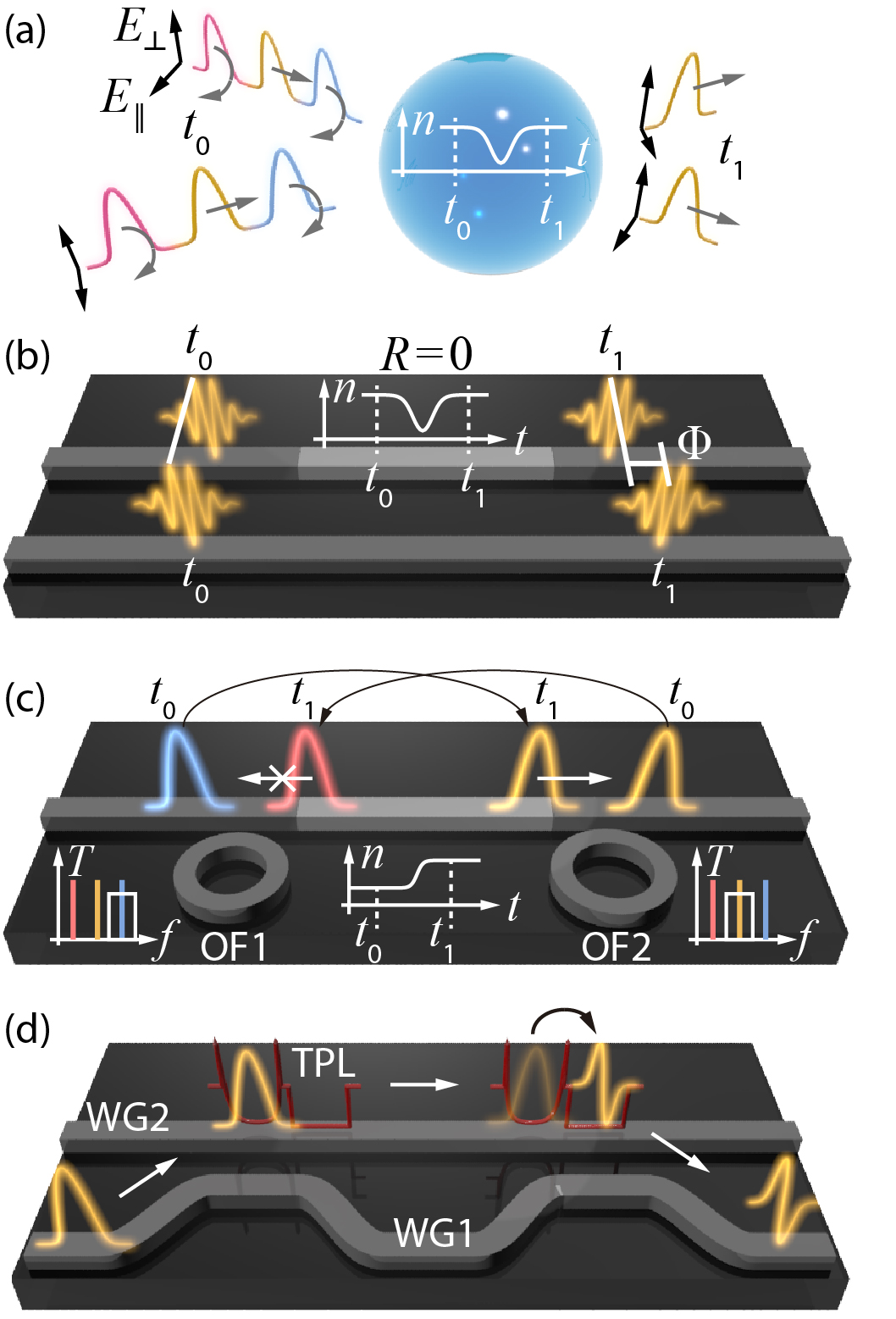}
\par\end{center}

\noindent \textbf{Figure\,1.\,Potential\,T-SUSY\,applications.}\,(a)\,Omnidirectional,\,isotropic,\,polarization-independent,
all-dielectric (all-magnetic), and 3D time-varying material that is
invisible in a given (reconfigurable) spectral region, allowing the
generation of frequency-selective transparent temporal windows. (b)\,Perfect
phase shifter: a short region in a waveguide (lighter grey) with a
fast T-SUSY time-varying index inducing a dynamically-reconfigurable
frequency-independent reflectionless phase shift $\Phi$ over an optical
pulse. (c) Optical isolator: another T-SUSY time-varying index shifts
the frequency of a right-propagating pulse (blue to yellow here),
which can traverse the optical filters OF1 and OF2. Any left-propagating
pulse is reflected at OF1 or OF2, protecting a left-side source from
external reflections. (d) Reconfigurable pulse-shape transformer:
an input pulse propagating along waveguide WG1 is spatially coupled
to a T-SUSY temporal photonic lantern (TPL, a moving index perturbation)
running along waveguide WG2. The pulse excites another TPL mode with
a different desired shape, coupled back to WG1.
\noindent \begin{center}
\newpage{}
\par\end{center}

\noindent \begin{center}
\includegraphics[width=12cm,height=15cm,keepaspectratio]{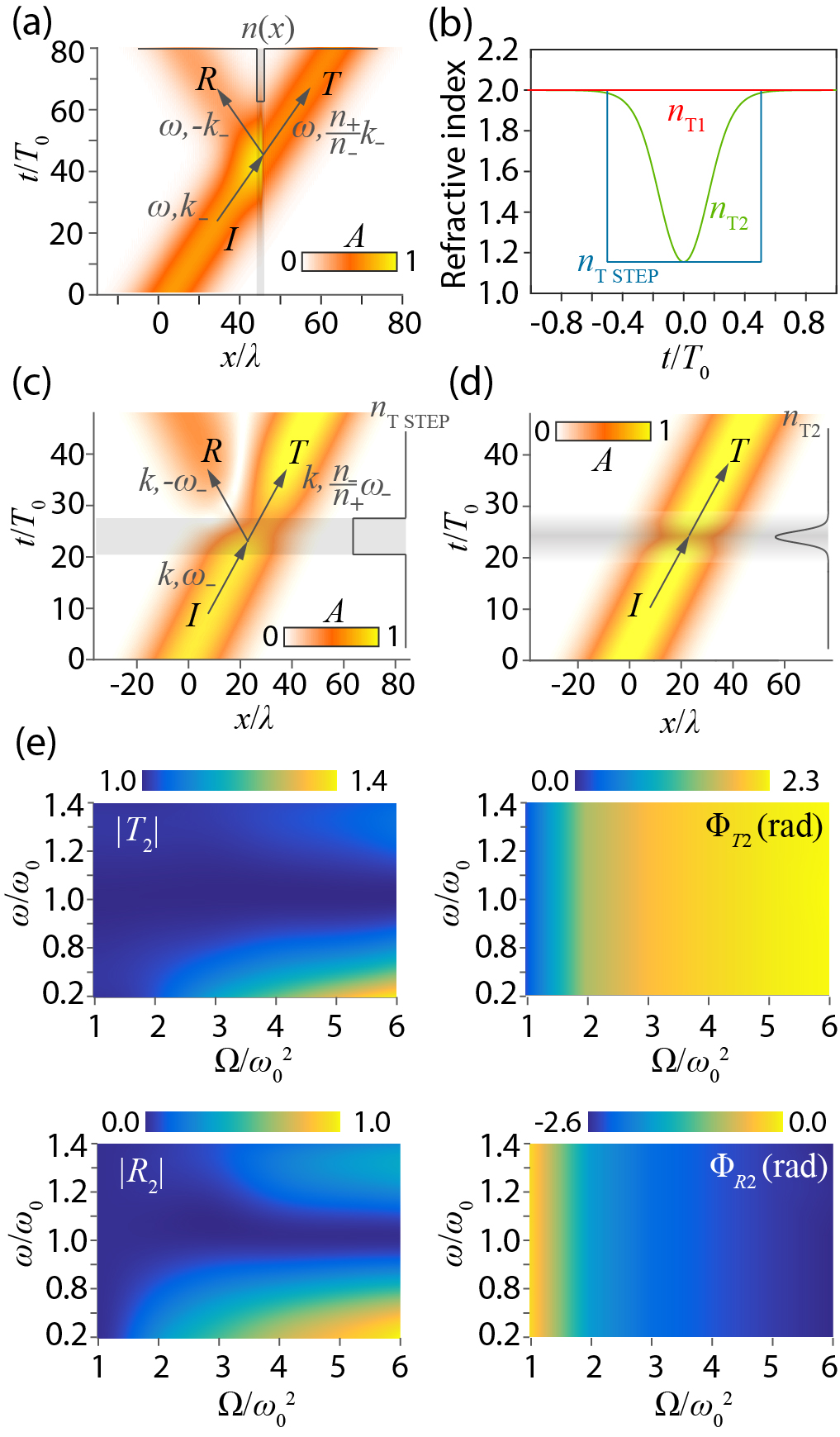}
\par\end{center}

\noindent \textbf{Figure 2.} \textbf{Reflectionless all-dielectric
(all-magnetic) T-SUSY time-varying optical media.} (a) Example of
spatial reflection for an optical beam propagating through a time-invariant
spatial step-index medium. (b) Index profiles of homogeneous media
characterized by a temporal step-index $n_{\textrm{T STEP}}\left(t\right)$
(reflective), a constant index $n_{\textrm{T}1}\left(t\right)$ (non-reflective),
and its T-SUSY partner $n_{\textrm{T}2}\left(t\right)$ (also non-reflective),
with $n_{2,-}=2$ and $\Omega=2\omega_{0}^{2}$ ($T_{0}=2\pi/\omega_{0}$).
(c,d) Pulse propagation evolution at $\omega=\omega_{0}$ through
the media with index profiles $n_{\textrm{T STEP}}\left(t\right)$
in (c) and $n_{\textrm{T}2}\left(t\right)$ in (d). Here, $\lambda=\lambda_{0}/n_{2,-}$
and $\lambda_{0}=2\pi c_{0}/\omega_{0}$. In contrast to (c), the
result in (d) demonstrates the reflectionless nature of $n_{\textrm{T}2}$.\linebreak{}
(e) Scattering coefficients $T_{2}=\left|T_{2}\right|\exp(\textrm{i}\Phi_{T_{2}})$
and $R_{2}=\left|R_{2}\right|\exp(\textrm{i}\Phi_{R_{2}})$ of $n_{\textrm{T}2}$
as a function of $\omega/\omega_{0}$ and $\Omega/\omega_{0}^{2}$.
\noindent \begin{center}
\newpage{}
\par\end{center}

\noindent \begin{center}
\includegraphics[width=9cm,height=14cm,keepaspectratio]{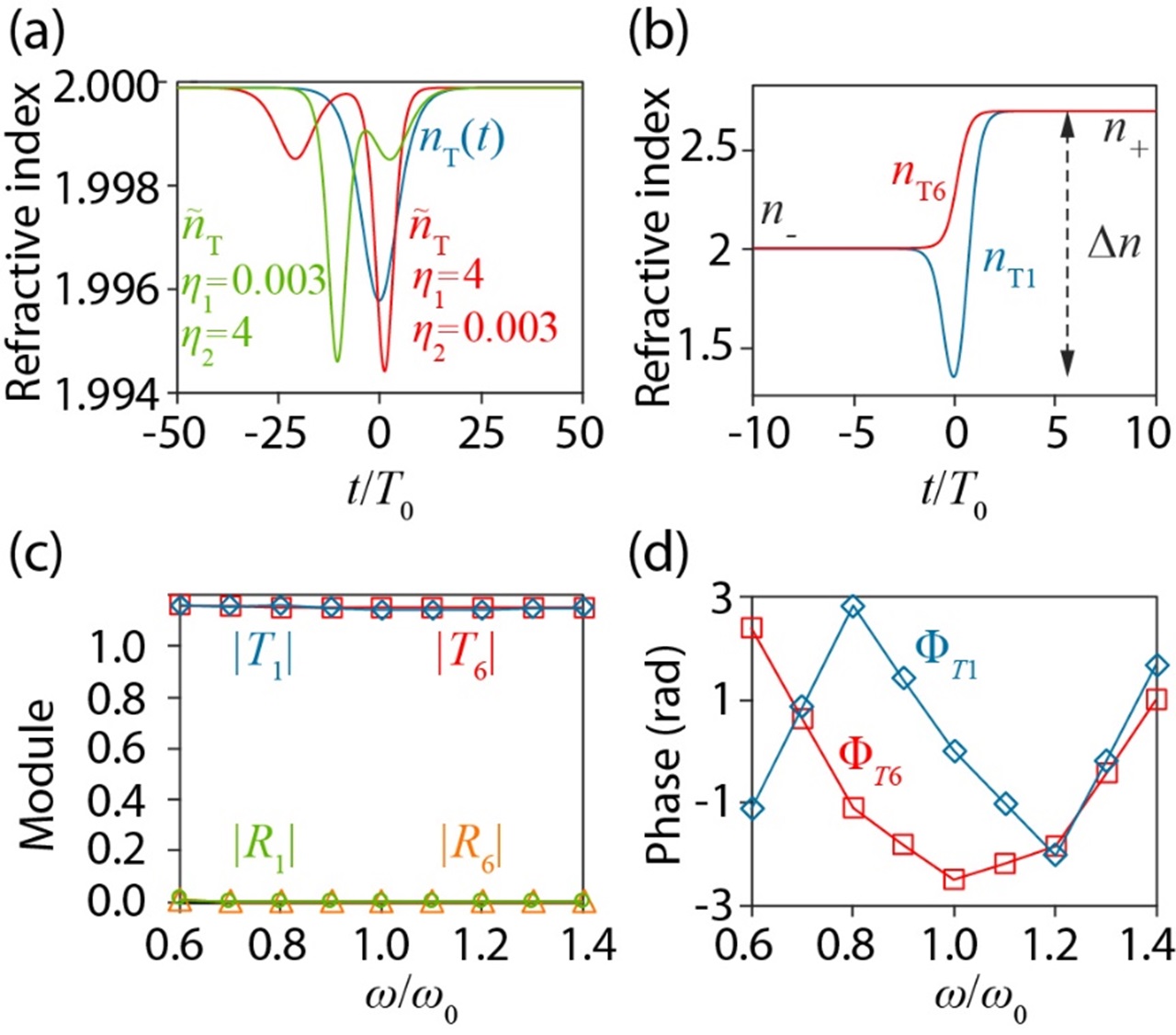}
\par\end{center}

\noindent \textbf{Figure 3.} \textbf{T-SUSY reflectionless isospectral
media and frequency converter. }(a) Two members of the 2-parameter
isospectral family $\widetilde{n}_{\textrm{T}}\left(t;\eta_{1},\eta_{2}\right)$
of the HRMII index $n_{\textrm{T}}\left(t\right)$. Numerical calculations
show that, in all cases, $R\left(\eta_{1},\eta_{2}\right)=0$ and
$T\left(\eta_{1},\eta_{2}\right)=e^{\textrm{i}0.23}$ (see Supplementary
Section\,3). (b) Supersymmetric refractive index profile $n_{\textrm{T}1}(t;a_{1})$
of equation (6) with $a_{1}=40$, $B=a_{1}^{2}/10$, $\alpha=10$,
$\omega_{0}=38$ rad/s and $n_{1,-}\left(a_{1}\right)=2$. The 6-th
order index of its corresponding SI chain $n_{\textrm{T}6}(t;a_{1})$
is also depicted. Both media induce a reflectionless frequency down-conversion
in any incident optical signal. (c,d) Module (c) and phase (d) of
the scattering coefficients of $n_{\textrm{T}1}(t;a_{1})$ and $n_{\textrm{T}6}(t;a_{1})$
as a function of frequency. The phase of $R_{1}$ and $R_{6}$ cannot
be estimated due to the non-reflecting behaviour of $n_{\textrm{T}1}(t;a_{1})$
and $n_{\textrm{T}6}(t;a_{1})$ in an extremely large optical bandwidth.
No reflected wave is observed in the numerical simulation when propagating
wide-band optical pulses through these time-varying media.
\noindent \begin{center}
\newpage{}
\par\end{center}

\noindent \begin{center}
\includegraphics[width=9cm,height=14cm,keepaspectratio]{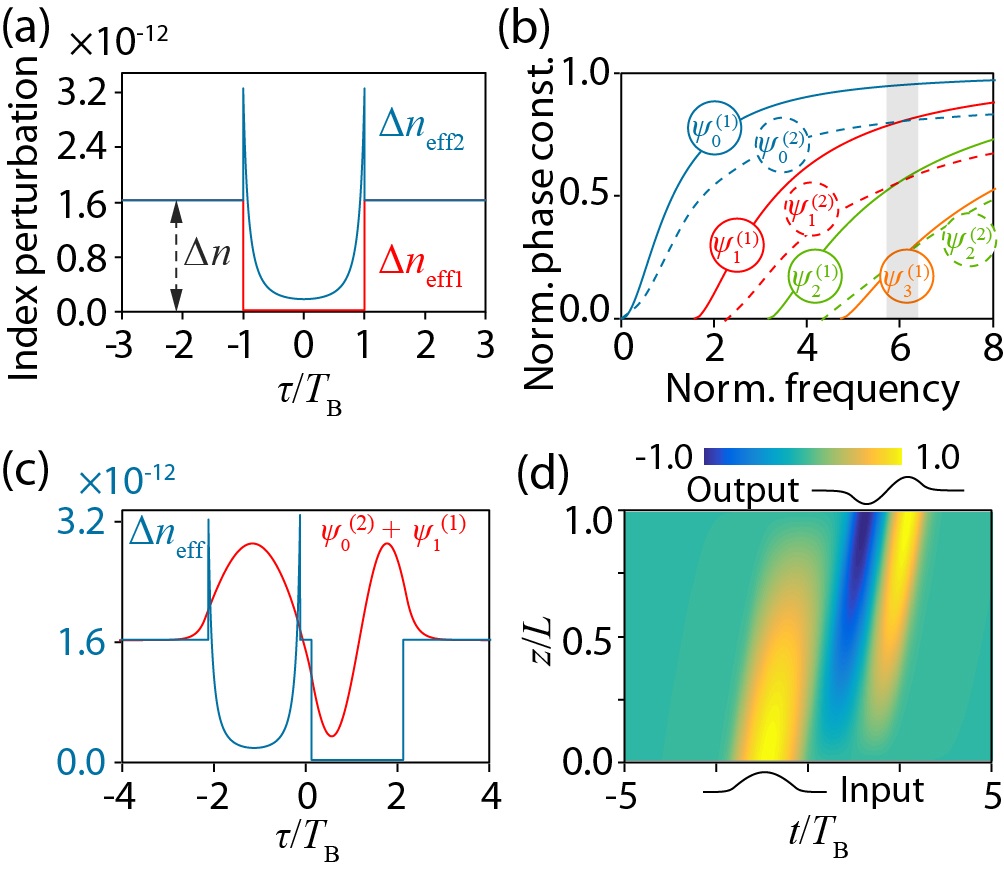}
\par\end{center}

\noindent \textbf{Figure 4.} \textbf{Supersymmetric TWGs and TPLs.
}(a) Temporal index perturbations $\Delta n_{\textrm{eff1}}$ and
$\Delta n_{\textrm{eff2}}$ of a step-index TWG ($2T_{\textrm{B}}=660$
ps, $\left|\Delta\beta_{1}\right|=10^{-3}$ ps/m and $\beta_{2}=0.06$
ps\textsuperscript{2}/m) and its T-SUSY partner. (b) Normalized dispersion
diagram $b$-$\nu$ of the temporal bound states $\psi_{n}^{\left(1\right)}$
and $\psi_{n}^{\left(2\right)}$ of both TWGs, where $b$ is the normalized
phase constant, defined for each $n$-th order mode as $b_{n}:=1-K_{n}/\Delta\beta-\Delta\beta_{1}^{2}/(2\beta_{2}\Delta\beta)$,
and $\nu$ is the normalized frequency, with $\nu^{2}:=2T_{\textrm{B}}^{2}\Delta\beta/\beta_{2}$
(here, $\nu=6$). $\psi_{n}^{\left(2\right)}$ exhibits a lower slope
(is less dispersive) than $\psi_{n+1}^{\left(1\right)}$. The grey
area is the phase-matching bandwidth, the interval $\Delta\nu$ where
$\Delta b\leq0.2$ between SUSY bound states. (c) Index perturbation
of a TPL (blue) constructed from two T-SUSY TWGs with a time separation
of $T_{\textrm{B}}/4$ ($2T_{\textrm{B}}=660$ ps), and its temporal
supermode (red), generated from the perfect phase-matching between
the states $\psi_{0}^{\left(2\right)}$ and $\psi_{1}^{\left(1\right)}$.
(d) Pulse shape transformation resulting from the energy transfer
between $\psi_{0}^{\left(2\right)}$ and $\psi_{1}^{\left(1\right)}$
in the TPL.
\noindent \begin{center}
\newpage{}
\par\end{center}

\noindent \begin{center}
\textbf{\LARGE{}Supplementary Information:}\\
\textbf{\LARGE{}}\\
\textbf{\LARGE{}Optical Supersymmetry in the Time Domain}
\par\end{center}{\LARGE \par}

\vspace*{0.5cm}
\begin{abstract}
{\normalsize{}In this supplementary information we explain }t{\normalsize{}he
theory of temporal supersymmetry in detail, we include additional
numerical examples and we provided further information on the methods
employed in this work. Equations and figures are denoted with a prefix
``S'' to distinguish them from the ones in the main text.}{\normalsize \par}
\end{abstract}
\vspace{1cm}

\tableofcontents{}

\newpage{}

\section{Temporal scattering: optical wave equation\label{sec:1}}

In general, the theory of temporal supersymmetry (T-SUSY) applied
to temporal scattering is valid for dielectric (or magnetic), linear,
anisotropic, heterogeneous, time-varying, and temporally non-dispersive
media. As demonstrated in this section, these scenarios (including
the particular situation in which a refractive index of the form given
by Eq.\,(2) is assumed) lead to the temporal Helmholtz Eq.\,(3)
of the paper.

Let us start by considering the all-dielectric case {[}$\boldsymbol{\upmu}_{\textrm{r}}\left(\mathbf{r},t\right)=\mathbf{I}_{3}${]}.
For such media, by combining Faraday's and Amp{\small{}\`e}re's
laws (applying the curl operator in Faraday's law and taking the
time derivative of Amp{\small{}\`e}re's law) it is straightforward
to demonstrate that the \emph{exact} time-domain vector wave equation
for the electric flux density $\mathbf{D}$ takes the form:
\begin{equation}
-\nabla\times\nabla\times\left[\boldsymbol{\upvarepsilon}_{\textrm{r}}^{-1}\left(\mathbf{r},t\right)\mathbf{D}\left(\mathbf{r},t\right)\right]=\frac{1}{c_{0}^{2}}\frac{\partial^{2}}{\partial t^{2}}\mathbf{D}\left(\mathbf{r},t\right),\tag{S1.1}\label{eq:S1.1}
\end{equation}
where $\boldsymbol{\upvarepsilon}_{\textrm{r}}\left(\mathbf{r},t\right)$
is the medium relative permittivity tensor. Let us now assume that
the relative permittivity can be expressed as:
\begin{equation}
\boldsymbol{\upvarepsilon}_{\textrm{r}}\left(\mathbf{r},t\right)=\varepsilon_{\textrm{T}}\left(t\right)\boldsymbol{\upvarepsilon}_{\textrm{S}}\left(\mathbf{r}\right),\tag{S1.2}\label{eq:S1.2}
\end{equation}
with $\boldsymbol{\upvarepsilon}_{\textrm{S}}\left(\mathbf{r}\right)$
being a tensor and $\varepsilon_{\textrm{T}}\left(t\right)$ a scalar.
It then follows that: 
\begin{equation}
-\nabla\times\nabla\times\left(\boldsymbol{\upvarepsilon}_{\textrm{S}}^{-1}\left(\mathbf{r}\right)\mathbf{D}\left(\mathbf{r},t\right)\right)=\frac{1}{c_{0}^{2}}\varepsilon_{\textrm{T}}\left(t\right)\frac{\partial^{2}}{\partial t^{2}}\mathbf{D}\left(\mathbf{r},t\right).\tag{S1.3}\label{eq:S1.3}
\end{equation}
The same wave equation applies to all-magnetic media {[}$\boldsymbol{\upvarepsilon}_{\textrm{r}}\left(\mathbf{r},t\right)=\mathbf{I}_{3}${]}
if $\mathbf{D}$ is replaced by the magnetic flux density $\mathbf{B}$
and $\boldsymbol{\upvarepsilon}_{\textrm{r}}\left(\mathbf{r},t\right)$
by the relative permeability tensor $\boldsymbol{\upmu}_{\textrm{r}}\left(\mathbf{r},t\right)=\mu_{\textrm{T}}\left(t\right)\boldsymbol{\upmu}_{\textrm{S}}\left(\mathbf{r}\right)$.

Applying separation of variables in the electromagnetic field under
analysis $\mathbf{F}\in\left\{ \mathbf{D},\mathbf{B}\right\} $: 
\begin{equation}
\mathbf{F}\left(\mathbf{r},t\right)=\psi\left(t\right)\boldsymbol{\Phi}\left(\mathbf{r}\right),\tag{S1.4}\label{eq:S1.4}
\end{equation}
Eq.\,(\ref{eq:S1.3}) becomes:
\begin{equation}
-\nabla\times\nabla\times\left(\boldsymbol{\upvarepsilon}_{\textrm{S}}^{-1}\left(\mathbf{r}\right)\bm{\Phi}\left(\mathbf{r}\right)\right)=\frac{\varepsilon_{\textrm{T}}\left(t\right)}{c_{0}^{2}}\frac{\ddot{\psi}\left(t\right)}{\psi(t)}\bm{\Phi}\left(\mathbf{r}\right),\tag{S1.5}\label{eq:S1.5}
\end{equation}
with $\ddot{\psi}\left(t\right)$ the second-order time derivative
of $\psi(t)$. Therefore, we must have:
\begin{equation}
\varepsilon_{\textrm{T}}\left(t\right)\frac{\ddot{\psi}\left(t\right)}{\psi(t)}=C,\tag{S1.6}\label{eq:S1.6}
\end{equation}
where $C$ is a constant. Defining $n_{\textrm{T}}^{2}\left(t\right):=\varepsilon_{\textrm{T}}\left(t\right)$
{[}or $n_{\textrm{T}}^{2}\left(t\right):=\mu_{\textrm{T}}\left(t\right)$
in the all-magnetic case{]} and assuming that $n_{-}:=n_{\textrm{T}}\left(t\to-\infty\right)$
is also a constant, we obtain $C=-\omega_{0}^{2}n_{-}^{2}$ for a
wave with a frequency $\omega_{0}$ at $t\to-\infty$, yielding Eq.\,(3)
of the main text. Note that, in the particular case of \emph{isotropic}
all-dielectric media, Eq.\,(\ref{eq:S1.1}) can be recast as (using
$\nabla\cdot\mathbf{D}=0$):
\begin{equation}
\triangle\left(\frac{1}{n^{2}\left(\mathbf{r},t\right)}\mathbf{D}\left(\mathbf{r},t\right)\right)-\nabla\left(\nabla\left(\frac{1}{n^{2}\left(\mathbf{r},t\right)}\right)\cdot\mathbf{D}\left(\mathbf{r},t\right)\right)-\frac{1}{c_{0}^{2}}\frac{\partial^{2}}{\partial t^{2}}\mathbf{D}\left(\mathbf{r},t\right)=\mathbf{0},\tag{S1.7}\label{eq:S1.7}
\end{equation}
with $n^{2}\left(\mathbf{r},t\right)=\varepsilon_{\textrm{r}}\left(\mathbf{r},t\right)$.
The same equation applies to isotropic all-magnetic media with $n^{2}\left(\mathbf{r},t\right)=\mu_{\textrm{r}}\left(\mathbf{r},t\right)$
and replacing $\mathbf{D}$ by $\mathbf{B}$. Obviously, proceeding
as in the general case by applying separation of variables to Eq.\,(\ref{eq:S1.7}),
Eq.\,(3) is again recovered. In the case of simultaneously dielectric
and magnetic materials {[}$n^{2}\left(\mathbf{r},t\right)=\varepsilon_{\textrm{r}}\left(\mathbf{r},t\right)\mu_{\textrm{r}}\left(\mathbf{r},t\right)${]},
it is also possible to obtain Eq.\,(3) for the electric (magnetic)
flux density if we assume a slowly-varying spatial and temporal evolution
in $\mu_{\textrm{r}}$ ($\varepsilon_{\textrm{r}}$).\footnote{The slowly-varying temporal evolution in a constitutive parameter,
e.g. $\mu_{\textrm{r}}$, requires to assume that $\left|\delta_{t}\mu_{\textrm{r}}\right|\ll\left|\mu_{\textrm{r}}\left(t\right)\right|$
in $\delta t\sim2\pi/\omega_{0}$, where $\delta_{t}\mu_{\textrm{r}}:=\mu_{\textrm{r}}\left(t+\delta t\right)-\mu_{\textrm{r}}\left(t\right)$.
In a similar way, the slowly-varying spatial evolution requires to
assume that $\left|\delta_{\mathbf{r}}\mu_{\textrm{r}}\right|\ll\left|\mu_{\textrm{r}}\left(\mathbf{r},t\right)\right|$
in $\left|\delta\mathbf{r}\right|\sim\lambda_{0}$, where $\delta_{\mathbf{r}}\mu_{\textrm{r}}:=\mu_{\textrm{r}}\left(\mathbf{r}+\delta\mathbf{r},t\right)-\mu_{\textrm{r}}\left(\mathbf{r},t\right)$
and $\lambda_{0}$ is the maximum wavelength of the problem.} In the isotropic homogeneous case {[}$n\left(\mathbf{r},t\right)=n_{\textrm{T}}\left(t\right)${]},
Eq.\,(\ref{eq:S1.7}) reduces to the familiar form of the wave equation:\newpage{}

\begin{equation}
\left(\triangle-\frac{n_{\textrm{T}}^{2}\left(t\right)}{c_{0}^{2}}\frac{\partial^{2}}{\partial t^{2}}\right)\mathbf{D}\left(\mathbf{r},t\right)=\mathbf{0}.\tag{S1.8}\label{eq:S1.8}
\end{equation}
As a final remark, although it is out of the scope of this work, it
can also be demonstrated that Eq.\,(3) also follows in the case of
nonlocal media by applying separation of variables in the electromagnetic
fields, provided that the nonlocal and time-varying nature of the
constitutive parameters can be decoupled.

\subsubsection*{Boundary conditions and polarization dependence}

The natural boundary conditions of Eq.\,(3) arise by noticing that
$\psi$ must be twice differentiable, i.e., such conditions are the
continuity of $\psi$ and $\psi^{\prime}$. These conditions are also
decoupled from the spatial part of the problem, governed by the wave
equation (in the isotropic case):
\begin{equation}
\triangle\left(\frac{1}{n_{\textrm{S}}^{2}\left(\mathbf{r}\right)}\boldsymbol{\Phi}\left(\mathbf{r}\right)\right)+k_{0}^{2}n_{-}^{2}\boldsymbol{\Phi}\left(\mathbf{r}\right)-\nabla\left(\nabla\left(\frac{1}{n_{\textrm{S}}^{2}\left(\mathbf{r}\right)}\right)\cdot\boldsymbol{\Phi}\left(\mathbf{r}\right)\right)=\mathbf{0}.\tag{S1.9}\label{eq:S1.9}
\end{equation}
Under the previous assumptions, the temporal and spatial wave equations
(as well as their corresponding boundary conditions) are completely
uncoupled. Since, in addition, the orientation of the fields is fully
determined by the spatial wave equation (as well as the response of
the medium as a function of the polarization state), the temporal
wave equation has no influence on the polarization dependence of the
optical system. That is, if a system is polarization independent (or
dependent), so will be its T-SUSY counterpart.

Finally, it is worth mentioning that T-SUSY systems share the same
spatial solution ($\boldsymbol{\Phi}^{\left(2\right)}=\boldsymbol{\Phi}^{\left(1\right)}$),
but possess a different temporal evolution ($\psi^{\left(2\right)}\neq\psi^{\left(1\right)}$).
Consequently, the eigenvalue degeneracy and the scattering properties
between both optical systems are preserved \emph{if and only if} both
$\psi^{\left(1\right)}$ and $\psi^{\left(2\right)}$ fulfil the aforementioned
temporal boundary conditions.

\newpage{}

\section{Temporal scattering: T-SUSY theory\label{sec:2}}

In all the aforementioned media, the temporal evolution of $\mathbf{D}$
($\mathbf{B}$), encoded by the $\psi$ function, obeys Eq.\,(3)
of the main text, reproduced here for clarity:
\begin{equation}
\left(\frac{\textrm{d}^{2}}{\textrm{d}t^{2}}+\omega_{0}^{2}N^{2}\left(t\right)\right)\psi\left(t\right)=0,\tag{S2.1}\label{eq:S2.1}
\end{equation}
where $N^{2}\left(t\right):=n_{-}^{2}/n_{\textrm{T}}^{2}\left(t\right)$,
$n_{-}:=n_{\textrm{T}}\left(t\rightarrow-\infty\right)$, and $\omega_{0}$
is the angular frequency (central angular frequency) of the electromagnetic
fields at $t\rightarrow-\infty$ in the monochromatic (non-monochromatic)
regime.

As discussed in the main text, Eq.\,(\ref{eq:S2.1}) matches the
1D time-independent Schr\"{o}dinger equation {[}Eq.\,(1){]} taking
$\alpha=1$, performing the relabelling $x\rightarrow t$, and identifying:
\begin{align}
\Omega-V\left(t\right) & \equiv\omega_{0}^{2}N^{2}\left(t\right).\tag{S2.2}\label{eq:S2.2}
\end{align}
In this way, assuming $\Omega$ {[}the eigenvalue in Eq.\,(1){]}
as a degree of freedom of the problem, we will be able to use the
algebraic transformations of 1D SUSYQM in the time domain, which will
give rise to time-varying refractive index profiles $n_{\textrm{T}1,2}\left(t\right)$
with similar scattering properties, provided that we use real $V_{1,2}$
potentials \cite{key-1,key-2}.

In order to have a well-defined \emph{temporal} scattering problem
in both superpartners, defined on the full line ($t\in\overline{\mathbb{R}}$),
we require that:\footnote{A spatial or temporal scattering problem is well-defined when we can
observe an incident wave and at least one reflected or transmitted
wave. In the temporal scattering case, the incident wave is always
defined at $t\rightarrow-\infty$ and the reflected and transmitted
waves are found at $t\rightarrow\infty$. Thus, we require a non-vanishing
wave function at $t\rightarrow\pm\infty$.} 
\begin{equation}
\psi^{\left(1,2\right)}\left(t\rightarrow\pm\infty\right)\neq0.\tag{S2.3}\label{eq:S2.3}
\end{equation}
A sufficient condition to satisfy Eq.\,(\ref{eq:S2.3}) is to consider
$V_{1,2}\left(t\rightarrow\pm\infty\right)<\infty$, which is fulfilled
by assuming: (\emph{i}) $W_{\pm}:=W\left(t\rightarrow\pm\infty\right)$
exists and is finite, and (\emph{ii}) $W^{\prime}$ is uniformly continuous
on the full line. Thus, from Barbalat's lemma \cite{key-3} and Riccati's
equation ($V_{1,2}=W^{2}\mp W^{\prime}$), we can infer that $W_{\pm}^{\prime}=0$
and $V_{1,\pm}=V_{2,\pm}=W_{\pm}^{2}<\infty$. To summarize, we will
have a well-defined scattering problem in both superpartners with:
\begin{equation}
\left.\begin{array}{c}
\left|W_{\pm}\right|<\infty\\
W^{\prime}\ \textrm{unif. cont.}
\end{array}\right\} \Rightarrow V_{\pm}\equiv V_{1,\pm}=V_{2,\pm}=W_{\pm}^{2}<\infty\Rightarrow\psi^{\left(1,2\right)}\left(t\rightarrow\pm\infty\right)\neq0.\tag{S2.4}\label{eq:S2.4}
\end{equation}
Hence, combining Eqs.\,(\ref{eq:S2.2}) and (\ref{eq:S2.4}), keeping
in mind that both superpartners share the same eigenvalue $\Omega$
\cite{key-1}, we infer that: 
\begin{equation}
N_{1}^{2}\left(t\rightarrow\pm\infty\right)=N_{2}^{2}\left(t\rightarrow\pm\infty\right)\equiv N_{\pm}^{2},\tag{S2.5}\label{eq:S2.5}
\end{equation}
with $N_{-}=1$ by definition. From the above equation, the following
remarks are in order:
\begin{itemize}
\item In spite of the fact that $N_{1,+}^{2}=N_{2,+}^{2}$, note that $n_{1,+}^{2}\neq n_{2,+}^{2}$
when $n_{1,-}^{2}\neq n_{2,-}^{2}$. This can be observed when the
SUSY systems are implemented over different background materials.
\item If we assume positive-real constitutive parameters, $N_{1,\pm}=N_{2,\pm}\equiv N_{\pm}$,
and then:
\begin{equation}
\frac{n_{1,-}}{n_{1,+}}=\frac{n_{2,-}}{n_{2,+}}.\tag{S2.6}\label{eq:S2.6}
\end{equation}
\end{itemize}
In order to connect the temporal scattering problem of both superpartners,
consider a plane wave in each system at $t\rightarrow-\infty$. In
this way, the asymptotic behaviour of $\psi^{\left(1,2\right)}$ at
$t\rightarrow-\infty$ is equivalent to:
\begin{equation}
\psi^{\left(1,2\right)}\left(t\right)\underset{t\rightarrow-\infty}{\sim}\psi_{-}^{\left(1,2\right)}\left(t\right)=\exp\left(\textrm{i}N_{-}\omega_{0}t\right).\tag{S2.7}\label{eq:S2.7}
\end{equation}
Next, after the interaction with the refractive index variations $n_{\textrm{T}1,2}\left(t\right)$,
the asymptotic behaviour of $\psi^{\left(1,2\right)}$ at $t\rightarrow\infty$
will be found to be equivalent to:
\begin{align}
\psi^{\left(1,2\right)}\left(t\right) & \underset{t\rightarrow\infty}{\sim}\psi_{+}^{\left(1,2\right)}\left(t\right)=R_{1,2}\exp\left(-\textrm{i}N_{+}\omega_{0}t\right)+T_{1,2}\exp\left(\textrm{i}N_{+}\omega_{0}t\right),\tag{S2.8}\label{eq:S2.8}
\end{align}
$R_{1,2}$ and $T_{1,2}$ being respectively the reflection and transmission
coefficients,\footnote{The scattering coefficients $R_{1,2}$ and $T_{1,2}$ are equal to
the complex amplitudes of the reflected and transmitted waves when
the amplitude of the incident wave is set to $1$.} and $\omega_{0}N_{\pm}=\sqrt{\Omega-W_{\pm}^{2}}$. In such a scenario,
using the SUSY relation $\psi^{\left(1\right)}=\xi\widehat{\textrm{A}}^{+}\psi^{\left(2\right)}$
\cite{key-1}, where $\xi\in\mathbb{C}$, $\widehat{\textrm{A}}^{\pm}:=\mp\textrm{d}/\textrm{d}t+W\left(t\right)$
are the SUSY operators and $W$ is the superpotential (a real-valued
function in our case), we can relate the asymptotic behaviours as:
\begin{equation}
\psi_{\pm}^{\left(1\right)}\left(t\right)=\xi\left(-\frac{\textrm{d}}{\textrm{d}t}+W_{\pm}\right)\psi_{\pm}^{\left(2\right)}\left(t\right).\tag{S2.9}\label{eq:S2.9}
\end{equation}
Thus, we have at $t\rightarrow-\infty$:
\begin{align}
\exp\left(\textrm{i}\omega_{0}t\right) & =\xi\left(-\textrm{i}\omega_{0}+W_{-}\right)\exp\left(\textrm{i}\omega_{0}t\right),\tag{S2.10}\label{eq:S2.10}
\end{align}
and at $t\rightarrow\infty$:
\begin{align}
R_{1}\exp\left(-\textrm{i}N_{+}\omega_{0}t\right)+T_{1}\exp\left(\textrm{i}N_{+}\omega_{0}t\right) & =\xi R_{2}\left(\textrm{i}N_{+}\omega_{0}+W_{+}\right)\exp\left(-\textrm{i}N_{+}\omega_{0}t\right)\nonumber \\
 & +\xi T_{2}\left(-\textrm{i}N_{+}\omega_{0}+W_{+}\right)\exp\left(\textrm{i}N_{+}\omega_{0}t\right).\tag{S2.11}\label{eq:S2.11}
\end{align}
Finally, equating terms with the same exponent in Eqs.\,(\ref{eq:S2.10})
and (\ref{eq:S2.11}), we find:
\begin{equation}
\begin{aligned}\frac{R_{1}}{R_{2}}=\frac{W_{+}+\textrm{i}N_{+}\omega_{0}}{W_{-}-\textrm{i}\omega_{0}}; & \ \ \ \frac{T_{1}}{T_{2}}=\frac{W_{+}-\textrm{i}N_{+}\omega_{0}}{W_{-}-\textrm{i}\omega_{0}}.\end{aligned}
\tag{S2.12}\label{eq:S2.12}
\end{equation}
Let us take a closer look at the above equations. In particular, it
should be remarked that:
\begin{enumerate}
\item The SUSY refractive index profiles have identical intensity scattering
behaviour: 
\begin{equation}
\left|R_{1}\right|^{2}=\left|R_{2}\right|^{2};\ \ \ \left|T_{1}\right|^{2}=\left|T_{2}\right|^{2},\tag{S2.13}\label{eq:S2.13}
\end{equation}
as a direct consequence of the fact that both systems share the same
eigenvalue $\Omega=\omega_{0}^{2}N_{\pm}^{2}+W_{\pm}^{2}$. 
\item Although we consider complex-valued wave functions, the superpotential
and the potentials must be real-valued functions to guarantee Eq.\,(\ref{eq:S2.13}).
Otherwise, the intensity scattering behaviour could be different in
each system \cite{key-2}.
\item Eq.\,(\ref{eq:S2.12}) does not depend on $n_{1,2,-}$. Consequently,
we can engineer time-varying refractive index profiles exhibiting
the same intensity scattering behaviour using the same ($n_{1,-}=n_{2,-}$)
or different ($n_{1,-}\neq n_{2,-}$) background materials.
\item Bearing in mind that $N_{-}=1$, we find that $\Omega=\omega_{0}^{2}+W_{-}^{2}$.
Therefore, we infer that $\Omega\geq\omega_{0}^{2}$.\label{enu:Omega mayor que}
\item The use of negative frequencies in Eq.\,(\ref{eq:S2.11}) allows
us, not only to describe adequately the temporal scattering problem
(as discussed in the main text), but also to decouple the ratios $R_{1}/R_{2}$
and $T_{1}/T_{2}$. If we used only positive frequencies and negative
wave numbers, we would find a scattering relation of the form:
\begin{equation}
\frac{R_{1}+T_{1}}{R_{2}+T_{2}}=\frac{W_{+}-\textrm{i}N_{+}\omega_{0}}{W_{-}-\textrm{i}\omega_{0}},\tag{S2.14}\label{eq:S2.14}
\end{equation}
that is, without the possibility of decoupling the reflected and transmitted
amplitudes.
\item The ratios of the scattering coefficients $\mathcal{R}_{i}$ and $\mathcal{T}_{i}$
of the electric (magnetic) field strength are analogous to those in
Eq.\,(\ref{eq:S2.12}), since $\mathcal{R}_{i}=N_{+}^{2}R_{i}$ and
$\mathcal{T}_{i}=N_{+}^{2}T_{i}$. Hence, we find that $\mathcal{R}_{1}/\mathcal{R}_{2}=R_{1}/R_{2}$
and $\mathcal{T}_{1}/\mathcal{T}_{2}=T_{1}/T_{2}$.\newpage{}
\item Note that $\left|W_{+}\right|=\left|W_{-}\right|\Leftrightarrow N_{+}=N_{-}=1\Leftrightarrow n_{i,+}=n_{i,-}$.
Such a situation takes place:
\begin{itemize}
\item If $W_{+}=-W_{-}$, in which case SUSY is said to be unbroken. Here,
we observe that $R_{1}=-R_{2}$. The reflected wave has an extra phase
shift of $\pi$ rad in the SUSY system.
\item If $W_{+}=W_{-}$, in which case SUSY is said to be broken. Here,
we find that $T_{1}=T_{2}$.
\end{itemize}
\item In a SUSY refractive index chain $\left\{ n_{\textrm{T}s}\left(t\right)\right\} _{s=1}^{m}$
where the scattering problem is well-defined, the value of the potentials
$\left\{ V_{s}\right\} _{s=1}^{m}$ and superpotentials $\left\{ W_{s}\right\} _{s=1}^{m-1}$
at $t\rightarrow\pm\infty$ is the same. That is:
\begin{align}
V_{1,\pm} & =V_{2,\pm}=\ldots=V_{m,\pm}=W_{1,\pm}^{2},\tag{S2.15}\label{eq:S2.15}
\end{align}
and $W_{1,\pm}=W_{2,\pm}=\ldots=W_{m-1,\pm}\equiv W_{\pm}$. Consequently,
the scattering coefficients of $n_{\textrm{T}m}$ and $n_{\textrm{T}1}$
are related by the following expressions: 
\begin{align}
\frac{R_{1}}{R_{m}}=\left(\frac{W_{+}+\textrm{i}N_{+}\omega_{0}}{W_{-}-\textrm{i}\omega_{0}}\right)^{m-1} & ;\ \ \ \frac{T_{1}}{T_{m}}=\left(\frac{W_{+}-\textrm{i}N_{+}\omega_{0}}{W_{-}-\textrm{i}\omega_{0}}\right)^{m-1}.\tag{S2.16}\label{eq:S2.16}
\end{align}
\end{enumerate}
On the other hand, given a refractive index $n_{\textrm{T}1}\left(t\right)=n_{1,-}/N_{1}\left(t\right)$,
its SUSY profile $n_{\textrm{T}2}\left(t\right)=n_{2,-}/N_{2}\left(t\right)$
can be directly found by combining Eq.\,(\ref{eq:S2.2}) and Riccati's
equation ($V_{1,2}=W^{2}\mp W^{\prime}$):
\begin{equation}
n_{\textrm{T}2}\left(t;\Omega,n_{2,-}\right)=\frac{n_{2,-}}{\sqrt{\frac{n_{1,-}^{2}}{n_{\textrm{T}1}^{2}\left(t\right)}-\frac{2}{\omega_{0}^{2}}W^{\prime}\left(t;\Omega\right)}},\tag{S2.17}\label{eq:S2.17}
\end{equation}
where the semicolon symbol is used to separate explicitly the system
parameters ($\equiv$degrees of freedom) from the time variable. As
mentioned above, $n_{2,-}$ allows us to change the background material
of $n_{\textrm{T}2}\left(t\right)$ while preserving the same intensity
scattering properties as the original modulation $n_{\textrm{T}1}\left(t\right)$,
and the $\Omega$ parameter can be employed to tailor different features
of $n_{\textrm{T}2}\left(t\right)$, such as its maximal excursion
(see Section \ref{sec:3}).

\subsubsection*{Shape Invariant Potentials (SIP)\label{subsec:2.1}}

Shape invariant potentials (SIP) are of great interest in quantum
mechanics to find new analytically solvable potentials \cite{key-1}.
In the framework of SUSY quantum mechanics, SIP allows us to: (\emph{i})
calculate the spectrum of a given potential and its (unbroken or broken)
SUSY Hamiltonian chain in a simple and elegant way, and (\emph{ii})
analyse and design the scattering properties of a large number of
potentials. In particular, we are interested in this second feature.

In general, we will say that two SUSY partner potentials $V_{1,2}$
are \emph{shape invariant} if they obey the relation \cite{key-1}:
\begin{equation}
V_{2}\left(t;\mathbf{a}_{1}\right)=V_{1}\left(t;\mathbf{a}_{2}\right)+M\left(\mathbf{a}_{1}\right),\tag{S2.18}\label{eq:S2.18}
\end{equation}
where $\left(\mathbf{a}_{1},\mathbf{a}_{2}\right)\in\mathbb{R}^{p}\times\mathbb{R}^{p}$
are a set of parameters related by a multivariate function $\mathbf{f}\in\mathcal{F}(\mathbb{R}^{p},\mathbb{R}^{p})$
of the form $\mathbf{a}_{2}=\mathbf{f}\left(\mathbf{a}_{1}\right)$,
and $M\in\mathcal{F}(\mathbb{R}^{p},\mathbb{R})$. In our numerical
examples (see Section \ref{sec:3}), we use $p=1$.
\noindent \begin{center}
\includegraphics[width=11.5cm,height=6cm,keepaspectratio]{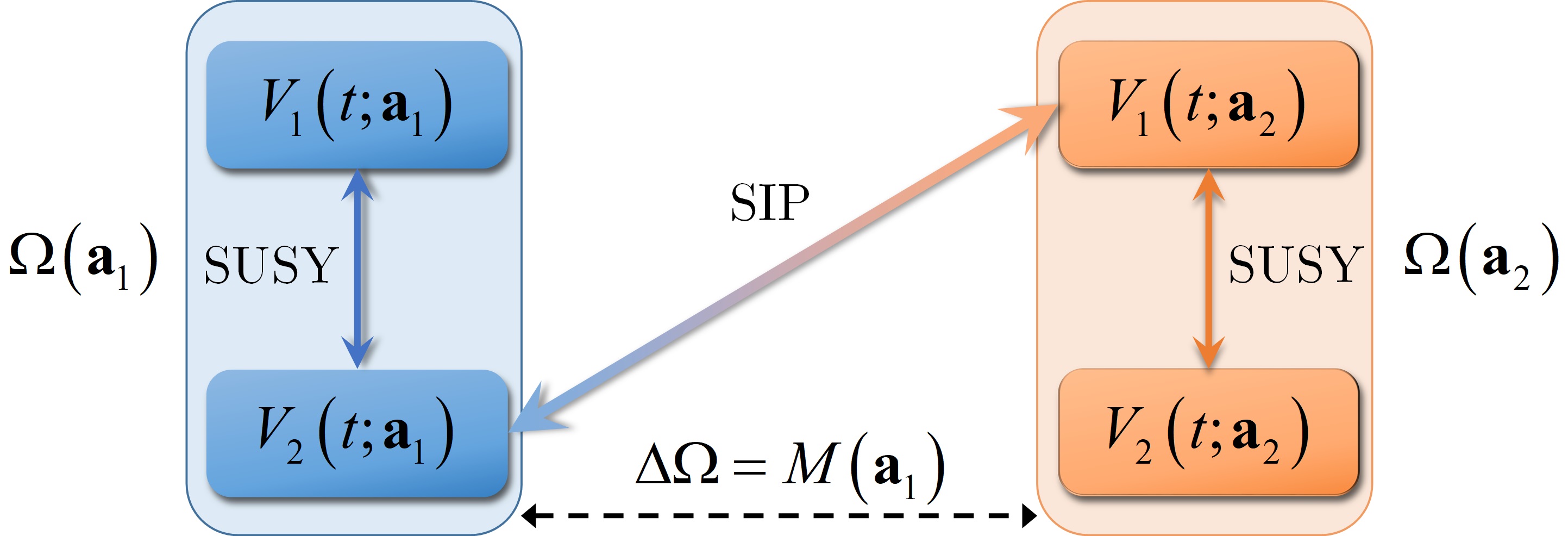}
\par\end{center}

\noindent \begin{center}
\textbf{\small{}Figure S2.1.}{\small{} Eigenvalue relation between
SIP superpartners supporting continuous spectra.\newpage{}}
\par\end{center}{\small \par}

From the above equation, we can infer the following properties of
the temporal scattering problem, with the QM superpartners supporting
a continuous spectrum:
\begin{itemize}
\item Eq.\,(\ref{eq:S2.18}) establishes an eigenvalue relation between
superpartners of the form (see Fig.\,S2.1):
\begin{equation}
\Omega\left(\mathbf{a}_{1}\right)=\Omega\left(\mathbf{a}_{2}\right)+M\left(\mathbf{a}_{1}\right),\tag{S2.19}\label{eq:S2.19}
\end{equation}
where $\Omega\left(\mathbf{a}_{i}\right)\in I_{i}\subset\mathbb{R}$
and $I_{i}$ is the eigenvalue spectrum of the SUSY partners $V_{1,2}\left(\mathbf{a}_{i}\right)$.
\item Since $V_{i,\pm}\left(\mathbf{a}_{j}\right)=W_{\pm}^{2}\left(\mathbf{a}_{j}\right)\ \forall\left(i,j\right)\in\left\{ 1,2\right\} ^{2}$,
then:
\begin{equation}
W_{\pm}^{2}\left(\mathbf{a}_{1}\right)=W_{\pm}^{2}\left(\mathbf{a}_{2}\right)+M\left(\mathbf{a}_{1}\right).\tag{S2.20}\label{eq:S2.20}
\end{equation}
Consequently, the superpotential depends on the SIP parameters at
$t\rightarrow\pm\infty$.
\item As demonstrated in \cite{key-1}, the wave functions are connected
as:
\begin{equation}
\psi^{\left(2\right)}\left(t;\mathbf{a}_{1}\right)=\psi^{\left(1\right)}\left(t;\mathbf{a}_{2}\right).\tag{S2.21}\label{eq:S2.21}
\end{equation}
\item Combining Eqs.\,(\ref{eq:S2.18}) and (\ref{eq:S2.19}) in $\Omega\left(\mathbf{a}_{1}\right)-V_{2}\left(t;\mathbf{a}_{1}\right)=\omega_{0}^{2}N_{2}^{2}\left(t;\mathbf{a}_{1}\right)$
we find that $N_{2}\left(t;\mathbf{a}_{1}\right)=N_{1}\left(t;\mathbf{a}_{2}\right)$,
and therefore:
\begin{equation}
n_{\textrm{T}2}\left(t;\mathbf{a}_{1}\right)=\frac{n_{2,-}\left(\mathbf{a}_{1}\right)}{n_{1,-}\left(\mathbf{a}_{2}\right)}n_{\textrm{T}1}\left(t;\mathbf{a}_{2}\right),\tag{S2.22}\label{eq:S2.22}
\end{equation}
with $n_{2,-}\left(\mathbf{a}_{1}\right)$ a degree of freedom of
the problem.
\item Interestingly, $N_{\pm}$ does not depend on the SIP parameters at
$t\rightarrow\pm\infty$. Using Eqs.\,(\ref{eq:S2.19}) and (\ref{eq:S2.20}):
\begin{align}
\omega_{0}^{2}N_{\pm}^{2}\left(\mathbf{a}_{1}\right) & =\Omega\left(\mathbf{a}_{1}\right)-W_{\pm}^{2}\left(\mathbf{a}_{1}\right)=\Omega\left(\mathbf{a}_{2}\right)+M\left(\mathbf{a}_{1}\right)-W_{\pm}^{2}\left(\mathbf{a}_{2}\right)-M\left(\mathbf{a}_{1}\right)\nonumber \\
 & =\Omega\left(\mathbf{a}_{2}\right)-W_{\pm}^{2}\left(\mathbf{a}_{2}\right)=\omega_{0}^{2}N_{\pm}^{2}\left(\mathbf{a}_{2}\right),\tag{S2.23}\label{eq:S2.23}
\end{align}
we verify that $N_{\pm}\left(\mathbf{a}_{1}\right)=N_{\pm}\left(\mathbf{a}_{2}\right)$.
\item From Eq.\,(\ref{eq:S2.21}), we can infer that $R_{2}\left(\mathbf{a}_{1}\right)=R_{1}\left(\mathbf{a}_{2}\right)$
and $T_{2}\left(\mathbf{a}_{1}\right)=T_{1}\left(\mathbf{a}_{2}\right)$.
As a result, Eq.\,(\ref{eq:S2.12}) can be restated as:
\begin{equation}
\begin{aligned}\frac{R_{1}\left(\mathbf{a}_{1}\right)}{R_{1}\left(\mathbf{a}_{2}\right)}=\frac{W_{+}\left(\mathbf{a}_{1}\right)+\textrm{i}N_{+}\omega_{0}}{W_{-}\left(\mathbf{a}_{1}\right)-\textrm{i}\omega_{0}}; & \ \ \ \frac{T_{1}\left(\mathbf{a}_{1}\right)}{T_{1}\left(\mathbf{a}_{2}\right)}=\frac{W_{+}\left(\mathbf{a}_{1}\right)-\textrm{i}N_{+}\omega_{0}}{W_{-}\left(\mathbf{a}_{1}\right)-\textrm{i}\omega_{0}}.\end{aligned}
\tag{S2.24}\label{eq:S2.24}
\end{equation}
\item In a SUSY refractive index chain with $m$ shape invariant potentials
in the continuum, Eq.\,(\ref{eq:S2.18}) can be generalized to:
\begin{equation}
V_{m}\left(t;\mathbf{a}_{1}\right)=V_{1}\left(t;\mathbf{a}_{m}\right)+\sum_{i=1}^{m-1}M\left(\mathbf{a}_{i}\right),\tag{S2.25}\label{eq:S2.25}
\end{equation}
with $\mathbf{a}_{i}=(\mathring{\mathbf{f}})^{i-1}\left(\mathbf{a}_{1}\right)$
{[}e.g., $\mathbf{a}_{3}=(\mathring{\mathbf{f}})^{2}\left(\mathbf{a}_{1}\right)=\left(\mathbf{f}\circ\mathbf{f}\right)\left(\mathbf{a}_{1}\right)=\mathbf{f}(\mathbf{f}(\mathbf{a}_{1}))${]}.
Hence, Eqs.\,(\ref{eq:S2.19})-(\ref{eq:S2.22}) become:
\begin{align}
\Omega\left(\mathbf{a}_{1}\right) & =\Omega\left(\mathbf{a}_{m}\right)+\sum_{i=1}^{m-1}M\left(\mathbf{a}_{i}\right);\tag{S2.26}\label{eq:S2.26}\\
W_{\pm}^{2}\left(\mathbf{a}_{1}\right) & =W_{\pm}^{2}\left(\mathbf{a}_{m}\right)+\sum_{i=1}^{m-1}M\left(\mathbf{a}_{i}\right);\tag{S2.27}\label{eq:S2.27}\\
\psi^{\left(m\right)}\left(t;\mathbf{a}_{1}\right) & =\psi^{\left(1\right)}\left(t;\mathbf{a}_{m}\right);\tag{S2.28}\label{eq:S2.28}\\
n_{\textrm{T}m}\left(t;\mathbf{a}_{1}\right) & =\frac{n_{m,-}\left(\mathbf{a}_{1}\right)}{n_{1,-}\left(\mathbf{a}_{m}\right)}n_{\textrm{T}1}\left(t;\mathbf{a}_{m}\right).\tag{S2.29}\label{eq:S2.29}
\end{align}
Thus, $R_{m}\left(\mathbf{a}_{1}\right)=R_{1}\left(\mathbf{a}_{m}\right)$,
$T_{m}\left(\mathbf{a}_{1}\right)=T_{1}\left(\mathbf{a}_{m}\right)$
and the scattering relations given by Eq.\,(\ref{eq:S2.16}) can
be recast as:
\begin{align}
\frac{R_{1}\left(\mathbf{a}_{1}\right)}{R_{1}\left(\mathbf{a}_{m}\right)}=\left(\frac{W_{+}\left(\mathbf{a}_{1}\right)+\textrm{i}N_{+}\omega_{0}}{W_{-}\left(\mathbf{a}_{1}\right)-\textrm{i}\omega_{0}}\right)^{m-1} & ;\ \ \ \frac{T_{1}\left(\mathbf{a}_{1}\right)}{T_{1}\left(\mathbf{a}_{m}\right)}=\left(\frac{W_{+}\left(\mathbf{a}_{1}\right)-\textrm{i}N_{+}\omega_{0}}{W_{-}\left(\mathbf{a}_{1}\right)-\textrm{i}\omega_{0}}\right)^{m-1}.\tag{S2.30}\label{eq:S2.30}
\end{align}
\end{itemize}

\subsubsection*{Isospectral Transformations\label{subsec:2.2}}

In the next lines, we will discuss the possibility of using SUSY transformations
in the time domain to construct, from a given refractive index $n_{\textrm{T}1}\left(t\right)$,
an $m$-parameter family of isospectral refractive index profiles
$\widetilde{n}_{\textrm{T}1}\left(t;\eta_{1},\ldots,\eta_{m}\right)$,
that is, time-varying optical systems with exactly the same scattering
properties in module and phase as the original one. 

The \emph{one-parameter} isospectral family $\widetilde{V}_{1}\left(x;\eta_{1}\right)$
of a given potential $V_{1}\left(x\right)$ can be calculated as indicated
in Section 7.1 of \cite{key-1}. In this vein, using our quantum-optical
analogy {[}Eq.\,(\ref{eq:S2.2}){]}, we find the one-parameter isospectral
family $\widetilde{n}_{\textrm{T}1}\left(t;\eta_{1}\right)$ of the
refractive index $n_{\textrm{T}1}\left(t\right)$ as:
\begin{equation}
\widetilde{n}_{\textrm{T}1}\left(t;\eta_{1}\right)=\frac{\widetilde{n}_{1,-}\left(\eta_{1}\right)}{\sqrt{\frac{n_{1,-}^{2}}{n_{\textrm{T}1}^{2}\left(t\right)}+\frac{2}{\omega_{0}^{2}}\frac{\textrm{d}^{2}}{\textrm{d}t^{2}}\ln\left[\eta_{1}+\int^{t}\exp\left(-2\int^{\alpha}W\left(\beta\right)\textrm{d}\beta\right)\textrm{d}\alpha\right]}},\tag{S2.31}\label{eq:S2.31}
\end{equation}
with $\eta_{1}$ and $\widetilde{n}_{1,-}\left(\eta_{1}\right)$ degrees
of freedom of the problem. Interestingly, the above family has exactly
the same scattering coefficients as the original modulation, provided
that we use a nonsingular superpotential family, which is fulfilled
by taking $\eta_{1}>0$. It is straightforward to prove this statement.
Let us denote the scattering coefficients of $\widetilde{n}_{\textrm{T}1}\left(t;\eta_{1}\right)$
as $R_{1}\left(\eta_{1}\right)$ and $T_{1}\left(\eta_{1}\right)$.
The ratios $R_{1}\left(\eta_{1}\right)/R_{2}$ and $T_{1}\left(\eta_{1}\right)/T_{2}$
can be expressed in the same form as Eq.\,(\ref{eq:S2.12}), but
replacing $W_{\pm}$ by $\widetilde{W}_{\pm}\left(\eta_{1}\right)$
and $N_{\pm}$ by $\widetilde{N}_{\pm}\left(\eta_{1}\right)$, where
$\widetilde{N}_{\pm}\left(\eta_{1}\right):=\widetilde{N}_{1}\left(t\rightarrow\pm\infty;\eta_{1}\right)=\widetilde{n}_{1,-}\left(\eta_{1}\right)/\widetilde{n}_{\textrm{T}1}\left(t\rightarrow\pm\infty;\eta_{1}\right)$
and:
\begin{align}
\widetilde{W}_{\pm}\left(\eta_{1}\right): & =\widetilde{W}\left(t\rightarrow\pm\infty;\eta_{1}\right)\nonumber \\
 & =\lim_{t\rightarrow\pm\infty}\left\{ W\left(t\right)+\frac{\textrm{d}}{\textrm{d}t}\ln\left[\eta_{1}+\int^{t}\exp\left(-2\int^{\alpha}W\left(\beta\right)\textrm{d}\beta\right)\textrm{d}\alpha\right]\right\} .\tag{S2.32}\label{eq:S2.32}
\end{align}
Concretely, $\widetilde{W}\left(t;\eta_{1}\right)$ is the family
of superpotentials connecting $\widetilde{V}_{1}\left(t;\eta_{1}\right)$
and $V_{2}\left(t\right)$ via the Riccati equation. In order to preserve
the scattering properties between both superpartners, the superpotential
family must be nonsingular (sufficient condition but not necessary)
\cite{key-1}. To this end, we set $\eta_{1}>0$. Finally, taking
into account that $\widetilde{W}_{\pm}\left(\eta_{1}\right)=W_{\pm}$
and $\widetilde{N}_{\pm}\left(\eta_{1}\right)=N_{\pm}$, we demonstrate
that $R_{1}\left(\eta_{1}\right)=R_{1}$ and $T_{1}\left(\eta_{1}\right)=T_{1}$.

In addition, it is worth highlighting the possibility of combining
the one-parameter isospectral transformation along with SIP. More
specifically, given two potentials $V_{i}\left(t;\mathbf{a}_{1}\right)$
and $V_{i+1}\left(t;\mathbf{a}_{1}\right)$ of a SUSY chain calculated
respectively with SIP from $V_{1}\left(t;\mathbf{a}_{i}\right)$ and
$V_{1}\left(t;\mathbf{a}_{i+1}\right)$ by using Eq.\,(\ref{eq:S2.25}),
we can obtain the superpotential $W_{i}\left(t;\mathbf{a}_{1}\right)$
from:
\begin{align}
W_{i}\left(t;\mathbf{a}_{1}\right) & =W_{-}\left(\mathbf{a}_{1}\right)+\frac{1}{2}\int_{-\infty}^{t}\left[V_{1}\left(\tau;\mathbf{a}_{i+1}\right)-V_{1}\left(\tau;\mathbf{a}_{i}\right)+M\left(\mathbf{a}_{i}\right)\right]\textrm{d}\tau,\tag{S2.33}\label{eq:S2.33}
\end{align}
and later calculate the family $\widetilde{W}_{i}\left(t;\mathbf{a}_{1},\eta_{1}\right)$
to construct the isospectral family $\widetilde{n}_{\textrm{T}i}\left(t;\mathbf{a}_{1},\eta_{1}\right)$
with the same scattering properties in module and phase as $n_{\textrm{T}i}\left(t;\mathbf{a}_{1}\right)$.
As seen, the temporal scattering properties of a large number of time-varying
optical systems can be analysed and designed by combining both strategies.

On the other hand, the \emph{multi-parameter} isospectral family $\widetilde{n}_{\textrm{T}1}\left(t;\eta_{1},\ldots,\eta_{m}\right)$
can be calculated from the multi-parameter Darboux procedure detailed
in Section 7.2 of \cite{key-1}. Here, we only describe the different
steps of this procedure applied to time-varying optical systems:\pagebreak{}
\begin{enumerate}
\item We start from a given refractive index profile $n_{\textrm{T}1}\left(t\right)$
associated with a QM potential $V_{1}\left(t\rightarrow x\right)$
via Eq.\,(\ref{eq:S2.2}). This potential must support bound states
and must satisfy the sufficient conditions detailed at the beginning
of this section to guarantee that the temporal scattering problem
is well-defined {[}Eq.\,(\ref{eq:S2.4}){]}.
\item Next, we generate the family $\widetilde{V}_{1}\left(x;\eta_{1},\ldots,\eta_{m}\right)$
from $V_{1}\left(x\right)$ by using the multi-parameter Darboux procedure.
\item Finally, performing the relabelling $\widetilde{V}_{1}\left(x\rightarrow t;\eta_{1},\ldots,\eta_{m}\right)$
we obtain the sought family:
\begin{equation}
\widetilde{n}_{\textrm{T}1}\left(t;\eta_{1},\ldots,\eta_{m}\right)=\frac{\widetilde{n}_{1,-}\left(\eta_{1},\ldots,\eta_{m}\right)}{\sqrt{1+\frac{1}{\omega_{0}^{2}}\left[V_{1,-}-\widetilde{V}_{1}\left(t;\eta_{1},\ldots,\eta_{m}\right)\right]}},\tag{S2.34}\label{eq:S2.34}
\end{equation}
where $\left\{ \eta_{1},\ldots,\eta_{m}\right\} \subset\mathbb{R}^{+}$
and $\widetilde{n}_{1,-}\left(\eta_{1},\ldots,\eta_{m}\right)$ are
degrees of freedom in the above equation.
\end{enumerate}
\newpage{}

\section{Temporal scattering: numerical examples\label{sec:3}}

In this section, we include additional numerical examples of the temporal
scattering problem in\linebreak{}
T-SUSY optical systems.

\subsubsection*{Transparent hyperbolic secant modulation}

In the main text, we analysed the SUSY refractive index profile $n_{2}\left(\mathbf{r},t\right)=n_{\textrm{T}2}\left(t\right)$
of a constant and, therefore, transparent refractive index $n_{1}\left(\mathbf{r},t\right)=n_{1,-}$.
In such a case, using Eq.\,(\ref{eq:S2.2}), we infer that the original
potential is:
\begin{equation}
V_{1}\left(t\right)=\Omega-\omega_{0}^{2},\tag{S3.1}\label{eq:S3.1}
\end{equation}
which leads to a superpotential of the form:
\begin{equation}
W\left(t\right)=-\sqrt{\Omega-\omega_{0}^{2}}\tanh\left(\sqrt{\Omega-\omega_{0}^{2}}\thinspace t\right).\tag{S3.2}\label{eq:S3.2}
\end{equation}
Finally, using Eq.\,(\ref{eq:S2.17}) we find the SUSY refractive
index:
\begin{equation}
n_{2}\left(\mathbf{r},t\right)=n_{\textrm{T}2}\left(t\right)=\frac{n_{2,-}}{\sqrt{1+\frac{2}{\omega_{0}^{2}}\left(\Omega-\omega_{0}^{2}\right)\textrm{sech}^{2}\left(\sqrt{\Omega-\omega_{0}^{2}}\thinspace t\right)}},\tag{S3.3}\label{eq:S3.3}
\end{equation}
with $\Omega\geq\omega_{0}^{2}$, as discussed on page \pageref{enu:Omega mayor que}.
In the following, we will first discuss the case $\Omega>\omega_{0}^{2}$,
and later, we will analyse the case $\Omega=\omega_{0}^{2}$.

Considering $\Omega>\omega_{0}^{2}$, we observe two degrees of freedom
in Eq.\,(\ref{eq:S3.3}): $n_{2,-}$ and $\Omega$. The former allows
us to implement the above refractive index modulation over different
background materials, and the latter can be employed to tailor its
maximal excursion ($\Delta n$) and its temporal width ($\Delta t$).\footnote{The parameter $\Delta n$ is defined as $\Delta n:=n_{2,-}-\min\left\{ n_{\textrm{T}2}\left(t\right)\right\} $
and the parameter $\Delta t$ is defined as the full-width at $1/\left(2e\right)$
maximum of the temporal profile $n_{2,-}-n_{\textrm{T}2}\left(t\right)$.} Figure S3.1 shows the refractive index $n_{\textrm{T}2}\left(t\right)$
for different values of the ratio $\Omega/\omega_{0}^{2}$. As seen,
the higher the value of $\Omega/\omega_{0}^{2}$ is, the higher $\Delta n$
and the lower $\Delta t$ are. Along this line, it can be noted that
the value of $\Omega/\omega_{0}^{2}$ also allows us to select the
phase shifting performed by $n_{2}$. Specifically, bearing in mind
that $R_{1}=0$ and $T_{1}=1$, we find from Eq.\,(\ref{eq:S2.12})
that $R_{2}=0$ and:
\begin{align}
T_{2} & =\left|T_{2}\right|\exp\left(\textrm{i}\Phi_{T_{2}}\right)=\exp\left[-\textrm{i}\left(\pi+2\arctan\frac{1}{\sqrt{\Omega/\omega_{0}^{2}-1}}\right)\right].\tag{S3.4}\label{eq:S3.4}
\end{align}
Fig.\,S3.2 compares Eq.\,(\ref{eq:S3.4}) with the numerical results
calculated by solving Eq.\,(\ref{eq:S1.8}) with COMSOL taking $n_{2,-}=2$,
$\omega_{0}=38$ rad/s and $c_{0}=1$ m/s to guarantee a low computational
time of the numerical simulations.\footnote{The conclusions detailed below and in the next numerical examples
are found to be valid for any value of $\omega_{0}$ and $c_{0}$.} We can note that the transmitted amplitude $T_{2}$ calculated with
T-SUSY is in good agreement with the numerical results of the wave
equation.

These graphics, along with Fig.\,2 of the paper, could be of great
interest to design and synthesize a\emph{ }perfect omnidirectional,
polarization-independent, transparent and reconfigurable phase shifter
using, e.g., electro-optic modulators at microwave frequencies ($T_{0}\in[3,3000]$
ps), where the required index modulation speed ($\Delta t\geq T_{0}$)
may be implementable with current electro-optic technology \cite{key-4,key-5}.
From the selected value $\Omega/\omega_{0}^{2}$, we can directly
estimate the performed phase shifting (Fig.\,S3.2), the required
$\Delta n$ and $\Delta t$ (Fig.\,S3.1), and the spectral band of
transparency (Fig.\,2). Remarkably, these results can be implemented
in all-dielectric, all-magnetic materials, or a combination of both.
In the latter case, despite the fact that we must assume a slowly-varying
temporal evolution in one of the constitutive parameters (see Section\,\ref{sec:1}),
$n_{\textrm{T}2}$ may also present rapidly-varying temporal fluctuations.
Moreover, note that these results are independent of the value of
$\omega_{0}$ and only depend on the ratio $\Omega/\omega_{0}^{2}$.
Accordingly, they can be directly extrapolated to a different angular
frequency.

\newpage{}
\noindent \begin{center}
\includegraphics[width=15cm,height=7cm,keepaspectratio]{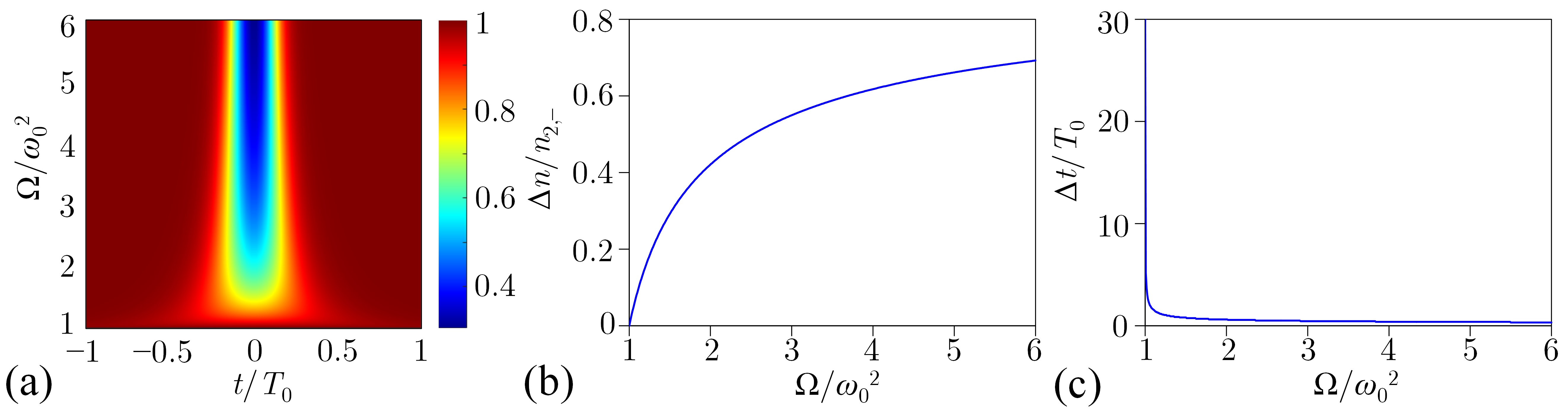}
\par\end{center}

\noindent \textbf{\small{}Figure S3.1.}{\small{} Hyperbolic secant
modulation Eq.\,(\ref{eq:S3.3}). (a) Normalized refractive index
profile $n_{\textrm{T}2}\left(t\right)/n_{2,-}$, (b) normalized maximal
excursion $\Delta n/n_{2,-}$, and (c) normalized temporal width $\Delta t/T_{0}$,
with $T_{0}=2\pi/\omega_{0}$. All graphics have been normalized to
guarantee the same results for any value of $\omega_{0}$ and $n_{2,-}$.\\}{\small \par}
\noindent \begin{center}
\includegraphics[width=9cm,height=6cm,keepaspectratio]{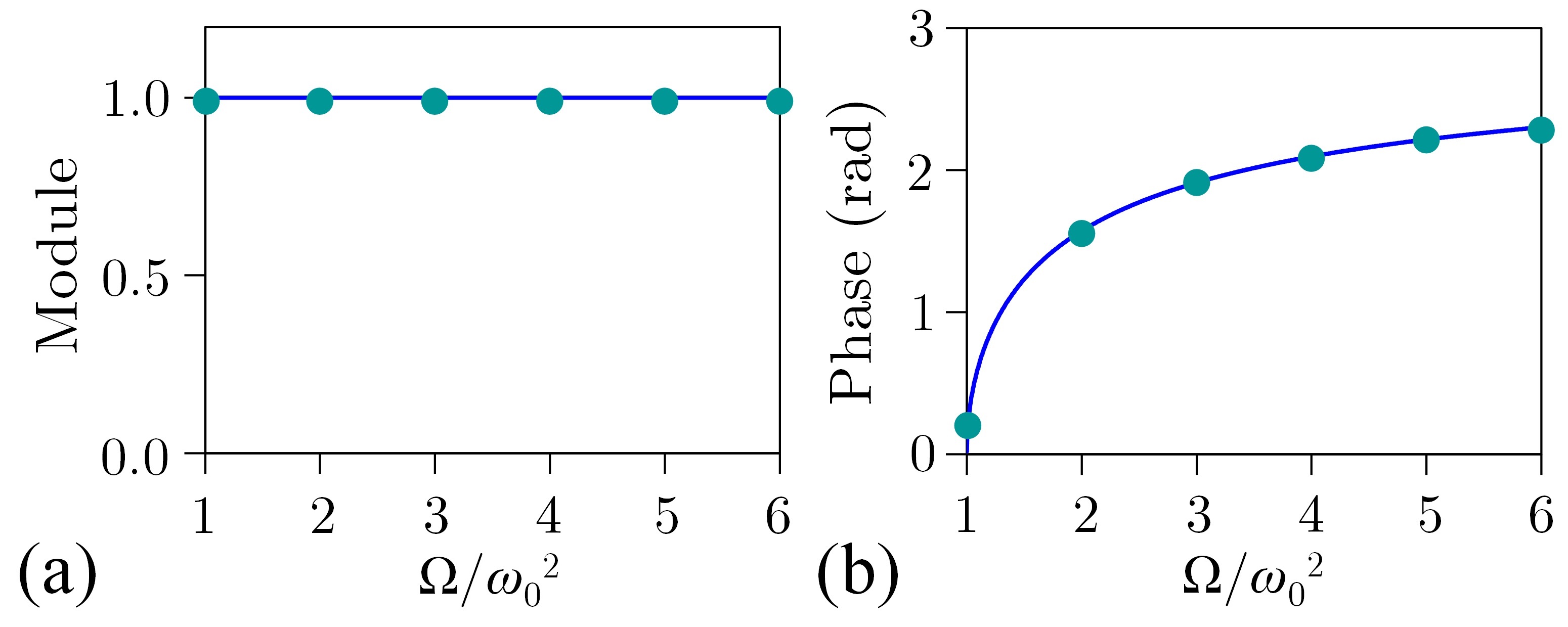}
\par\end{center}

\noindent \textbf{\small{}Figure S3.2.}{\small{} Scattering coefficient
$T_{2}$ given by Eq.\,(\ref{eq:S3.4}) (blue line) and calculated
numerically from Eq.\,(\ref{eq:S1.8}) using COMSOL Multiphysics
(dots). (a) Module and (b) phase as a function of the ratio $\Omega/\omega_{0}^{2}$.\\}{\small \par}

In addition, as seen in Fig.\,2(e), $\Phi_{T_{2}}$ shows a flat
frequency response. Consequently, in our phase shifter, the reconfigurable
and frequency-independent response of the phase could be of extreme
utility for wavelength-division multiplexing (WDM) transmissions of
narrow-band signals to generate the same phase shifting in each WDM
channel. In such a scenario, if we use electro-optic modulators with
a reduced $\Delta n$ excursion ($\Delta n\sim10^{-3}$), we must
operate in the range $\Omega/\omega_{0}^{2}<1.01$, where $\Phi_{T_{2}}<1$
rad. In these circumstances, it would be of interest to us to concatenate
a chain of non-reflecting hyperbolic secant modulations to increase
the phase shifting induced by a single hyperbolic secant. Concretely,
the hyperbolic secant chain (HSC) can be described as:
\begin{equation}
n_{\textrm{HSC}}\left(\mathbf{r},t\right)=\sum_{k=0}^{N_{\textrm{HSC}}-1}n_{\textrm{T}2}\left(t+\left(\frac{N_{\textrm{HSC}}-1}{2}-k\right)\mathcal{T}_{\textrm{HSC}}\right)-\left(N_{\textrm{HSC}}-1\right)n_{2,-},\tag{S3.5}\label{eq:S3.5}
\end{equation}
where $\mathcal{T}_{\textrm{HSC}}$ is the fundamental period of the
chain and $N_{\textrm{HSC}}$ is the number of fundamental periods
{[}see Fig.\,S3.3(a){]}. In order to analyse the scattering behaviour
of $n_{\textrm{HSC}}$, we numerically calculate the scattering coefficients
$R_{\textrm{HSC}}$ and $T_{\textrm{HSC}}$ as a function of the ratio
$\mathcal{T}_{\textrm{HSC}}/T_{\textrm{P}}$, where $T_{\textrm{P}}$
is the full-width at $1/\left(2e\right)$ of the peak power of the
incident pulse. Figure S3.3 shows the numerical results of $R_{\textrm{HSC}}$
and $T_{\textrm{HSC}}$ taking: $\Omega/\omega_{0}^{2}=1.01$ {[}Fig.\,S3.3(b){]}
and $\Omega/\omega_{0}^{2}=6$ {[}Fig.\,S3.3(c){]} in Eq.\,(\ref{eq:S3.5}),
$N_{\textrm{HSC}}\in\left\{ 2,3,4\right\} $, $\mathcal{T}_{\textrm{HSC}}/T_{\textrm{P}}\in\left(0,2\right]$,
and $T_{\textrm{P}}=1$ s. We can observe that the module and phase
of $R_{\textrm{HSC}}$ and $T_{\textrm{HSC}}$ have a flat behaviour
as a function of $\mathcal{T}_{\textrm{HSC}}/T_{\textrm{P}}$, even
if $\mathcal{T}_{\textrm{HSC}}<T_{\textrm{P}}$. Hence, as expected,
we can use an HSC to increase the phase shifting of the original hyperbolic
secant {[}$\Phi_{T_{\textrm{HSC}}}\left(N_{\textrm{HSC}}\right)=N_{\textrm{HSC}}\Phi_{T_{2}}${]}
maintaining its non-reflecting nature. \newpage{}
\noindent \begin{center}
\includegraphics[width=12cm,height=10cm,keepaspectratio]{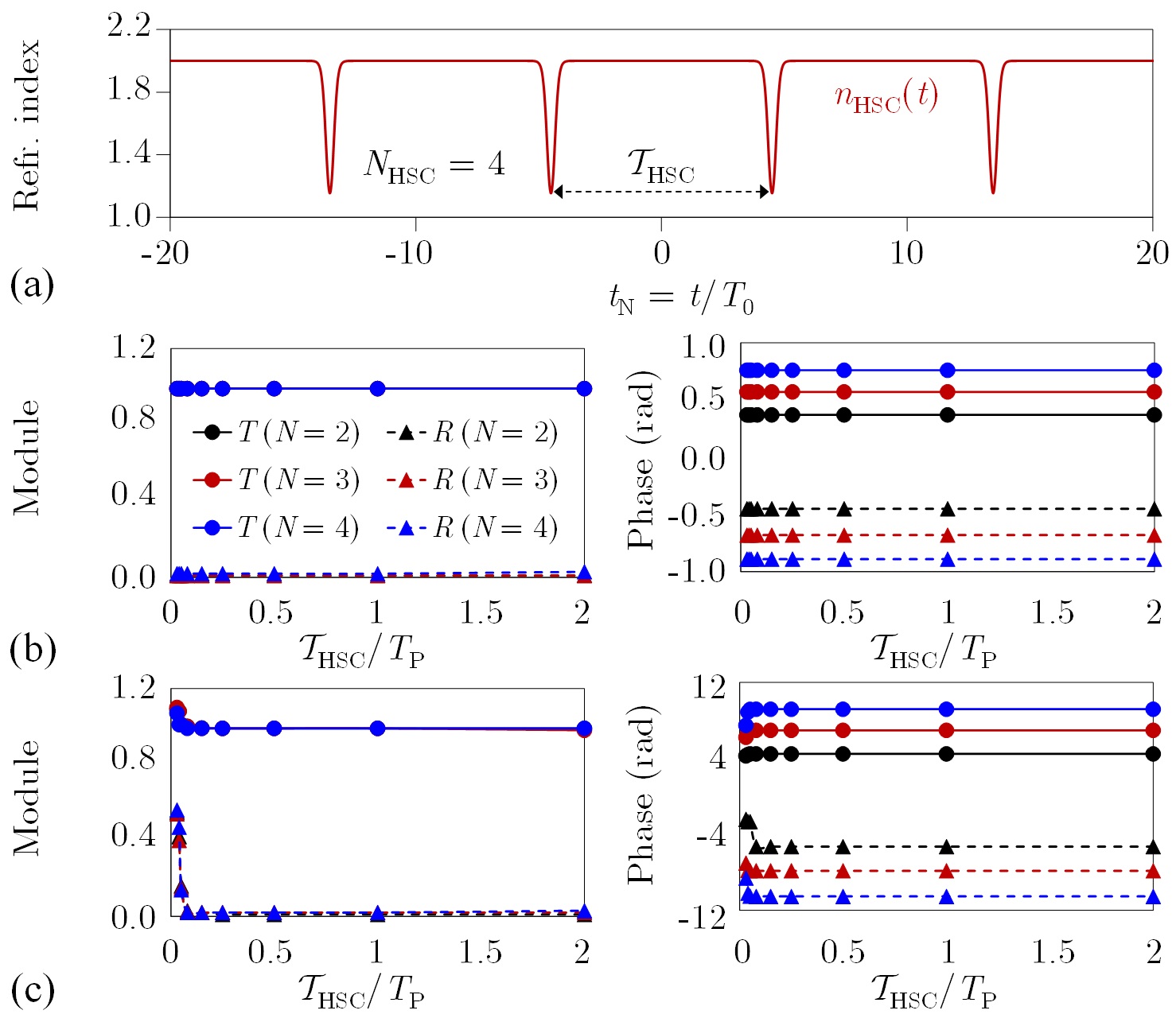}
\par\end{center}

\noindent \textbf{\small{}Figure S3.3.}{\small{} Chain of non-reflecting
hyperbolic secant refractive index profiles. (a) Refractive index
profile $n_{\textrm{HSC}}$. Scattering coefficients $T_{\textrm{HSC}}$
and $R_{\textrm{HSC}}$ calculated numerically as a function of the
ratio $\mathcal{T}_{\textrm{HSC}}/T_{\textrm{P}}$, where $T_{\textrm{P}}$
is the full-width at $1/\left(2e\right)$ of the peak power of the
incident pulse for: (b) $\Omega/\omega_{0}^{2}=1.01$ and (c)\,$\Omega/\omega_{0}^{2}=6$.
The number of fundamental periods $N_{\textrm{HSC}}$ of the chain
ranges from $2$ to $4$. The legend of (b) also applies to (c). }\linebreak{}
{\small{}We omit the subscripts of $T_{\textrm{HSC}}$, $R_{\textrm{HSC}}$
and $N_{\textrm{HSC}}$ in the legend due to space constraints.}{\small \par}
\noindent \begin{center}
\includegraphics[width=11.5cm,height=6cm,keepaspectratio]{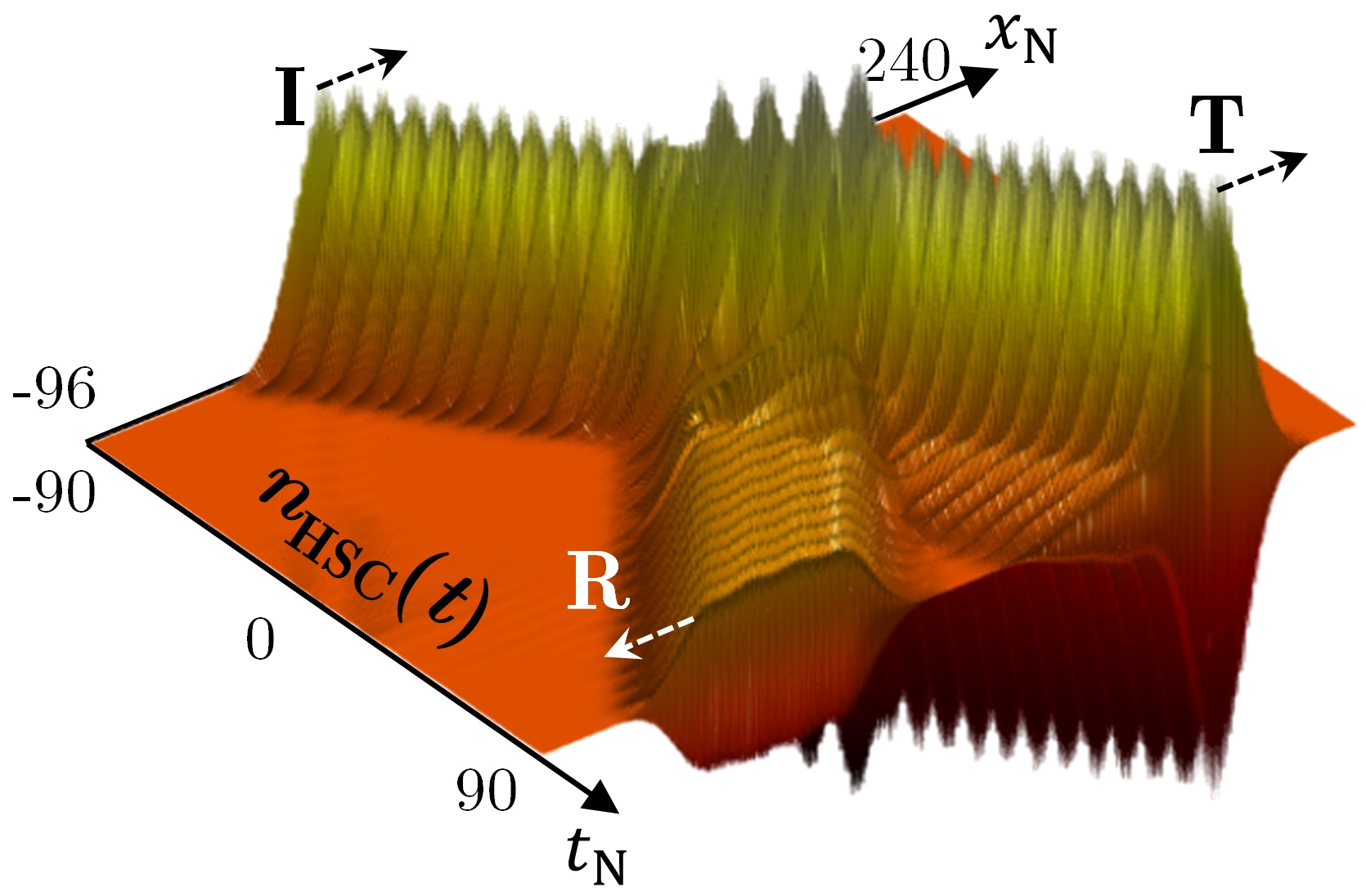}
\par\end{center}

\noindent \textbf{\small{}Figure S3.4.}{\small{} Pulse propagation
evolution through the time-varying medium described by $n_{\textrm{HSC}}(t)$.
Reflected (R) and transmitted (T) optical pulses generated from the
interaction of an incident (I) optical pulse with an hyperbolic secant
chain operating at $\omega=1.4\omega_{0}$. The transmitted pulse
has a higher peak power than the incident pulse because the energy
is not conserved in a time-varying system ($\textrm{I}\neq\textrm{R}+\textrm{T}$).
The temporal axis is normalized as $t_{\textrm{N}}=t/T_{0}$ with
$T_{0}=2\pi/\omega_{0}$. The $x$-axis is normalized as $x_{\textrm{N}}=x/\lambda$,
with $\lambda=\lambda_{0}/n_{2,-}$ and $\lambda_{0}=2\pi c_{0}/\omega_{0}$.}{\small \par}

\noindent {\small{}\newpage{}}{\small \par}

On the other hand, the reflecting behaviour of $n_{\textrm{T}2}\left(t\right)$
at $\omega\neq\omega_{0}$ and the nonlinear nature of $\Phi_{T_{2}}\left(\omega\right)$
{[}see Fig.\,2(e){]} can be exploited to implement pulse shaping
operations in reflection and transmission. Specifically, in reflection,
we can build a flat-top optical pulse using the HSC. The chain generates
$N_{\textrm{HSC}}$ reflected pulses of temporal width $T_{\textrm{P}}$
and separated $\mathcal{T}_{\textrm{HSC}}$ in time. Hence, selecting
$T_{\textrm{P}}\sim\mathcal{T}_{\textrm{HSC}}$, we will obtain a
flat-top optical pulse emerging from the superposition of all the
reflected pulses. Figure S3.4 illustrates this basic idea. We build
an HSC with $\Omega/\omega_{0}^{2}=6$, $\omega=1.4\omega_{0}$, $\mathcal{T}_{\textrm{HSC}}/T_{\textrm{P}}=1.2$,
$T_{\textrm{P}}=1$\,s and $N_{\textrm{HSC}}=4$. As seen, a flat-top
optical pulse can be observed in reflection at the end of the chain.
Moreover, the tail of the transmitted pulse has been distorted due
to the nonlinear nature of $\Phi_{T_{2}}\left(\omega\right)$.\footnote{The nonlinear frequency dependence of the phase $\Phi_{T_{2}}\left(\omega\right)$
implies that each spectral component of $\mathbf{D}\left(\mathbf{r},t\right)$
is phase shifted by a different quantity. Hence, the envelope of an
incident wide-band pulse will be distorted in transmission.} This can be further investigated in future works to generate transmitted
optical pulses with an exotic shape, e.g., for optical wavelet transforms,
coherent laser control of physicochemical processes, or spectrally
selective nonlinear microscopy among other application areas \cite{key-6,key-7,key-8}.
On the contrary, the pulse distortion induced by the nonlinear frequency
response of $\Phi_{T_{2}}\left(\omega\right)$ can be reduced if we
operate with narrow-band incident pulses.

So far, we have extensively evaluated the SUSY refractive index of
$n_{1}\left(\mathbf{r},t\right)=n_{1,-}$ by considering $\Omega>\omega_{0}^{2}$.
Now, we focus our attention on the case $\Omega=\omega_{0}^{2}$.
In such circumstances, $V_{1}\left(t\right)=0$, $W\left(t\right)=1/\left(C-t\right)$
and:
\begin{equation}
n_{2}\left(\mathbf{r},t\right)=n_{\textrm{T}2}\left(t\right)=\frac{n_{2,-}}{\sqrt{1-\frac{2}{\omega_{0}^{2}\left(C-t\right)^{2}}}},\tag{S3.6}\label{eq:S3.6}
\end{equation}
where $C$ is an integration constant arising from Riccati's equation.
Figure S3.5 illustrates the SUSY refractive index profile given by
the above equation. It is fundamental to note that now we have a singular
superpotential that may break the degeneracy of the eigenvalue spectra
between both superpartners ($\Omega^{\left(1\right)}\neq\Omega^{\left(2\right)}$).
More precisely, note that $\mathbf{D}^{\left(2\right)}\propto\left(\partial_{t}+W\left(t\right)\right)\mathbf{D}^{\left(1\right)}$
is not a continuous function at $t=C$. Therefore, $\mathbf{D}^{\left(2\right)}$
cannot be a solution of Maxwell's equations. The boundary conditions
of the temporal wave equation {[}Eq.\,(3){]} are not satisfied ($\mathbf{D}^{\left(2\right)}$
and $\partial_{t}\mathbf{D}^{\left(2\right)}$ must be necessarily
continuous functions in the temporal variable). As a result, we cannot
guarantee the same intensity scattering properties for $n_{1}$ and
$n_{2}$.\\
\noindent \begin{center}
\includegraphics[width=14cm,height=7cm,keepaspectratio]{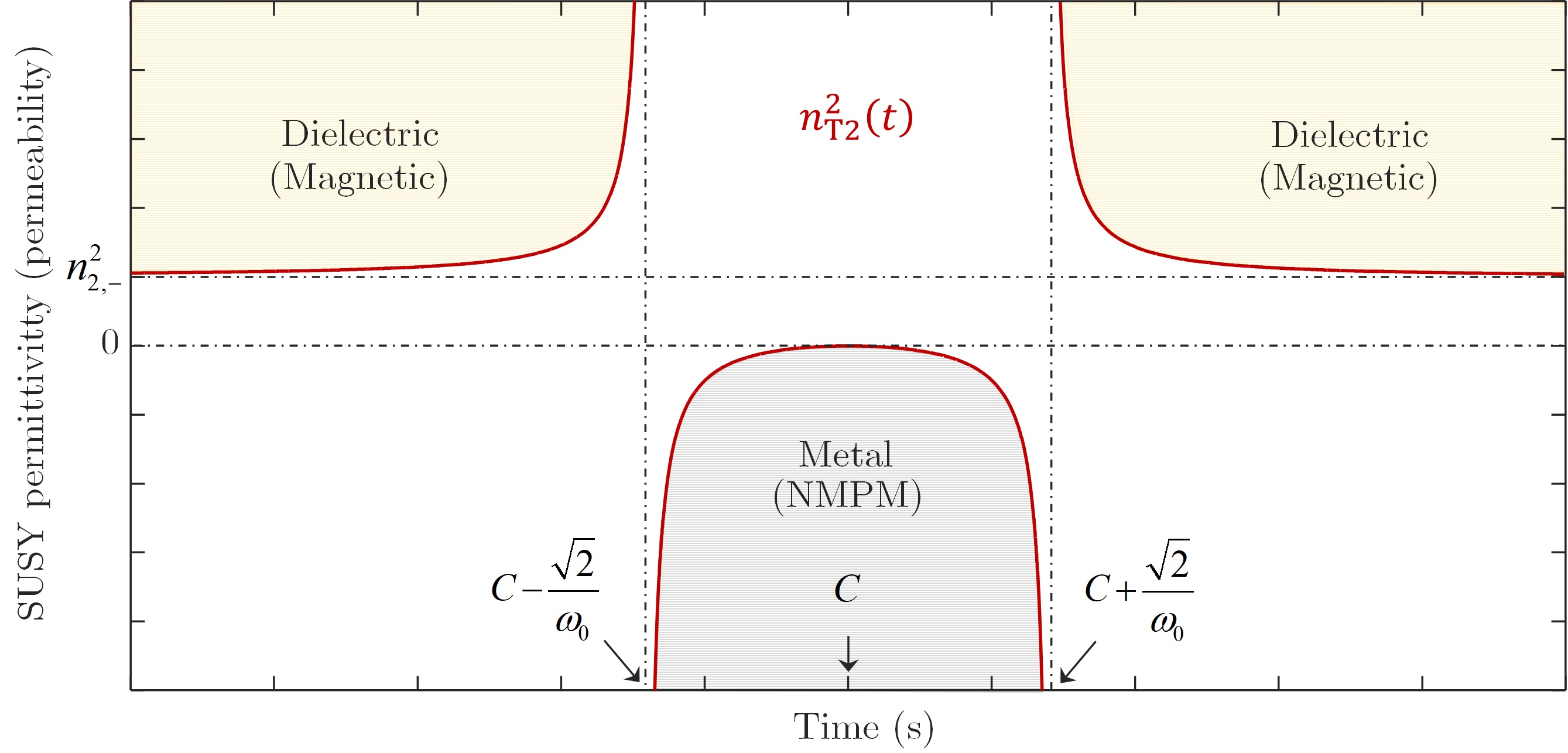}
\par\end{center}

\noindent \textbf{\small{}Figure S3.5.}{\small{} SUSY permittivity
(permeability) $n_{\textrm{T}2}^{2}$ {[}Eq.\,(\ref{eq:S3.6}){]}
of a constant permittivity (permeability) $n_{1,-}^{2}$ when $\Omega=\omega_{0}^{2}$.
(NMPM: negative magnetic permeability material). \\}{\small \par}

\pagebreak{}

\subsubsection*{Hyperbolic Rosen-Morse II potential: general expressions}

In the next two numerical examples, we will analyse the SUSY refractive
index profiles arising from the shape invariant hyperbolic Rosen-Morse
II (HRMII) potentials, described by the following equations:
\begin{align}
W\left(t\right) & =A\tanh\left(\alpha t\right)+B/A;\tag{S3.7}\label{eq:S3.7}\\
V_{1}\left(t\right) & =A^{2}+B^{2}/A^{2}-A\left(A+\alpha\right)\textrm{sech}^{2}\left(\alpha t\right)+2B\tanh\left(\alpha t\right);\tag{S3.8}\label{eq:S3.8}\\
V_{2}\left(t\right) & =A^{2}+B^{2}/A^{2}-A\left(A-\alpha\right)\textrm{sech}^{2}\left(\alpha t\right)+2B\tanh\left(\alpha t\right),\tag{S3.9}\label{eq:S3.9}
\end{align}
with $A$, $B$ and $\alpha$ real parameters and $B<A^{2}$. From
the above expressions, we can verify that the SIP condition {[}Eq.\,(\ref{eq:S2.18}){]}
is fulfilled with $a_{1}\equiv A$, $a_{2}=f\left(a_{1}\right)=a_{1}-\alpha$
and $M\left(a_{1}\right)=a_{1}^{2}-\left(a_{1}-\alpha\right)^{2}+B^{2}[1/a_{1}^{2}-1/\left(a_{1}-\alpha\right)^{2}]$.
Taking $\Omega\left(a_{1}\right)=\omega_{0}^{2}+W_{-}^{2}=\omega_{0}^{2}+\left(B/a_{1}-a_{1}\right)^{2}$,
the corresponding refractive index profiles are:
\begin{align}
n_{\textrm{T}1}\left(t;a_{1}\right) & =\frac{\omega_{0}n_{1,-}\left(a_{1}\right)}{\sqrt{\omega_{0}^{2}-2B+a_{1}\left(a_{1}+\alpha\right)\textrm{sech}^{2}\left(\alpha t\right)-2B\tanh\left(\alpha t\right)}};\tag{S3.10}\label{eq:S3.10}\\
n_{\textrm{T}2}\left(t;a_{1}\right) & =\frac{\omega_{0}n_{2,-}\left(a_{1}\right)}{\sqrt{\omega_{0}^{2}-2B+a_{1}\left(a_{1}-\alpha\right)\textrm{sech}^{2}\left(\alpha t\right)-2B\tanh\left(\alpha t\right)}}.\tag{S3.11}\label{eq:S3.11}
\end{align}
Equation (\ref{eq:S3.10}) is Eq.\,(6) of the paper. We also observe
that $\left|W_{-}\right|=\left|W_{+}\right|$ ($\left|W_{-}\right|\neq\left|W_{+}\right|$)
if $B=0$ ($B\neq0$). Hence, we have the ability of constructing
time-varying optical systems with $\omega_{i,-}=\omega_{i,+}$ ($\omega_{i,-}\neq\omega_{i,+}$).
In the case $B\neq0$, we should take into account that:
\begin{itemize}
\item The scattering coefficients of the electric (magnetic) flux density
$R_{i}$ and $T_{i}$ are found to be different from the scattering
coefficients of the electric (magnetic) field strength $\mathcal{R}_{i}$
and $\mathcal{T}_{i}$. More precisely, $\mathcal{R}_{i}=N_{+}^{2}R_{i}$
and $\mathcal{T}_{i}=N_{+}^{2}T_{i}$, with $N_{+}^{2}=1-4B/\omega_{0}^{2}\in\left(0,1\right)$.
Nonetheless, $\mathcal{R}_{1}/\mathcal{R}_{2}=R_{1}/R_{2}$ and $\mathcal{T}_{1}/\mathcal{T}_{2}=T_{1}/T_{2}$.
\item From the previous point, we deduce that $0<N_{+}<1$, $n_{i,-}<n_{i,+}$
and $B<\omega_{0}^{2}/4$.
\item From the temporal version of Snell's law, we infer that the frequency
of the transmitted signal is lower than that of the incident signal,
with $\omega_{i,+}=N_{+}\omega_{i,-}$.
\end{itemize}
In both cases ($B=0$ and $B\neq0$), the scattering relations {[}Eq.\,(\ref{eq:S2.12}){]}
become:
\begin{align}
\frac{R_{1}\left(a_{1}\right)}{R_{2}\left(a_{1}\right)} & =\frac{R_{1}\left(a_{1}\right)}{R_{1}\left(a_{2}\right)}=\frac{a_{1}+B/a_{1}+\textrm{i}\sqrt{\omega_{0}^{2}-4B}}{-a_{1}+B/a_{1}-\textrm{i}\omega_{0}};\tag{S3.12}\label{eq:S3.12}\\
\frac{T_{1}\left(a_{1}\right)}{T_{2}\left(a_{1}\right)} & =\frac{T_{1}\left(a_{1}\right)}{T_{1}\left(a_{2}\right)}=\frac{a_{1}+B/a_{1}-\textrm{i}\sqrt{\omega_{0}^{2}-4B}}{-a_{1}+B/a_{1}-\textrm{i}\omega_{0}}.\tag{S3.13}\label{eq:S3.13}
\end{align}
Figure S3.6 shows the above ratios as a function of $\omega_{0}$
and $a_{1}$ setting $B=0.4a_{1}^{2}$. As seen, the module is equal
to $1$ when $N_{+}\in\mathbb{R}$ and the phase has a low dependence
on $\omega_{0}$. Furthermore, the lower the value of $a_{1}$ is,
the lower the frequency dependence of the phase and the area where
$N_{+}\in\mathbb{C}$ are.
\noindent \begin{center}
\includegraphics[width=15cm,height=8.7cm,keepaspectratio]{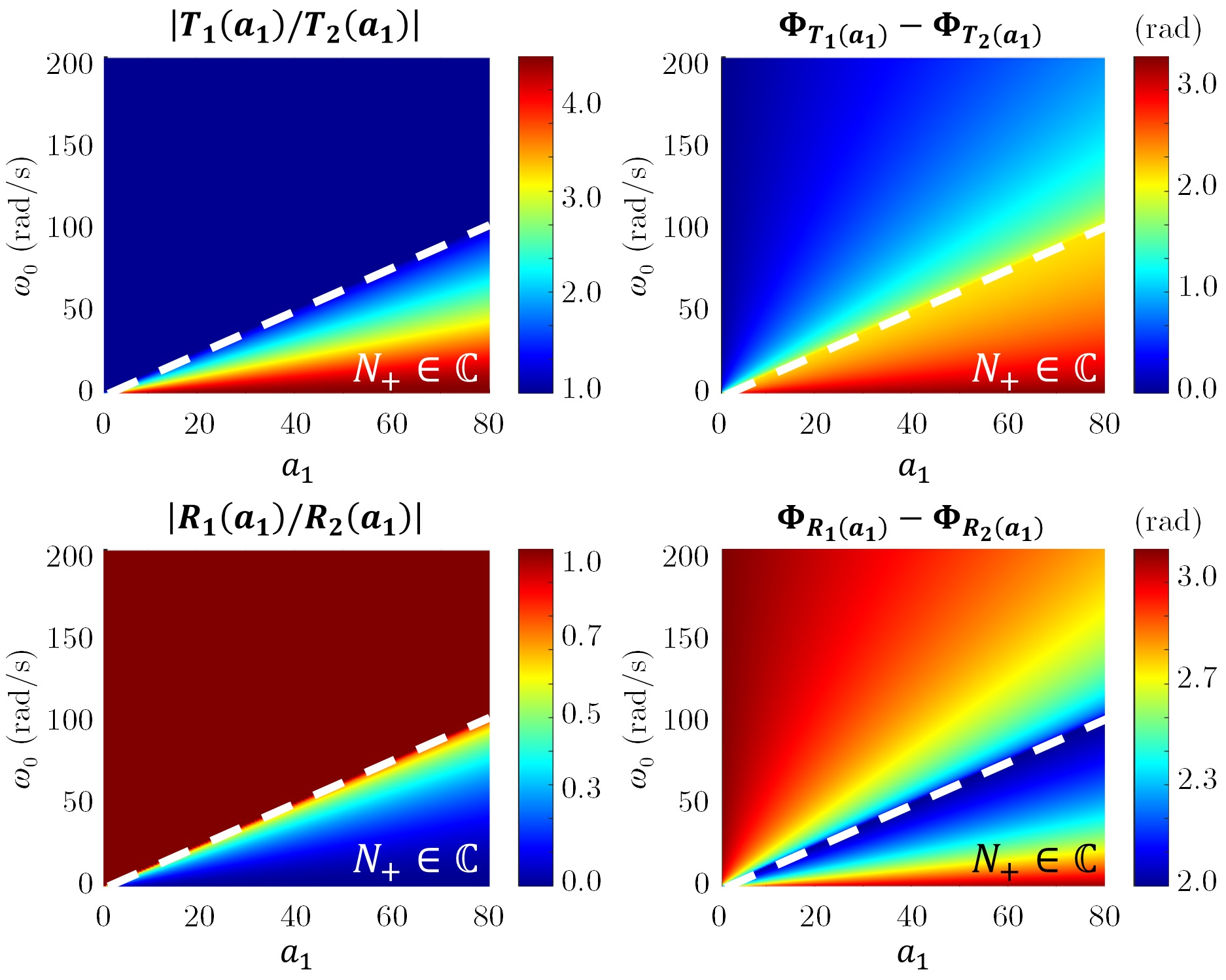}
\par\end{center}

\noindent \textbf{\small{}Figure S3.6.}{\small{} Ratios of the scattering
coefficients in the HRMII potential. The dashed white line separates
the allowed ($N_{+}\in\mathbb{R}$, top) and forbidden ($N_{+}\in\mathbb{C}$,
bottom) regions.}{\small \par}

\subsubsection*{Hyperbolic Rosen-Morse II potential: case $\boldsymbol{B}=\boldsymbol{0}$\label{subsec:HRMII B=00003D0}}

The case $B=0$ leads to reciprocal, omnidirectional, polarization-independent
and transparent phase shifters such as the hyperbolic secant profile
of the first numerical example. Specifically, taking $\alpha=a_{1}$\linebreak{}
in Eq.\,(\ref{eq:S3.10}) we retrieve the aforementioned refractive
index profile {[}Eq.\,(\ref{eq:S3.3}){]}. Nevertheless, setting
$\alpha=1$ and $a_{1}=m\in\mathbb{Z}$ we will find new refractive
index modulations with an extremely large transparent optical bandwidth.
In such a scenario, Eqs.\,(\ref{eq:S3.10}) and (\ref{eq:S3.11})
are reduced to:
\begin{align}
n_{\textrm{T}1}\left(t;m\right) & =\frac{\omega_{0}n_{1,-}\left(m\right)}{\sqrt{\omega_{0}^{2}+m\left(m+1\right)\textrm{sech}^{2}t}};\tag{S3.14}\label{eq:S3.14}\\
n_{\textrm{T}2}\left(t;m\right) & =\frac{\omega_{0}n_{2,-}\left(m\right)}{\sqrt{\omega_{0}^{2}+m\left(m-1\right)\textrm{sech}^{2}t}},\tag{S3.15}\label{eq:S3.15}
\end{align}
obeying the SIP relation Eq.\,(\ref{eq:S2.22}), which can be rewritten
as:
\begin{equation}
n_{\textrm{T}2}\left(t;m\right)=\frac{n_{2,-}\left(m\right)}{n_{1,-}\left(m-1\right)}n_{\textrm{T}1}\left(t;m-1\right).\tag{S3.16}\label{eq:S3.16}
\end{equation}
From this recurrence relation, we can infer the transparent behaviour
of these systems taking into account that they are SUSY-connected
with the potential of the free particle ($m=1$). Nevertheless, in
contrast to the first numerical example {[}Eq.\,(\ref{eq:S3.3}){]},
we can note that the HRMII has the advantage of allowing independent
design control over $\Delta n\sim n_{1,2,-}\left|1-\omega_{0}/\sqrt{\omega_{0}^{2}+m\left(m\pm1\right)}\right|$
for a fixed $\Delta t\sim20T_{0}$, enabling a technology-oriented
adjustment of the index modulation contrast. Figure S3.7(a) shows
the SUSY refractive index profiles for the case $m=30$, and Fig.\,S3.7(b)
depicts their scattering coefficients as a function of the frequency.
As commented above, we can observe an extremely large spectral band
of transparency.

\newpage{}
\noindent \begin{center}
\includegraphics[width=12cm,height=10cm,keepaspectratio]{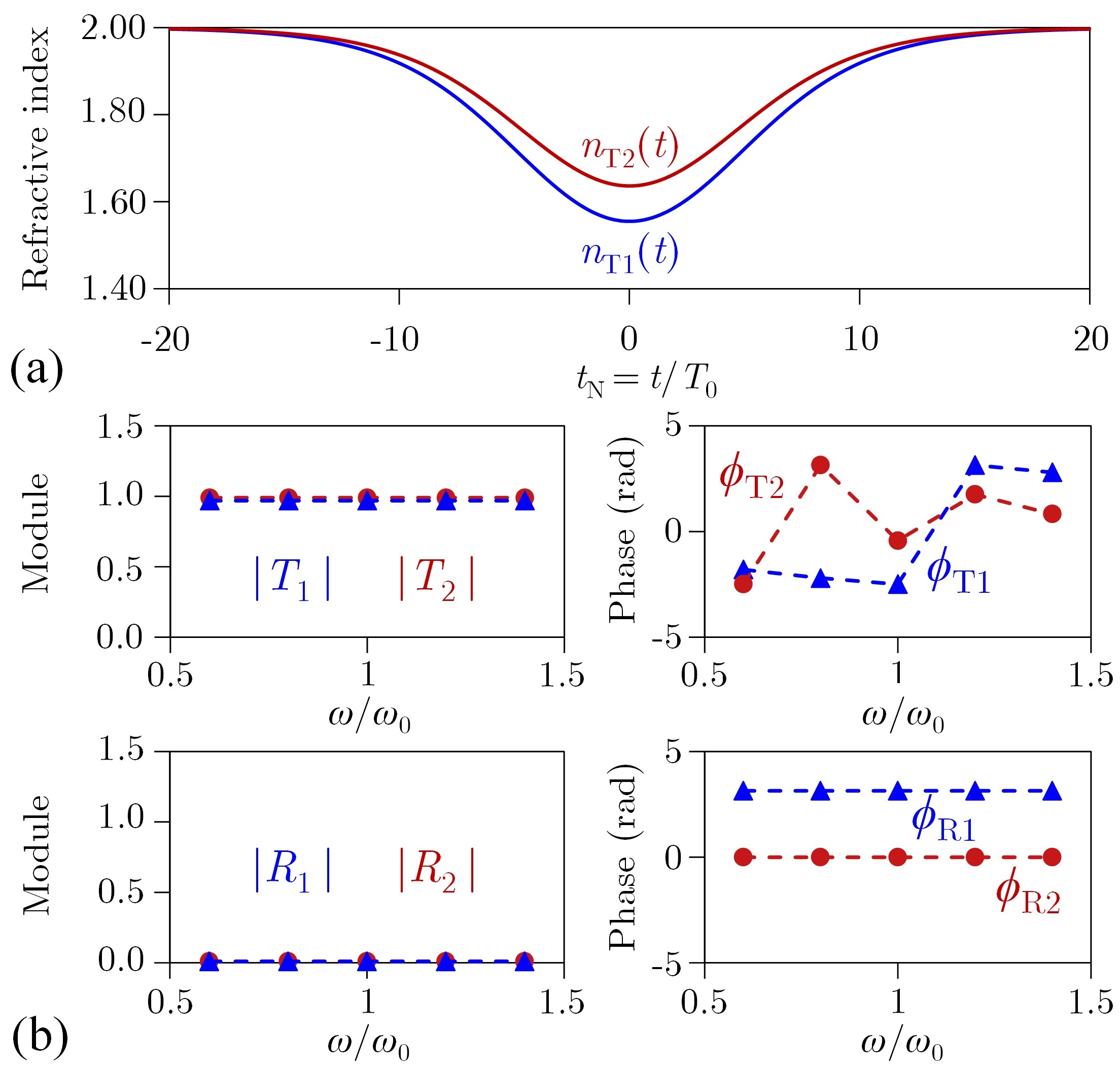}
\par\end{center}

\noindent \textbf{\small{}Figure S3.7.}{\small{} (a) SUSY refractive
index profiles given by Eqs.\,(\ref{eq:S3.14}) and (\ref{eq:S3.15})
with $m=30$. (b) Scattering coefficients calculated numerically as
a function of the ratio $\omega/\omega_{0}$.\\}{\small \par}

On the other hand, in this case we can report an analytic solution
for $T_{1}\left(m\right)$, which can be calculated in the same way
as the transmitted probability amplitude associated with the potential
$V_{1}\left(x;m\right)=m^{2}+\omega_{0}^{2}[1-N_{1}^{2}\left(t\rightarrow x;m\right)]$
in quantum mechanics (see page 299 of \cite{key-1}):
\begin{equation}
T_{1}\left(m\right)=\frac{\Gamma\left(-m-\textrm{i}\omega_{0}\right)\Gamma\left(m+1-\textrm{i}\omega_{0}\right)}{\Gamma\left(-\textrm{i}\omega_{0}\right)\Gamma\left(1-\textrm{i}\omega_{0}\right)},\tag{S3.17}\label{eq:S3.17}
\end{equation}
where $\Gamma$ is the Gamma function. Figure S3.8 shows an excellent
fitting between the theoretical and numerical transmission coefficient
associated with $n_{\textrm{T}1}\left(t;m\right)$. Strikingly, Eq.\,(\ref{eq:S3.17})
can be combined with Eq.\,(\ref{eq:S3.13}) to solve straightforwardly
the temporal scattering problem in this family of time-varying systems
without using Maxwell's equations. Furthermore, we can select the
desired phase of the transmitted wave via the $m$ parameter.
\noindent \begin{center}
\includegraphics[width=10cm,height=6cm,keepaspectratio]{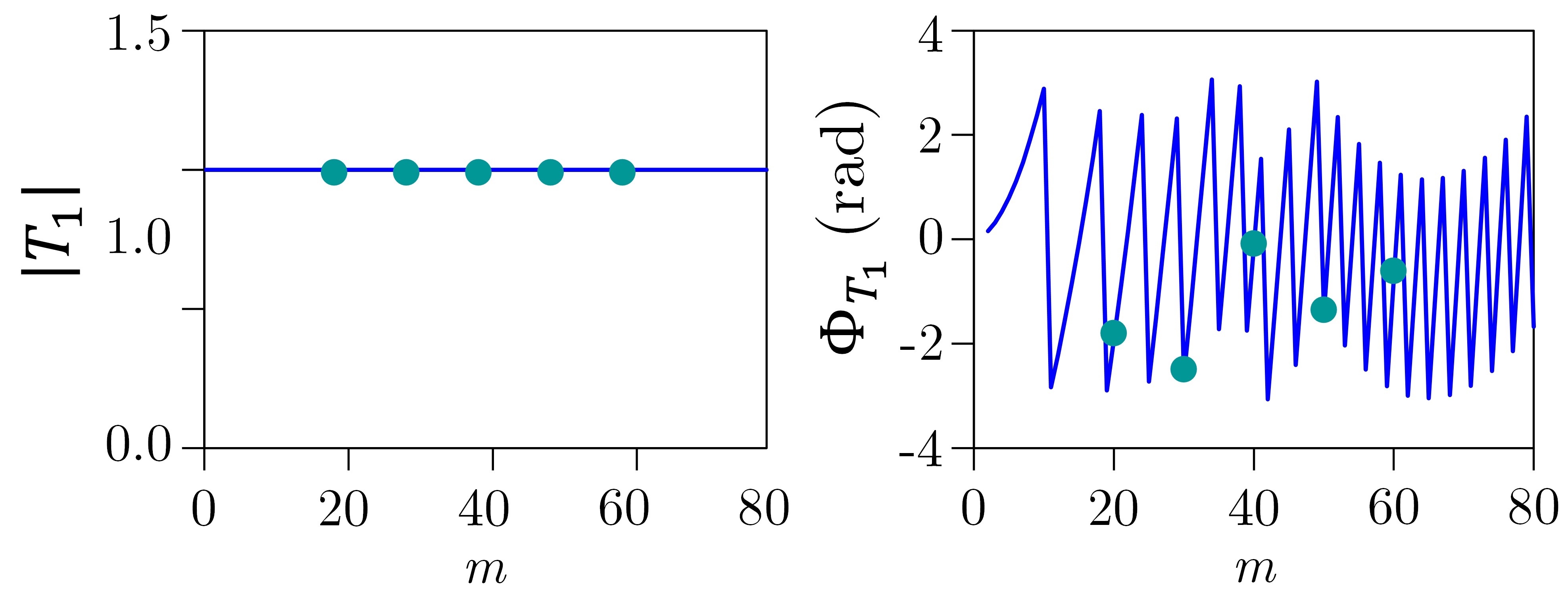}
\par\end{center}

\noindent \textbf{\small{}Figure S3.8.}{\small{} Transmission coefficient
associated with the refractive index profile given by Eq.\,(}\ref{eq:S3.14}{\small{})
as a function of $m$. {[}Solid line: Eq.\,(\ref{eq:S3.17}). Dots:
numerical results{]}.}{\small \par}

\newpage{}

\paragraph{\emph{Isospectral two-parameter family.} }

\noindent Figure 3(a) of the paper depicts the two-parameter isos-pectral
family $\widetilde{n}_{\textrm{T}1}\left(t;\eta_{1},\eta_{2}\right)$
of the HRMII index given by Eq.\,(\ref{eq:S3.14}). Here, we detail
how we have calculated $\widetilde{n}_{\textrm{T}1}\left(t;\eta_{1},\eta_{2}\right)$.
Taking $\Omega=\omega_{0}^{2}$ in Eq.\,(\ref{eq:S2.2}), the corresponding
quantum potential is of the form (step\,1 of the isospectral theory,
see page \pageref{eq:S2.34}):
\begin{equation}
V_{1}\left(t\rightarrow x\right)=\Omega-\omega_{0}^{2}\frac{n_{1,-}^{2}}{n_{\textrm{T}1}^{2}\left(t\rightarrow x;m\right)}=-m\left(m+1\right)\textrm{sech}^{2}x.\tag{S3.18}\label{eq:S3.18}
\end{equation}
In particular, $V_{1}$ holds $m$ bound states and the scattering
problem is well-defined ($V_{1,\pm}<\infty$). For the case $m=2$,
the two-parameter isospectral family $\widetilde{V}_{1}\left(x;\eta_{1},\eta_{2}\right)$
can be calculated from $V_{1}\left(x\right)$ by using the multi-parameter
Darboux procedure (step 2):
\begin{equation}
\widetilde{V}_{1}\left(x;\eta_{1},\eta_{2}\right)=-12\frac{3+4\cosh\left(2x-2\Lambda_{2}\right)+\cosh\left(4x-2\Lambda_{1}\right)}{\left[\cosh\left(3x-\Lambda_{2}-\Lambda_{1}\right)+3\cosh\left(x+\Lambda_{2}-\Lambda_{1}\right)\right]^{2}},\tag{S3.19}\label{eq:S3.19}
\end{equation}
with $\Lambda_{1,2}=-0.5\ln\left(1+1/\eta_{1,2}\right)$ and $\eta_{1,2}>0$.
Finally, performing the relabelling $\widetilde{V}_{1}\left(x\rightarrow t;\eta_{1},\eta_{2}\right)$
and using Eq.\,(\ref{eq:S2.34}), we obtain the two-parameter isospectral
family of time-varying optical systems (step\,3):
\begin{equation}
\widetilde{n}_{\textrm{T}1}\left(t;\eta_{1},\eta_{2}\right)=\frac{\widetilde{n}_{1,-}\left(\eta_{1},\eta_{2}\right)}{\sqrt{1-\frac{1}{\omega_{0}^{2}}\widetilde{V}_{1}\left(t;\eta_{1},\eta_{2}\right)}}.\tag{S3.20}\label{eq:S3.20}
\end{equation}
Figure 3(a) of the main text shows the refractive index profile of
different optical systems of the family taking $\widetilde{n}_{1,-}\left(\eta_{1},\eta_{2}\right)=n_{1,-}=2$.
We have numerically calculated the scattering coefficients of these
refractive index profiles and we have found the same reflected and
transmitted coefficients in module and phase as those of the original
modulation $n_{\textrm{T}1}\left(t;m=2\right)$ at $\omega=\omega_{0}$:
$R_{1}\left(\eta_{1},\eta_{2}\right)=0$ and $T_{1}\left(\eta_{1},\eta_{2}\right)=\exp\left(\textrm{i}0.23\right)$.

\subsubsection*{Hyperbolic Rosen-Morse II potential: case $\boldsymbol{B}\protect\neq\boldsymbol{0}$
(optical isolator)}

The case $B\neq0$ is of great interest to us given that it builds
a bridge to design non-reciprocal optical systems using time-varying
refractive index modulations, as commented above and in the main text.
In particular, from the temporal SIP theory developed above, the value
of $n_{\textrm{T}6}$ employed for the optical isolator demonstrated
in the paper can be obtained by replacing $a_{1}$ by $a_{6}=a_{1}-5\alpha$
in Eq.\,(\ref{eq:S3.10}). The frequency down-conversion between
the incident and transmitted waves, with angular frequencies $\omega_{0}\equiv\omega_{6,-}$
and $\omega_{6,+}$, respectively, can be calculated from the temporal
version of Snell's law $\omega_{6,-}n_{6,-}=\omega_{6,+}n_{6,+}$.
Hence, the ratio $\omega_{6,+}/\omega_{6,-}$ obeys the relation:
\begin{equation}
\frac{\omega_{6,+}}{\omega_{6,-}}=\frac{n_{6,-}}{n_{6,+}}=N_{+}=\sqrt{1-4B/\omega_{0}^{2}}.\tag{S3.21}\label{eq:S3.21}
\end{equation}
Figure S3.9 depicts the ratio $\omega_{6,+}/\omega_{6,-}$ for different
values of $\omega_{0}$ and $B$. The dashed line separates the allowed
($N_{+}\in\mathbb{R}$) and forbidden ($N_{+}\in\mathbb{C}$) areas.
In our case, we operate at $\omega_{6,+}/\omega_{6,-}\simeq0.7$ with
$\omega_{0}\equiv\omega_{6,-}=38$ rad/s. If we are interested in
synthesizing the optical isolator of the paper for a different angular
frequency, we must select an adequate value of $B$ that preserves
the same ratio in Eq.\,(\ref{eq:S3.21}).

On the other hand, the ratios $R_{1}\left(a_{1}\right)/R_{6}\left(a_{1}\right)$
and $T_{1}\left(a_{1}\right)/T_{6}\left(a_{1}\right)$ can be calculated
by combining Eq.\,(\ref{eq:S2.30}) with Eqs.\,(\ref{eq:S3.12})
and (\ref{eq:S3.13}). In particular, we find that:
\begin{equation}
\frac{T_{1}\left(a_{1}\right)}{T_{6}\left(a_{1}\right)}=\frac{T_{1}\left(a_{1}\right)}{T_{1}\left(a_{6}\right)}=\left(\frac{a_{1}+B/a_{1}-\textrm{i}\sqrt{\omega_{0}^{2}-4B}}{-a_{1}+B/a_{1}-\textrm{i}\omega_{0}}\right)^{5}=\exp\left(\textrm{i}2.55\right),\tag{S3.22}\label{eq:S3.22}
\end{equation}
which is in good agreement with Fig.\,3(c) at $\omega=\omega_{0}$.
The ratio $R_{1}\left(a_{1}\right)/R_{6}\left(a_{1}\right)$ could
not be numerically estimated because the non-reflecting behaviour
of $n_{\textrm{T}1}\left(t;a_{1}\right)$ and $n_{\textrm{T}6}\left(t;a_{1}\right)$
has a flat frequency response in an extremely large optical bandwidth
(see Fig.\,3). We could not observe any reflected wave in the numerical
simulation when propagating wide-band optical pulses through the above
media.
\noindent \begin{center}
\includegraphics[width=8cm,height=5cm,keepaspectratio]{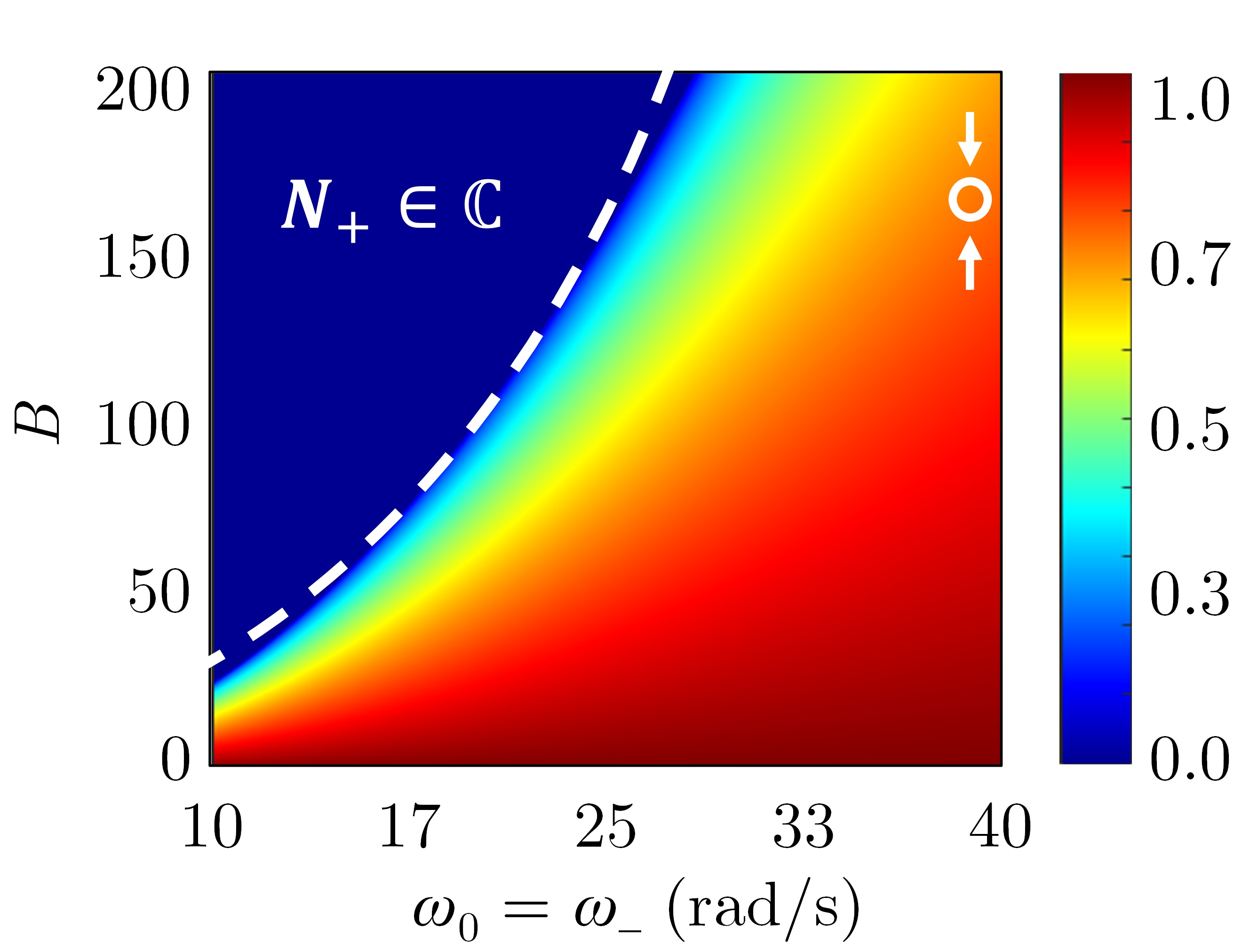}
\par\end{center}

\noindent \textbf{\small{}Figure S3.9.}{\small{} Frequency down-conversion
ratio $\omega_{6,+}/\omega_{6,-}$ as a function of $\omega_{0}$
and $B$. The dashed white line separates the allowed ($N_{+}\in\mathbb{R}$)
and forbidden ($N_{+}\in\mathbb{C}$) areas. The hollow circle indicates
the operation point of the optical isolator of the main text.\\}{\small \par}
\noindent \begin{center}
\includegraphics[width=13cm,height=7cm,keepaspectratio]{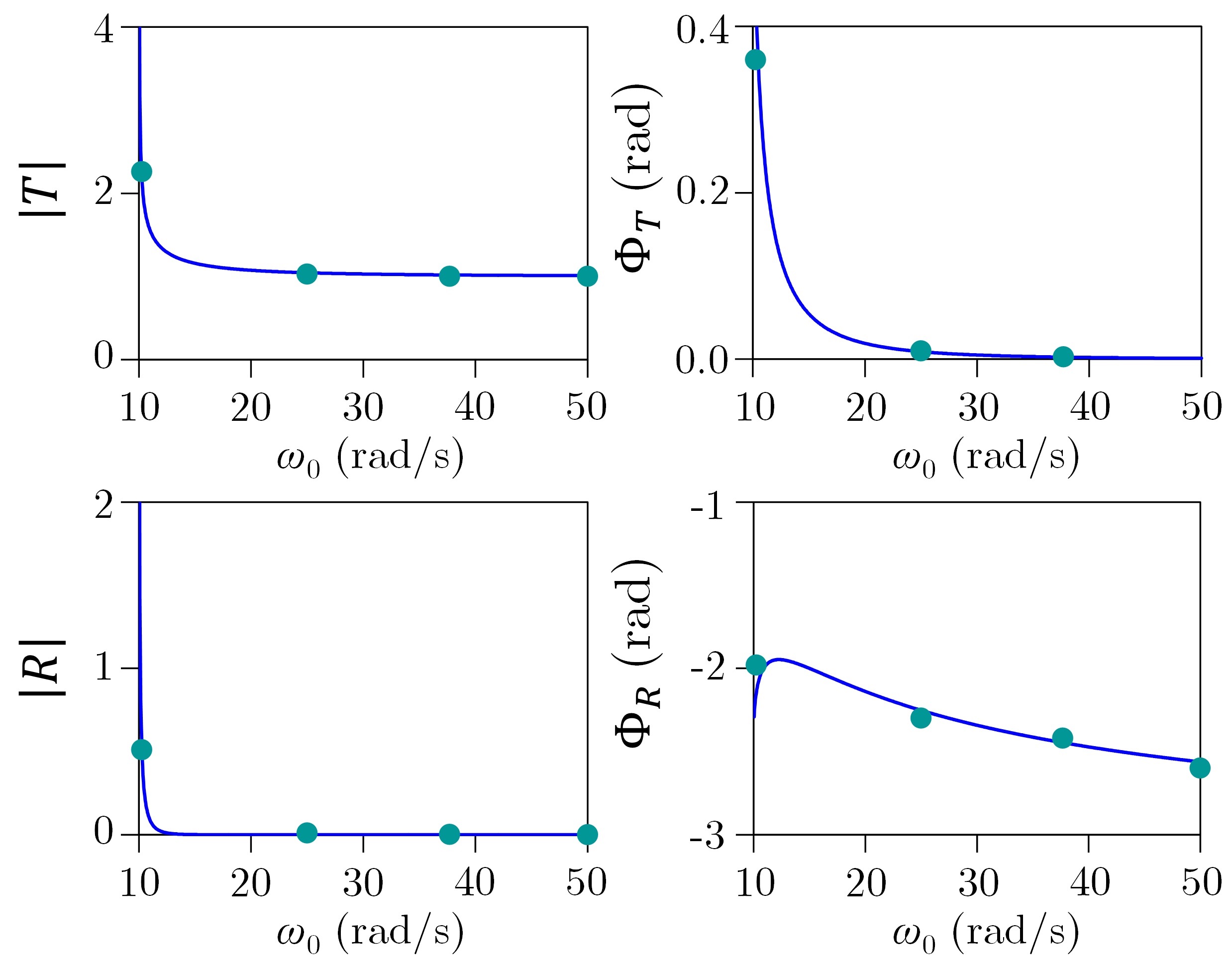}
\par\end{center}

\noindent \textbf{\small{}Figure S3.10.}{\small{} Scattering coefficients
of the hyperbolic step-index profile {[}Eq.\,(\ref{eq:S3.10}) taking
$\alpha=A=B^{1/2}${]} as a function of $\omega_{0}$ for the case
$\alpha=5$. {[}Solid line: Eqs.\,(\ref{eq:S3.23}) and (\ref{eq:S3.24}).
Dots: numerical results{]}.\\}{\small \par}

Interestingly, for the hyperbolic step-index profile with $\alpha=A=B^{1/2}$,
we can find an analytic expression for the scattering coefficients
(the mathematical derivation of these expressions is detailed below):
\begin{align}
R & =-\frac{1}{N_{+}}\frac{\Gamma\left(\textrm{i}\omega_{0}/\alpha\right)\Gamma\left(1-\textrm{i}\omega_{0}N_{+}/\alpha\right)}{\Gamma\left(\textrm{i}\omega_{0}\left(1-N_{+}\right)/2\alpha\right)\Gamma\left(1+\textrm{i}\omega_{0}\left(1-N_{+}\right)/2\alpha\right)};\tag{S3.23}\label{eq:S3.23}\\
T & =\frac{1}{N_{+}}\frac{\Gamma\left(\textrm{i}\omega_{0}/\alpha\right)\Gamma\left(1+\textrm{i}\omega_{0}N_{+}/\alpha\right)}{\Gamma\left(\textrm{i}\omega_{0}\left(1+N_{+}\right)/2\alpha\right)\Gamma\left(1+\textrm{i}\omega_{0}\left(1+N_{+}\right)/2\alpha\right)}.\tag{S3.24}\label{eq:S3.24}
\end{align}
Figure S3.10 demonstrates a perfect fitting between the above expressions
and the numerical results for a particular value of $\alpha$. In
this example, we can observe the non-reflecting behaviour of the hyperbolic
step-index profile when $\omega_{0}>10$ rad/s. In conclusion, we
can combine the theoretical tools provided by T-SUSY, SIP and Eqs.\,(\ref{eq:S3.23})
and (\ref{eq:S3.24}) to design broadband polarization-independent
optical isolators.\newpage{}

\paragraph{\emph{Derivation of Eqs.\,}(\ref{eq:S3.23}) \emph{and} (\ref{eq:S3.24}). }

\noindent Following a similar strategy as in \cite{key-9}, our first
goal is to analyse the asymptotic behaviour of the general solution
of the time-independent Schr\"{o}dinger equation when considering
the hyperbolic step potential:
\begin{equation}
V\left(x\right)=\frac{1}{2}V_{0}\left(1+\tanh\left(\frac{x}{2\widetilde{\alpha}}\right)\right),\tag{S3.25}\label{eq:S3.25}
\end{equation}
with $V_{0}>0$ and $\widetilde{\alpha}>0$. The general solution
to the equation $\psi^{\prime\prime}\left(x\right)+\left(E-V\left(x\right)\right)\psi\left(x\right)=0$
is given by Eq.\,(9) of \cite{key-9}, where $k=\sqrt{E}$ and $k^{\prime}=\sqrt{E-V_{0}}$.
Thus, using the asymptotic behaviour of the hypergeometric functions,
we find that {[}$\psi\left(x\right)\underset{x\rightarrow-\infty}{\sim}\psi_{-}\left(x\right)$
and $\psi\left(x\right)\underset{x\rightarrow\infty}{\sim}\psi_{+}\left(x\right)${]}:
\begin{align}
\psi_{-}\left(x\right) & =\left(C\Gamma_{1}\left(\mu,\nu\right)+D\Gamma_{2}\left(\mu,\nu\right)\right)\exp\left(\textrm{i}kx\right)+\left(C\Gamma_{3}\left(\mu,\nu\right)+D\Gamma_{4}\left(\mu,\nu\right)\right)\exp\left(-\textrm{i}kx\right)\nonumber \\
 & \equiv A\exp\left(\textrm{i}kx\right)+B\exp\left(-\textrm{i}kx\right);\tag{S3.26}\label{eq:S3.26}\\
\psi_{+}\left(x\right) & =C\exp\left(\textrm{i}k^{\prime}x\right)+D\exp\left(-\textrm{i}k^{\prime}x\right),\tag{S3.27}\label{eq:S3.27}
\end{align}
with $C$ and $D$ integration constants. The functions $\Gamma_{i}\left(\mu,\nu\right)$
can be found by identifying our Eq.\,(\ref{eq:S3.26}) with Eq.\,(11)
of \cite{key-9}.

Along these lines, it is important to note that the sign convention
employed in \cite{key-9} is the same as that in Cooper's tutorial
\cite{key-1}, and it is analogous to the sign convention employed
in the temporal scattering problem for the forward and backward plane
waves. In this way, one could expect that relabelling $k\rightarrow\omega_{0}$
and $k^{\prime}\rightarrow N_{+}\omega_{0}$ in the expressions of
$R$ and $T$ of \cite{key-9}, we would be able to obtain the temporal
scattering reflection and transmission coefficients. Unfortunately,
this procedure does not allow us to derive closed-form expressions
of $R$ and $T$ for the temporal scattering problem because the equations
connecting $R_{1}$ ($T_{1}$) and $R_{2}$ ($T_{2}$) in S-SUSY and
T-SUSY are not analogous. More specifically, note that the asymptotic
behaviours of the supersymmetric wave functions in the \emph{spatial
scattering problem} are S-SUSY-connected in \cite{key-1} as:
\begin{align}
\exp\left(\textrm{i}kx\right)+R_{1}\exp\left(-\textrm{i}kx\right) & =\xi\left(-\frac{\textrm{d}}{\textrm{d}x}+W_{-}\right)\left(\exp\left(\textrm{i}kx\right)+R_{2}\exp\left(-\textrm{i}kx\right)\right);\tag{S3.28}\label{eq:S3.28}\\
T_{1}\exp\left(\textrm{i}k^{\prime}x\right) & =\xi\left(-\frac{\textrm{d}}{\textrm{d}x}+W_{+}\right)T_{2}\exp\left(\textrm{i}k^{\prime}x\right).\tag{S3.29}\label{eq:S3.29}
\end{align}
In contrast, the asymptotic behaviours of the supersymmetric wave
functions in the\emph{ temporal scattering problem} are T-SUSY-connected
as {[}we reproduce Eqs.\,(\ref{eq:S2.10}) and (\ref{eq:S2.11})
for clarity{]}:
\begin{align}
\exp\left(\textrm{i}\omega_{0}t\right) & =\xi\left(-\frac{\textrm{d}}{\textrm{d}t}+W_{-}\right)\exp\left(\textrm{i}\omega_{0}t\right);\tag{S3.30}\label{eq:S3.30}\\
R_{1}\exp\left(-\textrm{i}N_{+}\omega_{0}t\right)+T_{1}\exp\left(\textrm{i}N_{+}\omega_{0}t\right) & =\xi\left(-\frac{\textrm{d}}{\textrm{d}t}+W_{+}\right)\left(R_{2}\exp\left(-\textrm{i}N_{+}\omega_{0}t\right)+T_{2}\exp\left(\textrm{i}N_{+}\omega_{0}t\right)\right).\tag{S3.31}\label{eq:S3.31}
\end{align}
If $R_{1}=R_{2}=0$, both systems are analogous and we can perform
the previous relabelling to find the temporal scattering reflection
and transmission coefficients. For instance, we calculated Eq.\,(\ref{eq:S3.17})
using this procedure. Nevertheless, $R_{1,2}\neq0$ in this case (see
Fig.\,S3.10). Consequently, we should recalculate the scattering
coefficients for the temporal problem from the beginning. 

Considering that $\psi_{-}\left(t\right)=\exp\left(\textrm{i}\omega_{0}t\right)$
and $\psi_{+}\left(t\right)=R\exp\left(-\textrm{i}N_{+}\omega_{0}t\right)+T\exp\left(\textrm{i}N_{+}\omega_{0}t\right)$,
we infer from Eqs.\,(\ref{eq:S3.26}) and (\ref{eq:S3.27}) that
$A=1$, $B=0$, $C=T$ and $D=R$. Now, using Eq.\,(14) of \cite{key-9},
we find that $C=t_{11}$ and $D=t_{21}$, where the transfer matrix
$t_{ij}$ is given by Eq.\,(16) of the aforementioned reference.
Therefore, performing the transformation of parameters $\alpha=1/2\widetilde{\alpha}$
to match the definition of the hyperbolic step potential of \cite{key-9}
with our definition of the hyperbolic step-index profile {[}Eq.\,(\ref{eq:S3.10})
taking $\alpha=A=B^{1/2}${]}, we finally obtain Eqs.\,(\ref{eq:S3.23})
and (\ref{eq:S3.24}).\newpage{}

\subsubsection*{Supersymmetric time-reversal modulations}

Interestingly, an \emph{even} superpotential $W\left(t\right)=W\left(-t\right)$
allows us to construct T-SUSY time-reversal refractive index profiles
of the form $n_{\textrm{T}2}\left(t\right)=n_{\textrm{T}1}\left(-t\right)$
with the same intensity scattering behaviour, certainly an unexpected
result taking into account that, in general, $n_{\textrm{T}1}\left(-t\right)$
has different scattering properties from those of the original modulation
$n_{\textrm{T}1}\left(t\right)$ (consider, e.g., the step-index case
\cite{key-10}). 

Let us analyse this scenario in more detail. To this end, consider
a given $W$ with even symmetry. In such a case, we can infer from
Riccati's equation that:
\begin{align}
V_{2}\left(t\right) & =W^{2}\left(t\right)+W^{\prime}\left(t\right)=W^{2}\left(-t\right)-W^{\prime}\left(-t\right)=V_{1}\left(-t\right),\tag{S3.32}\label{eq:S3.32}
\end{align}
and then, we find that $n_{\textrm{T}2}\left(t\right)=n_{\textrm{T}1}\left(-t\right)$
by setting $n_{2,-}=n_{1,-}$. Now, rewriting Eq.\,(\ref{eq:S3.32})
as $V_{2}\left(\mathbf{a}_{1}t\right)=V_{1}\left(\mathbf{a}_{2}t\right)$
and comparing this expression with the SIP condition {[}Eq.\,(\ref{eq:S2.18}){]},
we find that $M=0$ and the set of parameters $\mathbf{a}_{1}$ and
$\mathbf{a}_{2}$ are real numbers related as $a_{2}=-a_{1}=-1$,
i.e., we have a scaling SIP relation between superpartners. In this
way, T-SUSY and SIP allow us to find time-reversal refractive index
modulations with the same intensity scattering properties.
\noindent \begin{center}
\includegraphics[width=12cm,height=10cm,keepaspectratio]{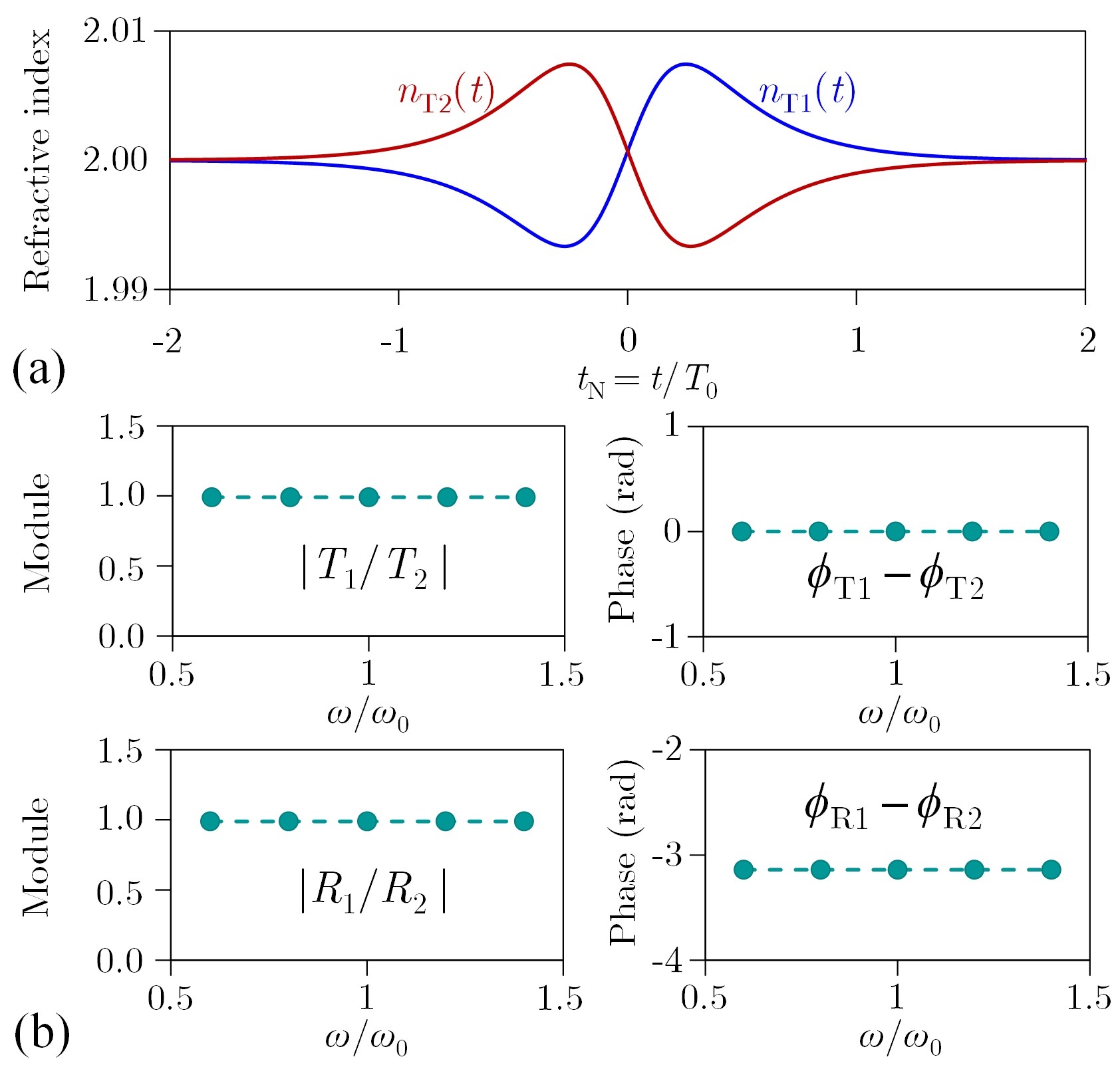}
\par\end{center}

\noindent \textbf{\small{}Figure S3.11.}{\small{} (a) SUSY time-reversal
refractive index profiles associated with the superpotential of Eq.\,(}\ref{eq:S3.33}{\small{})
and (b) ratios of the corresponding scattering coefficients $R_{1}/R_{2}$
and $T_{1}/T_{2}$ as a function of the ratio $\omega/\omega_{0}$.\\}{\small \par}

As an example, consider an even superpotential of the form:\footnote{This superpotential is a particular case of the hyperbolic Scarf II
superpotential $W\left(t\right)=A\tanh\left(\alpha t\right)+B\thinspace\textrm{sech}\left(\alpha t\right)$.}
\begin{equation}
W\left(t\right)=B\thinspace\textrm{sech}\left(\alpha t\right),\tag{S3.33}\label{eq:S3.33}
\end{equation}
where $B$ and $\alpha$ are real parameters. The SUSY refractive
index profiles $n_{\textrm{T}1,2}\left(t\right)$ can be calculated
by combining Riccati's equation and Eq.\,(\ref{eq:S2.2}) bearing
in mind that $\Omega=\omega_{0}^{2}+W_{-}^{2}=\omega_{0}^{2}$. The
corresponding profiles are shown in Fig.\,S3.11(a), and the behaviour
of the scattering coefficients as a function of the frequency {[}numerically
calculated from the wave equation Eq.\,(\ref{eq:S1.8}){]} is depicted
in Fig.\,S3.11(b). Outstandingly, we can see that both time-reversal
refractive index profiles have the same intensity scattering properties,
not only at $\omega=\omega_{0}$, but also at $\omega\neq\omega_{0}$.
In other words, the ratios $R_{1}/R_{2}$ and $T_{1}/T_{2}$ have
a flat frequency response. In any case, T-SUSY only provides control
at the design frequency.\footnote{We analysed other even superpotentials and we found that $R_{1}/R_{2}$
and $T_{1}/T_{2}$ also presented a flat frequency response. Nevertheless,
we cannot extrapolate this singular feature as a general rule in SUSY
time-reversal modulations. At least, the demonstration is not straightforward.}

\newpage{}

\section{Temporal waveguides: theory\label{sec:4}}

Here, we first detail the theory of unbroken SUSY transformations
in temporal waveguides (TWGs) and, secondly, we derive a coupled-mode
theory (CMT) for serial TWGs moving with the same speed and direction
in a given spatial waveguide.

\subsection{T-SUSY in temporal waveguides}

The unbroken SUSY relation between two quantum-mechanical superpartners
$V_{1,2}$ is given by the expression (we set $\hbar^{2}/2m\equiv1$
for simplicity) \cite{key-1}:
\begin{equation}
V_{2}\left(x\right)=V_{1}\left(x\right)-2\frac{\textrm{d}^{2}}{\textrm{d}x^{2}}\ln\psi_{0}^{\left(1\right)}\left(x\right),\tag{S4.1}\label{eq:S4.1}
\end{equation}
where $\psi_{0}^{\left(1\right)}$ is the ground state of $V_{1}$.
As detailed in the main text, we can identify an effective potential
in Eq.\,(7) of the form:
\begin{equation}
V_{i}\left(\tau\right)\equiv\frac{2}{\beta_{2}}\beta_{\textrm{B}i}\left(\tau\right).\tag{S4.2}\label{eq:S4.2}
\end{equation}
Hence, combining Eqs.\,(\ref{eq:S4.1}) and (\ref{eq:S4.2}) we infer
that:
\begin{equation}
\beta_{\textrm{B}2}\left(\tau\right)=\beta_{\textrm{B}1}\left(\tau\right)-\beta_{2}\frac{\textrm{d}^{2}}{\textrm{d}\tau^{2}}\ln\psi_{0}^{\left(1\right)}\left(\tau\right),\tag{S4.3}\label{eq:S4.3}
\end{equation}
where $\psi_{0}^{\left(1\right)}$ is the ground state (fundamental
mode) of $\beta_{\textrm{B}1}$ in this case.

\subsection{Coupled-mode theory for serial temporal waveguides\label{subsec:4.2}}

Consider two serial TWGs $a$ and $b$ constructed from two different
temporal perturbations $\beta_{\textrm{B},a}\left(t-z/v_{\textrm{B}}\right)$
and $\beta_{\textrm{B},b}\left(t-z/v_{\textrm{B}}\right)$ of temporal
width $2T_{\textrm{B},a}$ and $2T_{\textrm{B},b}$. Furthermore,
both TWGs are moving with the same speed $v_{\textrm{B}}$ through
the longitudinal $z$-axis of a given spatial waveguide and are separated
$T_{ab}$ in time and $v_{\textrm{B}}T_{ab}$ in space. Figure\textcolor{black}{{}
S4.1} illustrates this scenario.
\noindent \begin{center}
\includegraphics[width=9cm,height=6cm,keepaspectratio]{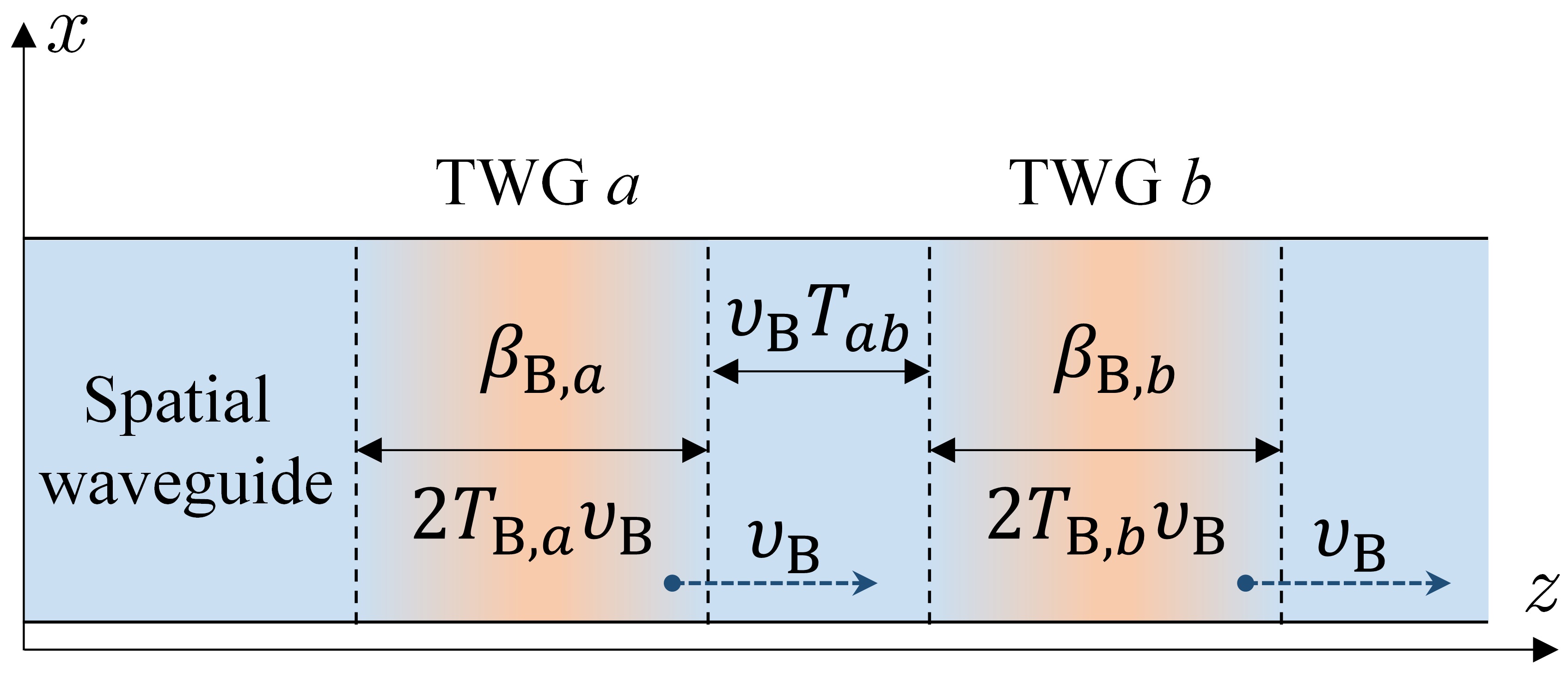}
\par\end{center}

\noindent \textbf{\small{}Figure S4.1.}{\small{} Serial temporal waveguides
(TWGs) moving with the same speed $v_{\textrm{B}}$ through the longitudinal
axis of a given spatial waveguide (blue area).\\}{\small \par}

For the sake of simplicity, let us first assume that both TWGs are
operating in the single-mode regime. The fundamental mode of each
TWG will be denoted with the subscript $a$ or $b$. In such a scenario,
the complex envelope of the global electric field of this optical
structure can be approximated using perturbation theory as ($\tau:=t-z/v_{\textrm{B}}$):
\begin{equation}
A\left(z,\tau\right)\simeq\sum_{m=a,b}\mathscr{A}_{m}\left(z\right)\psi_{m}\left(\tau\right)\exp\left(\textrm{i}\frac{\Delta\beta_{1}}{\beta_{2}}\tau\right)\exp\left(\textrm{i}K_{m}z\right),\tag{S4.4}\label{eq:S4.4}
\end{equation}
where $\mathscr{A}_{m}$ is the complex amplitude of each mode. In
isolated conditions (i.e., when each TWG is uncoupled from the other
one), we find that $\textrm{d}\mathscr{A}_{m}/\textrm{d}z=0$ and
$\psi_{m}$ must fulfil the temporal Helmholtz equation {[}Eq.\,(7)
of the paper{]}:
\begin{equation}
\left[\frac{\textrm{d}^{2}}{\textrm{d}\tau^{2}}+2\frac{K_{m}}{\beta_{2}}+\left(\frac{\Delta\beta_{1}}{\beta_{2}}\right)^{2}-2\frac{\beta_{\textrm{B},m}\left(\tau\right)}{\beta_{2}}\right]\psi_{m}\left(\tau\right)=0.\tag{S4.5}\label{eq:S4.5}
\end{equation}
Nonetheless, if the two TWGs are in close proximity (serially coupled),
the longitudinal dependence of $\mathscr{A}_{m}$ accounts for the
power exchange between modes,\footnote{The mode-coupling between temporal modes of serial TWGs takes place
through their evanescent tails.} $\psi_{m}$ also fulfils Eq.\,(\ref{eq:S4.5}), and the complex
envelope $A$ given by Eq.\,(\ref{eq:S4.4}) must satisfy the time-domain
equation \cite{key-11}:
\begin{equation}
\left(\frac{\partial}{\partial z}+\Delta\beta_{1}\frac{\partial}{\partial\tau}+\textrm{i}\frac{1}{2}\beta_{2}\frac{\partial^{2}}{\partial\tau^{2}}-\textrm{i}\beta_{\textrm{B}}\left(\tau\right)\right)A\left(z,\tau\right)=0,\tag{S4.6}\label{eq:S4.6}
\end{equation}
where $\beta_{\textrm{B}}\left(\tau\right)=\beta_{\textrm{B},a}\left(\tau\right)+\beta_{\textrm{B},b}\left(\tau\right)$.
Note that the above equation plays the same role as the wave equation
in optical couplers based on parallel spatial waveguides. Hence, substituting
Eq.\,(\ref{eq:S4.4}) into Eq.\,(\ref{eq:S4.6}) and using Eq.\,(\ref{eq:S4.5}),
we find after some algebra (we omit the independent variables for
simplicity):
\begin{equation}
\sum_{m=a,b}\frac{\textrm{d}\mathscr{A}_{m}}{\textrm{d}z}\psi_{m}\exp\left(\textrm{i}K_{m}z\right)-\textrm{i}\left(\beta_{\textrm{B}}-\beta_{\textrm{B},m}\right)\mathscr{A}_{m}\psi_{m}\exp\left(\textrm{i}K_{m}z\right)=0.\tag{S4.7}\label{eq:S4.7}
\end{equation}
From the above equation, we can find the coupled-mode equations describing
the power exchange between modes of both serial TWGs. For instance,
the coupled-mode equation governing the mode-coupling from mode $b$
to mode $a$ is found: (\emph{i}) multiplying Eq.\,(\ref{eq:S4.7})
by $\psi_{a}\exp\left(-\textrm{i}K_{a}z\right)$, (\emph{ii}) integrating
in $\tau\in\left(-\infty,\infty\right)$, and (\emph{iii}) writing
the first-order derivative of $\mathscr{A}_{a(b)}$ at the left-hand
side. In this way, we obtain:
\begin{equation}
\frac{\textrm{d}\mathscr{A}_{a}}{\textrm{d}z}=\textrm{i}c_{a}\mathscr{A}_{a}+\exp\left(\textrm{i}\Delta K_{b,a}z\right)\left(\textrm{i}\kappa_{a,b}-\chi_{a,b}\frac{\textrm{d}}{\textrm{d}z}\right)\mathscr{A}_{b},\tag{S4.8}\label{eq:S4.8}
\end{equation}
where $\Delta K_{b,a}:=K_{b}-K_{a}$. A similar coupled-mode equation
describing the coupled power from mode\,$a$ to mode $b$ can be
found by exchanging the subscripts in the above equation. The mode-coupling
coefficients (MCCs), accounting for the mode overlapping between $\psi_{a}$
and $\psi_{b}$, are\,defined\,as: 
\begin{align}
\chi_{a,b} & :=\frac{1}{N_{a}}\int_{-\infty}^{\infty}\psi_{b}\left(\tau\right)\psi_{a}\left(\tau\right)\textrm{d}\tau;\tag{S4.9}\label{eq:S4.9}\\
c_{a} & :=\frac{1}{N_{a}}\int_{-\infty}^{\infty}\beta_{\textrm{B},b}\left(\tau\right)\psi_{a}^{2}\left(\tau\right)\textrm{d}\tau=\frac{1}{N_{a}}\int_{\left\langle 2T_{\textrm{B},b}\right\rangle }\beta_{\textrm{B},b}\left(\tau\right)\psi_{a}^{2}\left(\tau\right)\textrm{d}\tau;\tag{S4.10}\label{eq:S4.10}\\
\kappa_{a,b} & :=\frac{1}{N_{a}}\int_{-\infty}^{\infty}\beta_{\textrm{B},a}\left(\tau\right)\psi_{b}\left(\tau\right)\psi_{a}\left(\tau\right)\textrm{d}\tau=\frac{1}{N_{a}}\int_{\left\langle 2T_{\textrm{B},a}\right\rangle }\beta_{\textrm{B},a}\left(\tau\right)\psi_{b}\left(\tau\right)\psi_{a}\left(\tau\right)\textrm{d}\tau,\tag{S4.11}\label{eq:S4.11}
\end{align}
with $N_{a}:=\int_{-\infty}^{\infty}\psi_{a}^{2}\left(\tau\right)\textrm{d}\tau$. 

It is worthy to note the complete analogy between the CMT of parallel
spatial waveguides (see e.g. Eq.\,(4.11) in \cite{key-12} or Eq.\,(5)
in \cite{key-13}) and the CMT of serial TWGs {[}Eq.\,(\ref{eq:S4.8}){]}.
In these references, it is demonstrated that $\chi_{a,b}$ can be
neglected. However, in serially-coupled TWGs, the MCC $\chi_{a,b}$
is generally higher than the MCCs $c_{a}$ and $\kappa_{a,b}$ and,
therefore, all the MCCs should be retained to guarantee a complete
description of the mode-coupling phenomenon. Along this line, note
that Eq.\,(\ref{eq:S4.8}) can be rewritten as:
\begin{equation}
\frac{\textrm{d}\mathscr{A}_{a}\left(z\right)}{\textrm{d}z}=\textrm{i}c_{a}^{\left(\textrm{eq}\right)}\mathscr{A}_{a}\left(z\right)+\textrm{i}\kappa_{a,b}^{\left(\textrm{eq}\right)}\exp\left(\textrm{i}\Delta K_{b,a}z\right)\mathscr{A}_{b}\left(z\right),\tag{S4.12}\label{eq:S4.12}
\end{equation}
where $c_{a}^{\left(\textrm{eq}\right)}:=(c_{a}-\chi_{a,b}\kappa_{b,a})/(1-\chi_{a,b}\chi_{b,a})$
and $\kappa_{a,b}^{\left(\textrm{eq}\right)}:=(\kappa_{a,b}-\chi_{a,b}c_{b})/(1-\chi_{a,b}\chi_{b,a})$.
A similar equation for $\textrm{d}\mathscr{A}_{b}\left(z\right)/\textrm{d}z$
can be obtained by exchanging the subscripts $a$ and $b$ in Eq.\,(\ref{eq:S4.12}).\pagebreak{}

\noindent Finally, for completeness, the following considerations
are in order:
\begin{itemize}
\item If we solve the CMT assuming that only the mode $a$ is excited at
$z=0$ {[}$\mathscr{A}_{b}\left(0\right)=0${]}, we find that:
\begin{equation}
\mathscr{A}_{b}\left(z\right)=\textrm{i}\frac{\kappa_{b,a}^{\left(\textrm{eq}\right)}}{\eta}\exp\left(-\textrm{i}\frac{\Delta K_{b,a}}{2}z\right)\sin\left(\eta z\right)\mathscr{A}_{a}\left(0\right),\tag{S4.13}\label{eq:S4.13}
\end{equation}
where:
\begin{equation}
\eta=\sqrt{\kappa_{a,b}^{\left(\textrm{eq}\right)}\kappa_{b,a}^{\left(\textrm{eq}\right)}+\frac{\left(\Delta K_{b,a}+c_{b}^{\left(\textrm{eq}\right)}-c_{a}^{\left(\textrm{eq}\right)}\right)^{2}}{4}}.\tag{S4.14}\label{eq:S4.14}
\end{equation}
Bearing in mind that $\psi_{0}^{\left(2\right)}$ and $\psi_{1}^{\left(1\right)}$
are degenerate modes in the TPL shown in Fig.\,4(c) of the paper
($\Delta K_{b,a}=0$), we conclude from Eqs.\,(\ref{eq:S4.13}) and
(\ref{eq:S4.14}) that they must exchange their optical power periodically
along the $z$-axis with a coupling length\footnote{The coupling length $L_{\textrm{C}}$ is the length that maximizes
the sinusoidal term of Eq.\,(\ref{eq:S4.13}).} $L_{\textrm{C}}=\pi/2\eta$ and a coupling efficiency $\left|\mathscr{A}_{b}\left(z=L_{\textrm{C}}\right)/\mathscr{A}_{a}\left(0\right)\right|^{2}=\left|\kappa_{b,a}^{\left(\textrm{eq}\right)}/\eta\right|^{2}$.
\item In the multi-mode regime, the modes $\left\{ a_{n}\right\} _{n=1}^{N}$
of TWG $a$ exchange optical power with the modes $\left\{ b_{n}\right\} _{n=1}^{N}$
of TWG $b$. In order to describe this situation, Eq.\,(\ref{eq:S4.4})
must be restated as:
\begin{equation}
A\left(z,\tau\right)\simeq\sum_{m=a,b}\sum_{n=1}^{N}\mathscr{A}_{mn}\left(z\right)\psi_{mn}\left(\tau\right)\exp\left(\textrm{i}\frac{\Delta\beta_{1}}{\beta_{2}}\tau\right)\exp\left(\textrm{i}K_{mn}z\right),\tag{S4.15}\label{eq:S4.15}
\end{equation}
and, consequently, the mode-coupling from the modes $\left\{ b_{n}\right\} _{n=1}^{N}$
to mode $a_{i}$ is governed by the coupled-mode equation:
\begin{align}
\frac{\textrm{d}\mathscr{A}_{ai}\left(z\right)}{\textrm{d}z} & =\textrm{i}c_{ai}\mathscr{A}_{ai}\left(z\right)+\sum_{n=1}^{N}\exp\left(\textrm{i}\Delta K_{bn,ai}z\right)\left(\textrm{i}\kappa_{ai,bn}-\chi_{ai,bn}\frac{\textrm{d}}{\textrm{d}z}\right)\mathscr{A}_{bn}\left(z\right).\tag{S4.16}\label{eq:S4.16}
\end{align}
We cannot observe internal mode-coupling among the modes of a given
TWG $m$ if we assume that its temporal perturbation profile $\beta_{\textrm{B},m}\left(\tau\right)$
is invariant during the propagation of the TWG along the longitudinal
axis of the spatial waveguide.\footnote{An internal mode-coupling among the modes of a TWG requires a temporal
perturbation with a varying shape along the $z$-axis. This can be
modelled by a profile of the form $\beta_{\textrm{B},m}\left(\tau;z\right)$.
The $\tau$ variable describes the ideal temporal profile of the TWG,
and the $z$ variable accounts for the fluctuation of its shape during
the propagation of the TWG along the longitudinal axis of the spatial
waveguide. This is analogous to an optical fibre with a refractive
index profile $n(r;z)$ which fluctuates along the $z$-axis due to
manufacturing imperfections, which induce mode-coupling between different
fibre modes.} Nevertheless, in practice, the temporal profile $\beta_{\textrm{B},m}$
may experience dispersion along the $z$-axis. The precise longitudinal
evolution of $\beta_{\textrm{B},m}$ depends on the exact physical
mechanism used to generate the temporal perturbation in the spatial
waveguide. For example, using the XPM with a pump-probe set-up, the
shape of the pump pulse will generally be affected by dispersion during
propagation \cite{key-11}. In our case, we could overcome this drawback
by selecting the pump wavelength at the zero-dispersion wavelength
of the spatial waveguide, provided that the higher-order dispersion
terms are negligible. This scenario requires to use, e.g., microstructured
optical fibres to tailor the material dispersion properties of the
spatial waveguide.
\item In serial TWGs with different speed or different propagation directions,
the MCCs are found to be space-dependent. Nevertheless, this scenario
is out of the scope of this work.
\end{itemize}
\newpage{}

\section{Temporal waveguides: numerical analysis\label{sec:5}}
\noindent \begin{center}
\includegraphics[width=15cm,height=12cm,keepaspectratio]{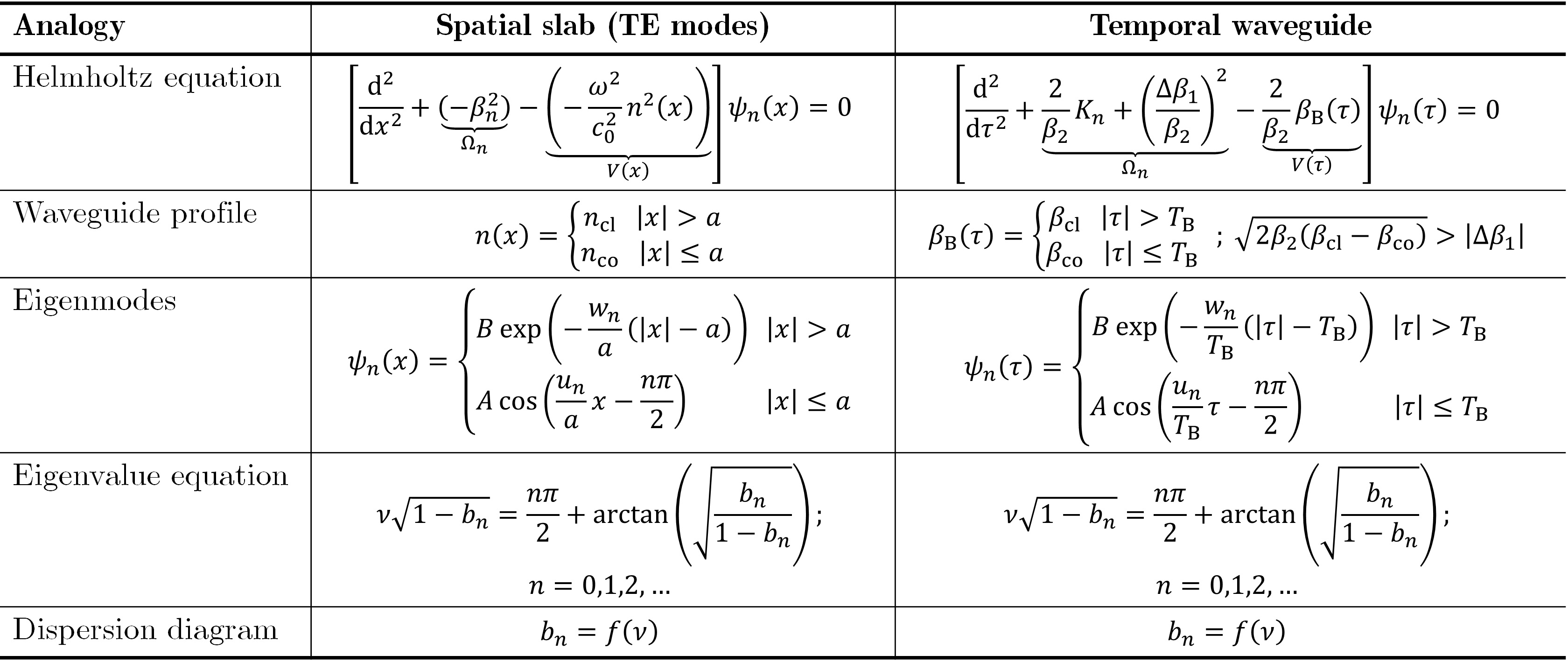}
\par\end{center}

\noindent \textbf{\small{}Table S1. }{\small{}Analogy between a step-index
dielectric slab waveguide \cite{key-12} and a step-index temporal
waveguide \cite{key-11}. The constants $A$ and $B$ of the eigenmodes
$\psi_{n}$ fulfil the relation $B=A\cos\left(u_{n}-n\pi/2\right)\textrm{sign}\left(\tau\right)^{n}$,
which arises from the continuity boundary condition of the eigenmodes.\\}{\small \par}

Table S1 summarizes the analogy between a step-index spatial dielectric
slab waveguide and a step-index TWG, reported in \cite{key-11}. Concretely,
the analogy only applies to the transversal electric (TE) modes of
the slab. As seen, the spatial and temporal Helmholtz equations are
analogous with an effective potential $V$ and eigenvalue $\Omega_{n}$
correspondence of the form indicated in Table S1. In addition, note
that the modal analysis of both structures involves the same eigenvalue
equation, where $b_{n}$ and $\nu$ are respectively the normalized
phase constant and normalized frequency, given by the expressions:
\begin{align}
\nu^{2} & =u^{2}+w^{2}=\frac{\omega^{2}}{c_{0}^{2}}a^{2}\left(n_{\textrm{co}}^{2}-n_{\textrm{cl}}^{2}\right)\equiv\frac{2}{\beta_{2}}T_{\textrm{B}}^{2}\left(\beta_{\textrm{cl}}-\beta_{\textrm{co}}\right);\tag{S5.1}\label{eq:S5.1}\\
u_{n} & =\nu\sqrt{1-b_{n}};\ \ \ \ \ w_{n}=\nu\sqrt{b_{n}};\tag{S5.2}\label{eq:S5.2}\\
b_{n} & =\frac{\beta_{n}^{2}/\left(\omega^{2}/c_{0}^{2}\right)-n_{\textrm{cl}}^{2}}{n_{\textrm{co}}^{2}-n_{\textrm{cl}}^{2}}\equiv\frac{\beta_{\textrm{cl}}-K_{n}-\left(\beta_{2}/2\right)\left(\Delta\beta_{1}/\beta_{2}\right)^{2}}{\beta_{\textrm{cl}}-\beta_{\textrm{co}}}.\tag{S5.3}\label{eq:S5.3}
\end{align}
 In the gradual-index case, the analogy is only preserved if the slowly-varying
condition of $n\left(x\right)$ is satisfied.\footnote{The slowly-varying condition of $n\left(x\right)$ requires that $\left|\delta_{x}n\right|\ll\left|n\left(x\right)\right|$
in $\delta x\sim\lambda_{0}/\overline{n}$, where $\delta_{x}n:=n\left(x+\delta x\right)-n\left(x\right)$,
$\lambda_{0}$ is the wavelength in vacuum and $\overline{n}$ is
the average value of $n\left(x\right)$ in $\delta x=2a$.} Otherwise, the TE solutions of the spatial slab, calculated from
Maxwell's equations, do not obey the spatial Helmholtz equation depicted
in Table S1 and, therefore, the analogy is broken.

Hence, in step- and gradual-index TWGs, the eigenfunctions $\psi_{n}$
and eigenvalues $\Omega_{n}$ of the temporal Helmholtz equation can
be calculated from the spatial Helmholtz equation of Table S1, provided
that the slowly-varying condition of the analogous $n\left(x\right)$
profile is satisfied (which is the case in all TWGs analysed in this
work). It is important to note that the analogy should be established
by selecting values of $\omega$ and $a$ that guarantee the same
normalized frequency in the spatial slab and in the TWG {[}Eq.\,(\ref{eq:S5.1}){]}.
In this work, we have solved the spatial Helmholtz equation of the
analogous slab with CST Microwave Studio. This software allows us
to calculate numerically the TE modes of interest directly from Maxwell's
equations, which coincide with those obtained from the spatial Helmholtz
equation for slowly-varying index profiles $n\left(x\right)$, as
mentioned above. Once we calculated the normalized dispersion diagram
$b_{n}=f\left(\nu\right)$ of the TE modes with CST, we verified in
MATLAB that the corresponding eigenfunctions and eigenvalues fulfil
the temporal Helmholtz equation of the TWG under analysis.

\subsubsection*{Temporal bound states of SUSY TWGs}

To complete the information provided by Fig.\,4(b) of the paper,
Fig.\,S5.1 shows the spatio-temporal profile of the temporal bound
states $\psi_{n}^{\left(1,2\right)}$ supported by both TWGs. As seen
in Fig.\,4(b) and Fig.\,S5.1, $\psi_{0}^{\left(1\right)}$ has no
SUSY counterpart in the eigenvalue spectrum of $\beta_{\textrm{B}2}$,
i.e., $\psi_{0}^{\left(1\right)}$ is not phase-matched with any temporal
bound state $\psi_{n}^{\left(2\right)}$. Contrariwise, the temporal
bound states $\psi_{n+1}^{\left(1\right)}$ are perfectly phase-matched
with the temporal bound states $\psi_{n}^{\left(2\right)}$.
\noindent \begin{center}
\includegraphics[width=13.5cm,height=12cm,keepaspectratio]{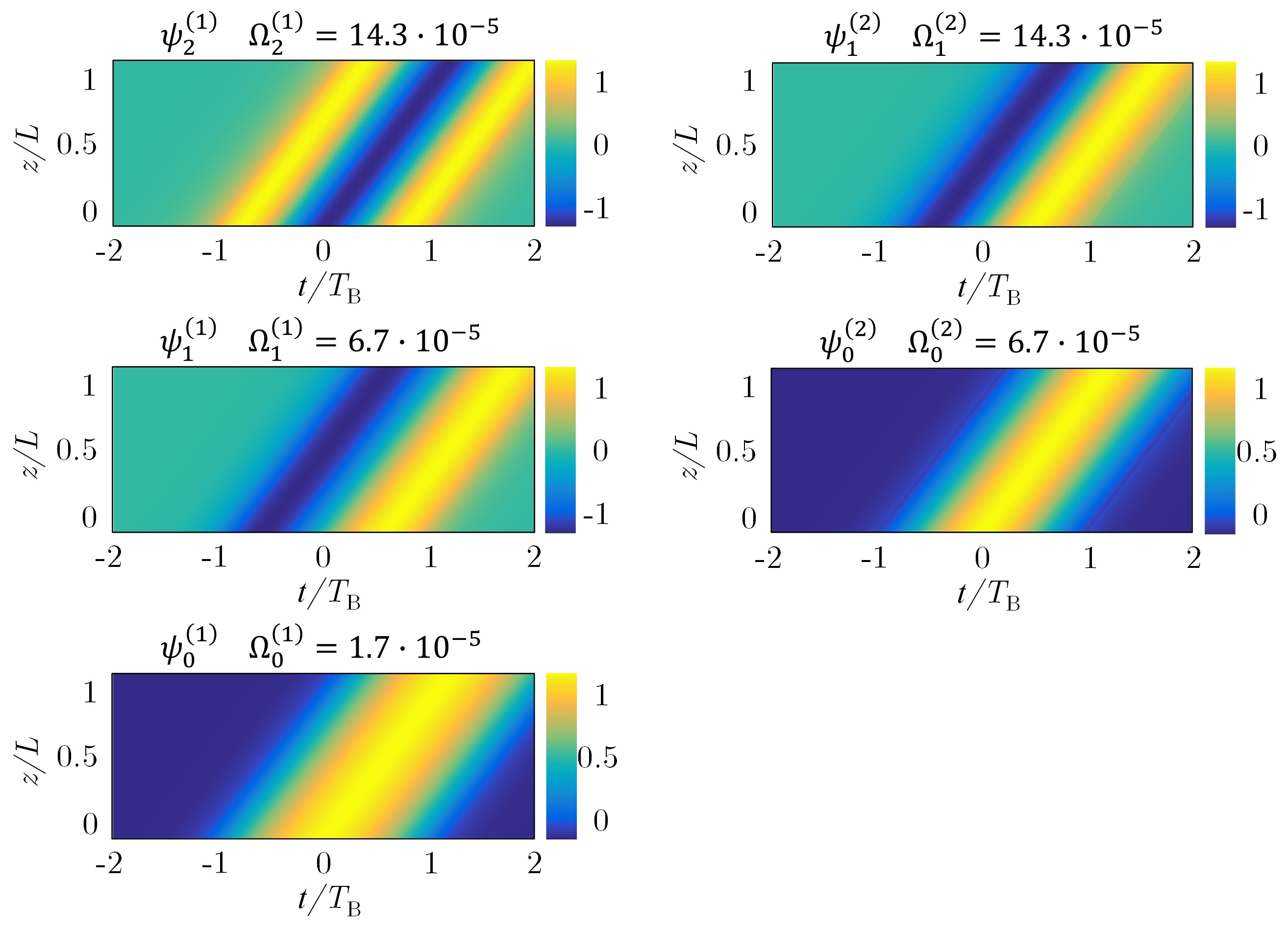}
\par\end{center}

\noindent \textbf{\small{}Figure S5.1.}{\small{} Spatio-temporal profile
$\psi_{n}^{\left(1,2\right)}(t-z/v_{\textrm{B}})$ and corresponding
eigenvalue $\Omega_{n}^{\left(1,2\right)}=2K_{n}^{\left(1,2\right)}/\beta_{2}+\Delta\beta_{1}^{2}/\beta_{2}^{2}$
in Eq.\,(7) of the paper. The temporal and spatial axes are normalized
as $t/T_{\textrm{B}}$ and $z/L$, where $L$ is the length of the
spatial waveguide over which the TWGs propagate. Note that $v_{\textrm{B}}$
and $L$ are arbitrary parameters in the numerical simulation. (Colorbar:
normalized amplitude).}{\small \par}

\subsubsection*{Note on the numerical analysis of the temporal photonic lantern using
the CMT}

The numerical simulation shown in Fig.\,4(d) has been performed in
MATLAB by using the CMT derived in Subsection \ref{subsec:4.2}. Aimed
to facilitate the comprehension of the results, we have taken $L=L_{\textrm{C}}$
keeping in mind that $L$ (the length of the spatial waveguide over
which the TPL propagates) is an arbitrary parameter. Additionally,
the parameter $L_{\textrm{C}}$ and the coupling efficiency of the
mode conversion can be calculated as indicated on page \pageref{eq:S4.14}.
In particular, the coupling efficiency is $0.96$.

\newpage{}

\section{The temporal acoustic Helmholtz equation\label{sec:6}}

In this section, a temporal Helmholtz equation formally equal to Eq.\,(3)
of the main text is derived for acoustic systems characterized by
space- and time-varying properties. Particularly, in the case of pressure
acoustics, the spatiotemporal evolution of the acoustic pressure $p(\mathbf{r},t)$
is governed by the following wave equation \cite{key-14}:

\begin{align}
-\frac{\partial^{2}p(\mathbf{r},t)}{\partial t^{2}}+B(\mathbf{r},t)\nabla\cdot\left(\rho^{-1}(\mathbf{r},t)\nabla p(\mathbf{r},t)\right) & =0,\tag{S6.1}\label{eq:S6.1}
\end{align}
where the medium bulk modulus $B(\mathbf{r},t)$ and mass density
$\rho(\mathbf{r},t)$ are, in general, functions of space and time.
Taking a medium for which $\rho$ depends only on time (a similar
result could be obtained for a medium with a slowly-varying spatial
dependence of $\rho$), it is possible to rewrite Eq.\,(\ref{eq:S6.1})
as: 
\begin{align}
-\frac{\partial^{2}p(\mathbf{r},t)}{\partial t^{2}}+\frac{B(\mathbf{r},t)}{\rho(t)}\triangle p(\mathbf{r},t) & =0.\tag{S6.2}\label{eq:S6.2}
\end{align}
Furthermore, if the bulk modulus can be expressed as $B(\mathbf{r},t)=B_{\textrm{S}}(\mathbf{r})B_{\textrm{T}}(t)$,
and applying separation of variables to the pressure as $p\left(\mathbf{r},t\right)=p_{\textrm{S}}\left(\mathbf{r}\right)p_{\textrm{T}}\left(t\right)$,
the previous equation can be recast as:
\begin{align}
\frac{\rho(t)}{B_{\textrm{T}}(t)}\frac{\ddot{p}_{\textrm{T}}\left(t\right)}{p_{\textrm{T}}(t)} & =B_{\textrm{S}}(\mathbf{r})\frac{\triangle p_{\textrm{S}}(\mathbf{r})}{p_{\textrm{S}}(\mathbf{r})}.\tag{S6.3}\label{eq:S6.3}
\end{align}
Once again, this is satisfied if and only if both sides of the equation
are equal to a constant. In analogy with the electromagnetic case,
defining $n_{\textrm{T}}^{2}(t):=\rho(t)/B_{\textrm{T}}(t)$ and assuming
that $n_{-}:=n_{\textrm{T}}\left(t\to-\infty\right)$ is also a constant,
the following temporal Helmholtz equation is readily obtained:
\begin{align}
\left(\frac{\textrm{d}^{2}}{\textrm{d}t^{2}}+\omega_{0}^{2}\frac{n_{-}^{2}}{n_{{\rm T}}^{2}\left(t\right)}\right)p_{\textrm{T}}\left(t\right) & =0,\tag{S6.4}\label{eq:S6.4}
\end{align}
that is, Eq.\,(3) of the main text.

\newpage{}

\end{document}